\documentclass[final]{siamltex} 
\usepackage{amsfonts}
\usepackage{amssymb,amsmath}
\usepackage{graphicx}
\usepackage{mymacros}
\usepackage{multirow}
\usepackage{booktabs}

\begin{document}

\title{Iterative bounds on effective transport for advection 
diffusion in periodic flow fields}

\author{N. B. Murphy, D. Hallman, E. Cherkaev, J. Xin, and K. M. Golden}
\date{}
\maketitle
\begin{abstract}
Over three decades ago a Stieltjes integral representation for the effective
diffusivity of a tracer in a steady fluid velocity field was developed,
involving the spectral measure of a compact self-adjoint operator and the
P{\'e}clet number of the flow. Rigorous bounds on the homogenized diffusivity
could then be obtained from knowledge of the moments of the spectral measure. A
recent extension to space-time periodic flows involves an unbounded self-adjoint
operator. Though Pad{\'e} approximants provide upper and lower bounds in terms
of the moments, the lack of a general method for calculating them has
significantly limited the utility of this approach. Here we develop an iterative
method that enables an arbitrary number of moments, hence bounds, to be
calculated analytically in closed form for spatially and space-time periodic
flows. The known behavior of the effective diffusivity for a 2D steady cellular
flow is accurately captured by high order upper and lower bounds. The bounds
extend to 3D steady and time periodic flow fields away from the advection
dominated regime where an open issue remains concerning the divergence of the
bounds. 
\end{abstract}
\bigskip

\section{Introduction}
The long-time large-scale behavior of transport of a passive tracer by an
incompressible fluid velocity field has been known for over a century to be
well-approximated by a diffusion process \cite{Taylor:1915:523} via an effective
diffusivity matrix $\Dg^*$  
\cite{Taylor:PRSL:196}. 
The effective diffusivity in general
depends on the local diffusivity of the fluid and the characteristics
of the velocity field such as its strength and geometry.
Estimating and computing $\Dg^*$ 
is an  
example of \textit{homogenization}, 
an interdisciplinary area of applied mathematics with impacts throughout
science, engineering, medicine, and industry.
Problems in homogenization \cite{Bensoussan:Book:1978}, such as the effective
electrical or thermal conductivity of a host medium 
with spheres inside, 
or the effective viscosity of a fluid with spherical particles,
have a long and storied history, with scientific giants such as 
Einstein and Maxwell having worked on them 
in the dilute limit.

A major advance in the theory of homogenization was the
introduction of variational methods to derive 
rigorous bounds on the homogenized coefficients for 
multiphase composite materials, given knowledge
of the coefficients of the constituents, their
relative volume fractions, and isotropy
of the composite microstructure \cite{Hashin:JAP-3125}. For two component media,
Hashin and Shtrikman \cite{Hashin:JAP-3125} showed that their bounds are optimal
for electrical/thermal 
conductivity, dielectric constant, 
magnetic permeability and diffusivity, and are attainable 
by coated sphere geometries
\cite{Hashin:JAP-3125,MILTON:2002:TC,Torquato:RHM-02}.
Other microstructures such as hierarchical laminates
can also 
attain the bounds 
\cite{MILTON:2002:TC,Torquato:RHM-02}.

When we consider a composite material interacting not
with a static applied field, but with a wave in the 
long wavelength or quasistatic limit, however, then the parameters
describing the medium and its response are complex,
such as the complex permittivity or complex viscoelasticity. 
The real bounds \cite{Hashin:JAP-3125}
no longer apply.
Extending rigorous bounds on homogenized
coefficients of composite media to the complex case 
led to another major advance. Independently,
Bergman \cite{Bergman:PRL-1285} and 
Milton \cite{Milton:APL-300} introduced a
representation formula for effective parameters 
which separated the component parameters from the 
microstructural geometry, and does not rely on a variational principle.
They obtained the first bounds on 
complex parameters for two phase composites,
finding complex versions of the 
bounds in \cite{Hashin:JAP-3125}, as well as simpler bounds that only assume
knowledge of the component parameters and the relative volume fractions
\cite{Bergman:PRL-1285,Milton:APL-300,MILTON:2002:TC,Torquato:RHM-02}.
Subsequently, using functional analysis and linear operator theory, Golden and
Papanicolaou \cite{Golden:CMP-473} laid the mathematical foundations of the
\textit{analytic continuation method}. They proved that the effective parameters
can be represented as Stieltjes integrals involving the spectral measure of a
self-adjoint operator which contains all the geometrical information about the
microstructure, and established a rigorous extremal framework for the bounding
procedure. In further work they obtained the first rigorous bounds on complex
parameters for multiphase materials using techniques of several complex
variables \cite{Golden:JSP-655,Golden:JMPS-333}, and developed an iterative
method for incorporating more and more geometrical information into tighter and
tighter bounds in the two component case \cite{Golden:JMPS-333,Milton:TC-571}.
We note that since the seminal findings in \cite{Hashin:JAP-3125}, 
the principal impact
of the works outlined above in applications 
have been the bounds on effective properties. Prior to these rigorous
results, scientists and engineers relied primarily on 
trial and error in the lab to figure out what the best possible
values for effective properties could be and what types of 
composite geometries could achieve these extremal properties. 

The Stieltjes framework in \cite{Golden:CMP-473}
was later adapted by Avellaneda and Majda to 
the effective diffusivity of a passive scalar in
an incompressible fluid flow
\cite{Avellaneda:PRL-753,Avellaneda:CMP-339}.
They systematically exploited analogies with the
theory of composite materials to guide their approach and results
\cite{Avellaneda:CMP-339}, and
introduced a Stieltjes integral representation 
for $\Dg^*$ in steady flows, via a spectral measure $\mu$ 
of a compact self-adjoint operator 
\cite{Avellaneda:PRL-753,Avellaneda:CMP-339},
which was further developed in \cite{Murphy:ADSRF-2019}. Recently this 
result was extended to space-time periodic flows, via a spectral 
measure for an unbounded self-adjoint operator \cite{Murphy:ADSTPF-2017}. The 
integral representation \emph{separates} the P{\'e}clet number from the 
geometry and dynamics of the fluid flow, which is encoded into the
spectral measure through its 
moments.

In \cite{Avellaneda:CMP-339} is   
an analog of the
Hashin-Shtrikman upper bound for velocity fields with isotropic statistics,
obtained from knowledge of only the mass of the measure, which is 1. 
A velocity field having structure analogous to a hierarchical laminate (or shear
structure) is argued to attain the isotropic bound. The theory of Pad{\'e}
approximants for Stieltjes functions provides a nested sequence of rigorous
bounds for the diagonal components of the matrix $\Dg^*$, given in terms of the
moments of the measure~\cite{Baker:1996:Book:Pade}. The bounds get tighter as
more moments are incorporated and can converge to the true value of $\Dg^*$ for
certain values of the P{\'e}clet number~\cite{Baker:1996:Book:Pade}. However,
while a sequence of tighter bounds on the effective diffusivity in terms of the
higher moments of the spectral measure are found in \cite{Avellaneda:CMP-339},
{\it the numerical values of these moments are assumed to be known, without
calculation}, and then the bounds are obtained in terms of these moments. 
In the theory of composite materials, the usefulness of
rigorous bounds in applications depends on
being able to compute the bounds for a class of microstructures
of interest. Similarly, the impact of Stieltjes integral
representations for the effective diffusivity on the broad range
of advection diffusion problems across science and engineering 
can be enhanced by the development of a systematic approach
to computing the moments of the spectral measure. 
For over three decades, 
the lack of a way to calculate the measure moments for general fluid 
velocity fields has hindered the progress of 
rigorous 
bounds for $\Dg^*$,
which can serve as
benchmarks in important areas of application.

Here, we develop an iterative method that, in principle, enables an arbitrary
number of measure moments to be calculated analytically in \textit{closed form}
for any spatially periodic or space-time periodic fluid velocity field
represented as a finite trigonometric Fourier series. This, in turn, enables an
arbitrary number of nested bounds to be calculated for such flows. {\it The
framework facilitates new analytical bounds that are not available from existing
numerical methods.} This iterative method is implemented into a numerical
algorithm using the Maple and Python-SymPy symbolic math toolboxes which can be
used to calculate measure moments in closed form for such flows up to a given
order only limited by computational resources. Moreover, we extend this
numerical algorithm to MATLAB, which enables hundreds of moments to be computed
using floating point arithmetic. We incorporate the moment values into an
existing numerical algorithm \texttt{padeapprox}
\cite{Gonnet:2013:55:1:101:110853236} which computes Pad{\'e} approximants in a
robust, stable way, and then compute several nested bounds for the diagonal
components of the matrix $\Dg^*$ for some model steady and space-time periodic
flows. High order bounds accurately capture the known
\cite{Fannjiang:1994:SIAM_JAM:333,Fannjiang:1997:1033} asymptotic behavior of
the effective diffusivity for a steady cell-flow as a function of P{\'e}clet
number in the advection-dominated
regime~\cite{Murphy:ADSTPF-2017,Murphy:ADSRF-2019}. Adding a space-time periodic
term to the fluid velocity field of this steady flow results in an appreciable 
enhancement of $\Dg^*$, shown both in the Pad{\'e} approximate bounds here and
in numerical results involving direct computation of the spectral measure
$\mu$~\cite{Murphy:ADSTPF-2017}. In the fluid setting, a notable weakness of
Pad{\'e} approximation is that in the advection-dominated regime, the upper and
lower bounds diverge from the $\Dg^*$ value, see Fig. 1 of
\cite{Avellaneda:CMP-339}. In other words, for a fixed number of spectral
moments, the bounds lose accuracy rapidly as the molecular diffusivity tends to
zero. The phenomenon also occurs in our numerical results here (see Fig.
\ref{fig:steady_2D_bounds}, Fig. \ref{fig:steady_3D_bounds}, and Fig.
\ref{fig:dynamic_bounds}). How to cure such divergence inherent in the bounds
remains an open problem.

The enhancement of diffusive transport of passive scalars by complex
fluid flow plays a key role in many important processes throughout
science and engineering.
Advection by geophysical
fluids intensifies the dispersion and large scale transport of
heat~\cite{Moffatt:RPP:621},
pollutants~\cite{Bilger:10.1175,Beychok:1994:9780964458802,Samson:1988:88009978},
 and
nutrients~\cite{Lorenzo:2013:26:4,Hofmann:ANS:2004:265075} diffusing
in their environments, such as porous media \cite{Tartakovsky_TPM_2019}.
Advective processes also enhance the large scale
transport of plankton~\cite{Hofmann:ANS:2004:265075}, which is an
important component of the food web that sustains life in the polar
oceans.  
%
In sea ice dynamics,
where the ice cover couples the atmosphere to the polar
oceans \cite{Washington:1986:9780935702521}, the transport of sea 
ice can also be enhanced by eddy
fluxes and large scale coherent structures in the ocean
\cite{Watanabe:2009JPO4010,Lukovich:AG:2015}.   
In sea ice thermodynamics, the temperature field of the 
atmosphere is coupled to the temperature field of the ocean through
sea ice, a composite of pure ice with brine inclusions whose
volume fraction and connectedness depend strongly on temperature
\cite{Thomas:2008:SI,Golden:GRL:L16501,Golden:NAMS:2009}. Convective brine
flow through the porous microstructure can enhance thermal transport
through the sea ice layer
\cite{Lytle:JGR-8853,Worster:PTRSA:2015,Kraitzman_PRSA_2024}.
In fact, in \cite{Kraitzman_PRSA_2024}
the first rigorous theory of the thermal conductivity of sea ice
that accounts for fluid convection, as well as diffusion, 
is developed, based on the Stieltjes integral
representation considered here.  
Bounds on the convection-enhanced thermal conductivity 
of sea ice as a function
of temperature are obtained by calculating the moments
of the spectral measure using the methods in this paper 
for cat's eye and BC flows, which
serve as models for convective velocity fields 
in sea ice. The bounds
capture Antarctic field data and numerical simulations. 

Finally, on the Lagrangian side, significant progress has been made in recent
years on stochastic particle methods (mesh-free and basis-free
structure-preserving schemes) for computing $\Dg^*$ in chaotic and stochastic
flows \cite{Lag18,Lag20,Lag21,Lag22}. Such methods are reliable and free from
the divergence issue encountered in Pad\'e bounds when computing $\Dg^*$ in the
advection-dominated regime. The related stochastic interacting particle methods
have turned out as efficient computing tools for reactive-transport and large
deviation rate functions, e.g., effective front speeds and chemotaxis
aggregation phenomena in complex fluid flows
\cite{LagKPP_22,LagKPP_25,DP_22,DP_24,GenIPM_25,BDP_26,IPMLD25}. For theoretical
advances using Lagrangian (game and control) representations for averaging
geometric level set equations arising in turbulent combustion, see
\cite{Games_24,BAMS_24}.




The organization of the paper is as follows. In Section \ref{sec:Eff_Trans} the
homogenization problem for the advection-diffusion equation is
reviewed~\cite{McLaughlin:SIAM_JAM:780,Fannjiang:1994:SIAM_JAM:333,Novikov:2005:CPAM:867,Majda:Kramer:1999:book}.
In Section \ref{app:Scalar_Fields} an abstract Hilbert space framework is
reviewed~\cite{Murphy:ADSTPF-2017} which is used in Section \ref{sec:Int_Repss}
to provide Stieltjes integral representations for the components $\Dg^*_{jk}$,
$j,k=1,\ldots,d$, of the effective diffusivity matrix $\Dg^*$ involving spectral
measures $\mu_{jk}$ of a self-adjoint
operator~\cite{Murphy:ADSTPF-2017,Murphy:ADSRF-2019}, where $d$ is the spatial
dimension of the system. This abstract framework is utilized in Section
\ref{sec:Iterative_moments} to develop an iterative method for calculating the
moments $\mu_{jk}^n$, $n=0,1,2,\ldots$, of $\mu_{jk}$ for spatially and
space-time periodic fluid velocity fields $\vecu$. A demonstration of the
iterative method is given in Appendix \ref{sec:moment_calculations_detailed},
with detailed calculations of all the moments for shear flow in Appendix
\ref{sec:Moments_shear_flow}, and the first few moments for a spatially and a
space-time periodic $\vecu$ for 2D in Appendix \ref{sec:moments_bc_flow} and
\ref{sec:moments_2_time_periodic_flow}, respectively. A numerical implementation
of the iterative method is discussed in Appendix
\ref{sec:moment_calculations_numerical}, where the method is applied to various
steady and dynamic flows in both 2D and 3D. Our numerical implementation of
Pad\'e approximants is described in Appendix
\ref{sec:numerical_implementation_pade}. Convergence and asymptotic analyses the
Pad\'{e} approximant bounds for the diagonal components of
$\Dg^*$~\cite{Baker:1996:Book:Pade} for each flow is given in Section
\ref{sec:bounds}. Concluding remarks are given in Section \ref{sec:conclusions}.

\section{Effective transport by
	advection-diffusion} \label{sec:Eff_Trans}    
The density $\phi$ of a cloud of passive tracer particles diffusing along
with molecular diffusivity $\varepsilon$ and being advected by an incompressible
velocity field $\vecu$ satisfies the advection-diffusion equation
\begin{align}\label{eq:ADE}
	\partial_t\phi(t,\vecx)
	=\vecu (t,\vecx)\bcdot\bnabla \phi(t,\vecx)+\varepsilon\Delta \phi(t,\vecx),
	\quad
	\phi(0,\vecx)=\phi_0(\vecx),  
\end{align}
for $t>0$ and $\vecx\in\mathbb{R}^d$. Here, the initial density $\phi_0(\vecx)$
and the fluid velocity field $\vecu$ are assumed given, and $\vecu$ satisfies
$\bnabla\bcdot\vecu=0$. In equation~\eqref{eq:ADE}, $\varepsilon>0$ is the
molecular diffusion constant, $\partial_t$ denotes partial differentiation with
respect to time $t$, and $\Delta=\bnabla\bcdot\bnabla =\nabla^2$ is the
Laplacian. Moreover,
$\vecpsi\bcdot\vecvarphi=\vecpsi^{\,T}\overline{\vecvarphi}$, where
$\vecpsi^{\,T}$ denotes transposition of the vector $\vecpsi$ and
$\overline{\vecvarphi}$ denotes component-wise complex conjugation, with
$\vecpsi\bcdot\vecpsi=|\vecpsi|^2$. Later, we will extensively use this form of
the dot product over complex fields, with built in complex conjugation. However,
we emphasize that all quantities considered in this section are
\emph{real-valued}.

In our analysis of the effective diffusivity matrix $\Dg^*$, it is beneficial to
use non-dimensional parameters. We therefore assume that equation~\eqref{eq:ADE}
has been non-dimensionalized as follows. Let $\ell$ and $\tau$ be typical length
and time scales associated with the problem of interest. Mapping to the
non-dimensional variables $t\mapsto t/\tau$ and $\vecx\mapsto \vecx/\ell$, one
finds that $\phi$ satisfies the advection diffusion equation in \eqref{eq:ADE}
with a non-dimensional molecular diffusivity and fluid velocity field,
\begin{align}\label{eq:Peclet_eps}
	\varepsilon\mapsto\tau\varepsilon/\ell^{\,2},
	\quad
	\vecu\mapsto\tau\,\vecu/\ell.    
\end{align}

This non-dimensionalization demonstrates that the
fluid velocity field $\vecu$ is divided by a quantity with dimensions
of velocity and the molecular
diffusivity is divided by a quantity with dimensions of velocity
multiplied by spatial length. A detailed discussion of 
various non-dimensionalizations involving the Strouhal number, the
P{\'e}clet number, and the periodic P{\'e}clet number is given
in~\cite{McLaughlin:Forest:PF:1999:880,Majda:Kramer:1999:book}.  It is 
convenient to choose the rescaled
$\vecu$ and $\varepsilon$ in a way that captures information about the
fluid velocity field. However, it is also convenient to choose
these rescaled variables in a way that \emph{separates} the rescaled
$\varepsilon$ from the \emph{geometry and dynamics} of $\vecu$; this leads to
mathematically and physically meaningful 
properties of rigorous bounds for $\Dg^*$ which follow from the
analytic structure of Stieltjes integral representations for
$\Dg^*$~\cite{Avellaneda:CMP-339,Baker:1996:Book:Pade} --- discussed in 
\secref{sec:Int_Repss} below.

We accomplish both of these goals as follows. Define the dimensional fluid
velocity field by $\vecu=u_0\vecv$, where the parameter $u_0$ has dimensions of
velocity and represents the ``\emph{flow strength}'' of $\vecu$ which is
independent of the geometry and dynamics of $\vecu$ which, in turn, is
encapsulated in the non-dimensional vector field $\vecv$. With these
definitions, we choose reference scales $\tau$ and $\ell$ in
equation~\eqref{eq:Peclet_eps} to satisfy $u_0=\ell/\tau$ so that
$\vecu\mapsto\vecv$ and $\varepsilon\mapsto\varepsilon/u_0\,\ell\,$. For
example, for $BC$-flow~\cite{Biferale:PF:2725}, we define the dimensional fluid
velocity field by $\vecu=u_0\,(C\cos{y},B\cos{x})$, where the flow strength
$u_0\in(0,\infty)$ is chosen to be independent of the non-dimensional parameters
$B,C\in[0,1]$ which determine the streamline geometry of $\vecu$ in
$\vecv=(C\cos{y},B\cos{x})$.

An example of a non-dimensional parameter that compares the rate of scalar
advection to the rate of diffusion is the P\'{e}clet number.
We define it by the ratio $\Pen=\ell u_0/\varepsilon$, although other
definitions have been
used~\cite{McLaughlin:Forest:PF:1999:880,Majda:Kramer:1999:book,Majda:Kramer:1999:book}.
Therefore, our choice of the rescaled $\varepsilon$ satisfies
$\Pen=1/\varepsilon$. The advection and diffusion dominated regimes are
characterized by $\Pen\gg1$ and $\Pen\ll1$, respectively.

The \emph{parameter separation} between $\Pen$ and the geometry of the flow is
important for  rigorous upper and lower Pad\'{e} approximant bounds for
$\Dg^*$~\cite{Avellaneda:CMP-339} discussed in Section \ref{sec:bounds}.
Pad\'{e} approximants of $\Dg^*$ are given in terms of ratios of
polynomials~\cite{Baker:1996:Book:Pade} $P(z)/Q(z)$, where $z=\Pen^2$,
$0<z<\infty$, and the coefficients of these polynomials depend on the moments of
a \emph{spectral measure} that, in turn, depend on the fluid velocity field
$\vecu$~\cite{Avellaneda:CMP-339}. For example, when $\vecu$ is given by
$BC$-flow the moments of the measure depend on the parameters $B$ and $C$. Our
numerical investigations have shown if the non-dimensionalization of
equation~\eqref{eq:ADE} is chosen in a way that the variable $z$ also depends on
the flow geometry through the ratio $B/C$, then this gives rise to
\emph{positive real} roots for the polynomials $P(z)$ and $Q(z)$. This, in turn,
gives rise to positive real roots and poles in the (rigorous) Pad\'{e}
approximant bounds for $\Dg^*$, which is not physically or mathematically
consistent with the known behavior of
$\Dg^*$~\cite{Fannjiang:1994:SIAM_JAM:333,Pavliotis:PHD_Thesis,Biferale:PF:2725,Majda:Kramer:1999:book}.
This demonstrates the importance of \emph{parameter separation} between $z$ and
the flow geometry for Pad\'{e} approximant bounds for $\Dg^*$.

This way of non-dimensionalizing equation~\eqref{eq:ADE} is also
convenient in the case of a time-dependent fluid velocity
field~\cite{Murphy:ADSTPF-2017}, where the parameter $u_0$ again
represents the flow strength and the vector field $\vecv$ encapsulates 
the \emph{geometric and dynamical} properties of the flow. For example, the
space-time periodic flow with velocity field
$\vecu=u_0(\,(C\cos{y},B\cos{x})+\theta\cos{t}\,(\sin{y},\sin{x})\,)$
has dynamical behavior exhibiting Lagrangian
chaos~\cite{Biferale:PF:2725,Murphy:ADSTPF-2017}. Here, the flow 
strength $u_0\in(0,\infty)$ is independent of the parameters
$B,C,\theta\in[0,1]$ which determine the geometric and dynamical
properties of $\vecu$. This choice of non-dimensionalization gives a
clearer interpretation of the advection and diffusion dominated
regimes in terms of $\Pen=1/\varepsilon$ than the non-dimensionalization 
given in~\cite{Murphy:ADSTPF-2017}.

We now discuss the effective transport properties of advection enhanced
diffusion, as described by the advection diffusion equation in~\eqref{eq:ADE}.
We will assume in this manuscript that the fluid velocity field $\vecu$ is
mean-zero in space for steady $\vecu=\vecu(\vecx)$ and mean-zero in space-time
when $\vecu=\vecu(t,\vecx)$ is space-time dependent (also
see~\cite{Pavliotis:Stuart:2008:Book}). 
The long time, large scale dispersion of diffusing tracers, such as heat or
pollutants, being advected by an incompressible fluid velocity field is
equivalent to an enhanced diffusion process~\cite{Taylor:PRSL:196} with an
effective diffusivity matrix $\Dg^*$. In recent decades, methods of
homogenization
theory~\cite{McLaughlin:SIAM_JAM:780,Fannjiang:1994:SIAM_JAM:333,Novikov:2005:CPAM:867,Majda:Kramer:1999:book}
have been used to provide an explicit representation for $\Dg^*$. In particular,
these methods have demonstrated that the averaged or \emph{homogenized} behavior
of the advection-diffusion equation in~\eqref{eq:ADE}, with space-time periodic
velocity field $\vecu$, is determined by a diffusion equation involving an
averaged scalar density $\bar{\phi}$ and an
effective diffusivity tensor
$\Dg^*$~\cite{Majda:Kramer:1999:book}       
\begin{align}\label{eq:phi_bar}
 \partial_t\bar{\phi}(t,\vecx)=\bnabla\bcdot[\Dg^*\bnabla \bar{\phi}(t,\vecx)], \quad
  \bar{\phi}(0,\vecx)=\phi_0(\vecx).
\end{align}

Equation~\eqref{eq:phi_bar} follows from the assumption that the initial tracer
density $\phi_0$ varies slowly relative to the variations of the fluid velocity
field
$\vecu$~\cite{McLaughlin:SIAM_JAM:780,Fannjiang:1997:1033,Majda:Kramer:1999:book}.
This information is incorporated into equation~\eqref{eq:ADE} by introducing a
small dimensionless parameter $\delta\ll1$ and
writing~\cite{McLaughlin:SIAM_JAM:780,Fannjiang:1997:1033,Majda:Kramer:1999:book}      
\begin{align}
  \phi(0,\vecx)=\phi_0(\delta\vecx). 
\end{align}
Anticipating that $\phi$ will have diffusive dynamics as $t\to\infty$, space and 
time are rescaled according to the standard diffusive relation
\begin{align}\label{eq:Fast_Vars}
  \vecxi=\vecx/\delta, \quad
  \tau= t/\delta^2.
\end{align}
The rescaled form of equation~\eqref{eq:ADE} is given
by~\cite{Majda:Kramer:1999:book}  
\begin{align}\label{eq:ADE_delta}
  \partial_t\phi^\delta(t,\vecx)=
  \delta^{-1}\vecu(t/\delta^2,\vecx/\delta)\bcdot\bnabla\phi^\delta(t,\vecx)
  +\varepsilon\Delta\phi^\delta(t,\vecx),
              \quad
             \phi^\delta(0,\vecx)=\phi_0(\vecx), 
\end{align}
where we have denoted $\phi^\delta(t,\vecx)=\phi(t/\delta^2,\vecx/\delta)$.
The convergence of $\phi^\delta$  to $\bar{\phi}$
 can be rigorously established in the following
sense~\cite{Majda:Kramer:1999:book}   
\begin{align}\label{eq:Homogenization_Theorem}
  \lim_{\delta\to0}\;\sup_{0\leq t\leq t_0}\,\sup_{\vecx\in\mathbb{R}^d}
  |\phi^\delta(t,\vecx)-\bar{\phi}(t,\vecx)| =0,
\end{align}
for every finite $t_0>0$, provided that $\phi_0$ and $\vecu$ obey some
mild smoothness and boundedness conditions.

An explicit representation of the
effective diffusivity tensor $\Dg^*$ is given in terms of the (unique)
mean zero, space-time periodic solution $\chi_j$ of the following
\emph{cell problem}~\cite{Biferale:PF:2725,Majda:Kramer:1999:book}, 
\begin{align}\label{eq:Periodic_Cell_Prob}
  \partial_\tau\chi_j(\tau,\vecxi)
  -\varepsilon\Delta_\xi\chi_j(\tau,\vecxi)
  -\vecu(\tau,\vecxi) \bcdot\bnabla_\xi \chi_j(\tau,\vecxi)
  =u_j(\tau,\vecxi),
\end{align}
where the subscript $\xi$ in $\Delta_\xi$ and $\bnabla_\xi$ indicates that
differentiation is with respect to the fast variable $\vecxi$ defined in
equation~\eqref{eq:Fast_Vars}.  The components $\Dg^*_{jk}$, $j,k=1,\ldots,d$,
of the matrix $\Dg^*$ are given
by~\cite{McLaughlin:SIAM_JAM:780,Fannjiang:1994:SIAM_JAM:333,Novikov:2005:CPAM:867,Majda:Kramer:1999:book}           
\begin{align}\label{eq:Djk}
  \Dg^*_{jk}=\varepsilon\delta_{jk}+\langle u_j\chi_k\rangle,
\end{align}
where $\delta_{jk}$ is the Kronecker delta and $u_j$ is the $j$th component of
the vector $\vecu$. The averaging $\langle\cdot\rangle$ in~\eqref{eq:Djk} is
with respect to the fast variables defined in equation~\eqref{eq:Fast_Vars}. The
averaging is over the bounded sets $\Tc\subset\mathbb{R}$ and
$\Vc\subset\mathbb{R}^d$, with $\tau\in\Tc$ and $\vecxi\in\Vc$, which define the
space-time period cell ($(d+1)$--torus) $\Tc\times\Vc$.

In general, the effective diffusivity tensor $\Dg^*$ has a symmetric
$\Sg^*$ and antisymmetric $\Ag^*$ part defined by 
\begin{align}\label{eq:Symm_Anti-Symm}
  \Dg^*=\Sg^*+\Ag^*,\qquad
  \Sg^*=\frac{1}{2}\left(\Dg^*+[\Dg^*]^{\,T}\right), \quad
  \Ag^*=\frac{1}{2}\left(\Dg^*-[\Dg^*]^{\,T}\right),
\end{align}
where $[\Dg^*]^{\,T}$ denotes transposition of the matrix $\Dg^*$. Denote by
$\Sg^*_{jk}$ and $\Ag^*_{jk}$, $j,k=1,\ldots,d$, the components of $\Sg^*$ and
$\Ag^*$ in~\eqref{eq:Symm_Anti-Symm}. When the fluid velocity field is mean-zero
and divergence-free, as discussed above, then
equation~\eqref{eq:Homogenization_Theorem} holds and the effective diffusivity
tensor $\Dg^*$ defined in~\eqref{eq:Djk} is
constant~\cite{Majda:Kramer:1999:book}. Consequently, only the symmetric part of
$\Dg^*$ plays a role in the effective transport equation shown
in~\eqref{eq:phi_bar}, as the antisymmetric part of $\Dg^*$ cancels out in the
sum $\sum_{ij}\Dg^*_{ij}\partial_i\partial_j\bar{\phi}$, where $\partial_i$
denotes differentiation in the $i$th spatial
direction~\cite{Pavliotis:PHD_Thesis}.

In \secref{sec:moments_2_time_periodic_flow} 
we consider the fluid velocity field $\vecu$ 
\begin{align}\label{eq:tdcell}
\vecu (t,\vecx)=(C\cos{y},B\cos{x}) + \theta\,\cos{t}\;(\sin{y},\sin{x}), 
\quad
\theta \in [0,1].
\end{align}
with temporal periodicity $\Tc=[0,2\pi]$ and spatial periodicity
$\Vc=[0,2\pi]^d$, with $d=2$. In the case of a time-dependent fluid velocity
field, $\langle\cdot\rangle$ denotes space-time averaging over $\Tc\times\Vc$.
In the special case of a time-independent fluid velocity field, the function
$\chi_j$ is time-independent and satisfies
equation~\eqref{eq:Periodic_Cell_Prob} with $\partial_\tau\chi_j\equiv0$, and
$\langle\cdot\rangle$ in~\eqref{eq:Djk} denotes spatial averaging over
$\Vc$~\cite{Fannjiang:1994:SIAM_JAM:333,Novikov:2005:CPAM:867,Majda:Kramer:1999:book}.

\subsection{Hilbert space}\label{app:Scalar_Fields}
In this section we provide an abstract Hilbert space formulation of
the effective parameter problem for advection-diffusion that was
proposed in~\cite{Pavliotis:PHD_Thesis}, based
on~\cite{Bhattacharya:AAP:1999:951}, and generalized to the setting
of a space-time periodic fluid velocity field in~\cite{Murphy:ADSTPF-2017}. 
To fix ideas, consider the following sets $\Tc=[0,T]$ and 
$\Vc=\otimes_{j=1}^d[0,L]$ which  define the space-time period cell
$\Tc\times\Vc$. Now consider the Hilbert spaces
$L^2(\Tc)$ and $L^2(\Vc)$ of Lebesgue measurable scalar functions over
the complex field $\mathbb{C}$ that are also square
integrable~\cite{Folland:99:RealAnalysis}. Define the associated
Hilbert spaces $\Hs_{\Tc}$, $\Hs_{\Vc}$, and
$\Hs_{\Tc\Vc}=\Hs_{\Tc}\otimes\Hs_{\Vc}$ of periodic functions, where  
\begin{align}\label{eq:Hilbert_Spaces_scalar}  
  \Hs_{\Tc}&=\big\{\psi\in L^2(\Tc) \, | \, \psi(t)=\psi(t+T)\big\},
  \\
  \Hs_{\Vc}&
  =\big\{\psi\in L^2(\Vc) \, | \, \psi(\vecx)=\psi(\vecx+L\vece_j), \ j=1,\ldots,d\big\},
  \notag
\end{align}
and the $\vece_j$ are standard basis
vectors.

More specifically, denote time average over $\Tc$ by
$\langle\cdot\rangle_{\Tc}$, space average over $\Vc$ by
$\langle\cdot\rangle_{\Vc}$, and space-time average over $\Tc\times\Vc$ by
$\langle\cdot\rangle$. The space-time average $\langle\cdot\rangle$, induces a
sesquilinear inner-product $\langle\cdot,\cdot\rangle$ given by
$\langle\psi,\varphi\rangle=\langle\psi\;\overline{\varphi}\rangle$, with
$\langle\varphi,\psi\rangle=\overline{\langle\psi,\varphi\rangle}$. This
$\Hs_{\Tc\Vc}$--inner-product, in turn, induces a norm $\|\cdot\|$ given by
$\|\psi\|=\langle\psi,\psi\rangle^{1/2}$~\cite{Folland:99:RealAnalysis}. The set
of space-time periodic Lebesgue measurable functions $\Hs_{\Tc\Vc}$ satisfying
$\|f\|<\infty$ is a (complete) Hilbert space~\cite{Folland:99:RealAnalysis}.
Similarly, the space and time averages, $\langle\cdot\rangle_{\Vc}$ and
$\langle\cdot\rangle_{\Tc}$, induce sesquilinear inner-products,
$\langle\cdot,\cdot\rangle_{\Vc}$ and $\langle\cdot,\cdot\rangle_{\Tc}$, that
induce norms, $\|\cdot\|_{\Vc}$ and $\|\cdot\|_{\Tc}$, associated with the
Hilbert spaces $\Hs_{\Vc}$ and $\Hs_{\Tc}$.

To treat temporal dependence, we define the space $\As_{\Tc}$ of functions that
are absolutely continuous~\cite{Stone:64,Royden:1988:RA} on the interval $\Tc$,
having derivative 
belonging to $L^2(\Tc)$, and the space $\tilde{\As}_{\Tc}$ of 
absolutely continuous $\Tc$--periodic functions with time derivatives
belonging to $L^2(\Tc)$, 
\begin{align}\label{eq:AC_BC}
  \tilde{\As}_{\Tc}&=\{\psi\in\As_{\Tc} \,|\, \psi(0)=\psi(T)\},
\end{align}
which is \emph{not} a Hilbert space but is
instead an everywhere dense subset of the Hilbert space
$\Hs_\Tc$~\cite{Stone:64}. To treat spatial dependence, we now define 
the Sobolev space $\Hs^{1,2}_{\Vc}$ which is itself a Hilbert
space~\cite{Bhattacharya:AAP:1999:951,Folland:95:PDEs,McOwen:2003:PDE},             
\begin{align}\label{eq:Sobolev}
  \Hs^{1,2}_{\Vc}=\big\{
    \psi\in \Hs_{\Vc} \ |\  \|\bnabla\psi\|_{\Vc}<\infty, \ \langle\psi\rangle_{\Vc}=0
  \big\}.
\end{align}
The condition $\langle\psi\rangle_{\Vc}=0$ in~\eqref{eq:Sobolev} is required to
eliminate non-zero constant $\psi$, which satisfies
$\|\bnabla\psi\|_{\Vc}=0$. The $\Hs^{1,2}_{\Vc}$--norm $\|\bnabla\cdot\|_{\Vc}$
is induced by the $\Hs^{1,2}_{\Vc}$--inner-product:
$\|\bnabla\psi\|_{\Vc}=\langle\bnabla\psi\bcdot\bnabla\psi\rangle_{\Vc}^{1/2}$. 


Finally, define the Hilbert space $\Hs$ and its everywhere dense subset $\Fs$
\begin{align}\label{eq:Function_Space_Scalar}
  \Hs=\Hs_{\Tc}\otimes\Hs^{1,2}_{\Vc}, \qquad
  \Fs=\tilde{\As}_{\Tc}\otimes\Hs^{1,2}_{\Vc}.
\end{align}
Due to the presence of $\tilde{\As}_{\Tc}$ in the definition of the function
space $\Fs$, it is \emph{not} a complete Hilbert space, and is instead an
everywhere dense subset of the complete Hilbert space $\Hs$. Recall that
$\langle\cdot\rangle$ denotes space-time average over $\Tc\times\Vc$ and
$\vecpsi\bcdot\veczeta=\vecpsi^T\,\overline{\veczeta}$. The sesquilinear   
$\Hs$--inner-product is given by
$\langle\psi,\varphi\rangle_{1,2}=\left\langle\bnabla \psi\bcdot\bnabla \varphi
\right\rangle$ with associated norm $\|\cdot\|_{1,2}$ given by $\|\psi\|_{1,2}=
\langle|\bnabla \psi|^2\rangle^{1/2}$. We emphasize that in the case of a
time-dependent fluid velocity field, it is necessary that $\psi\in\Hs$ satisfy
$\langle\psi\rangle_{\Vc}=0$, as required by the definition of $\Hs^{1,2}_{\Vc}$
in~\eqref{eq:Sobolev}. Otherwise,
$\|\cdot\|_{1,2}=|\Tc\times\Vc|^{-1}\int_{\Tc\times\Vc}\d
t\,\d\vecx\,|\bnabla\cdot|^2$ is not a norm, since a strictly positive function
$\psi(t,\vecx)=\psi(t)$ on $\Tc\times\Vc$ satisfies $\|\psi\|_{1,2}=0$, where
$|\Tc\times\Vc|$ denotes Lebesgue measure of the set $\Tc\times\Vc$. In the case
of a time-independent fluid velocity field $\vecu=\vecu(\vecx)$ we set
$\Hs\equiv\Fs\equiv\Hs^{1,2}_{\Vc}$, and
$\langle\cdot\rangle=\langle\cdot\rangle_{\Vc}$.

%
%

\subsection{Integral representations for the effective
  diffusivity}\label{sec:Int_Repss}

In this section we summarize the results of~\cite{Murphy:ADSTPF-2017}, which
provides Stieltjes integral representations for both the symmetric $\Sg^*$ and
antisymmetric $\Ag^*$ parts of $\Dg^*$. Since the analysis in this section
involves only the fast variables $(\tau,\vecxi)$ defined in
equation~\eqref{eq:Fast_Vars}, for notational simplicity, we will drop the
subscripts $\xi$ shown in equation~\eqref{eq:Periodic_Cell_Prob} and use
$\partial_t$ to denote $\partial_\tau$.

Inserting the expression for $u_j$ on
the right side of~\eqref{eq:Periodic_Cell_Prob}
into equation~\eqref{eq:Djk} leads to the following functional
representations for the components $\Sg^*_{jk}$ and $\Ag^*_{jk}$,
$j,k=1,\ldots,d$, of $\Sg^*$ and $\Ag^*$~\cite{Pavliotis:PHD_Thesis}      
\begin{align}\label{eq:Eff_Diffusivity_Sobolev}
  \Sg^*_{jk}=\varepsilon(\delta_{jk}+\langle\chi_j,\chi_k\rangle_{1,2}),
  \quad
  \Ag^*_{jk}=\langle A\chi_j,\chi_k\rangle_{1,2}\,,
  \quad
  A=(-\Delta)^{-1}(\partial_t-\vecu \bcdot\bnabla)\,.
\end{align}
Here, $\langle f,h\rangle_{1,2}=\langle\bnabla f\bcdot\bnabla h\rangle$ is a
Sobolev-type \emph{sesquilinear} inner-product~\cite{McOwen:2003:PDE} and the
operator $(-\Delta)^{-1}$ is based on convolution with respect to the Green's
function for the Laplacian $\Delta$~\cite{Stakgold:BVP:2000,Folland:95:PDEs}.
Since the function $\chi_j$ is \emph{real-valued} we have
$\langle\chi_j,\chi_k\rangle_{1,2}=\langle\chi_k,\chi_j\rangle_{1,2}$, which
implies that $\Sg^*$ is a symmetric matrix. The function $A\chi_j$ is also
real-valued. The operator $A$ is skew-adjoint on the Hilbert space
$\Hs$~\cite{Murphy:ADSTPF-2017}, which implies that $\Ag^*_{kj}=\langle
A\chi_k,\chi_j\rangle_{1,2}=-\langle\chi_k,A\chi_j\rangle_{1,2}=-\langle
A\chi_j,\chi_k\rangle_{1,2}=-\Ag^*_{jk}$ which, in turn, implies that $\Ag^*$ is
an antisymmetric matrix, hence $\Ag^*_{kk}=\langle
A\chi_k,\chi_k\rangle_{1,2}=0$.

Applying the linear operator $(-\Delta)^{-1}$ to both sides of the cell
problem in equation~\eqref{eq:Periodic_Cell_Prob} yields the following
resolvent formula for $\chi_j$ 
\begin{align}\label{eq:Resolvent_Rep_Scalar}
  \chi_j=(\varepsilon+A)^{-1}g_j, \qquad 
  g_j=(-\Delta)^{-1}u_j.
\end{align}
From equations~\eqref{eq:Eff_Diffusivity_Sobolev}
and~\eqref{eq:Resolvent_Rep_Scalar} we have the following functional
formulas for $\Sg^*_{jk}$ and $\Ag^*_{jk}$ involving the
skew-adjoint operator $A$
\begin{align}\label{eq:Eff_Diff_Resolvent_Sobolev}
 \Sg^*_{jk}&=\varepsilon\left(\delta_{jk}
  +\langle(\varepsilon+A)^{-1}g_j,(\varepsilon+A)^{-1}g_k\rangle_{1,2}\right), 
  \\
 \Ag^*_{jk}&=\langle A(\varepsilon+A)^{-1}g_j,(\varepsilon+A)^{-1}g_k\rangle_{1,2}.
 \notag
\end{align}
Since $A$ is a skew-adjoint operator, it can be written as $A=\imath M$
where $M$ is a symmetric operator~\cite{Stone:64}. In~\cite{Murphy:ADSTPF-2017}
it is shown that $M$ is \emph{self-adjoint} on the Hilbert space $\Hs$.

The spectral theorem for self-adjoint operators states that there is a
one-to-one correspondence between the self-adjoint operator $M$ and a family of
self-adjoint projection operators $\{Q(\lambda)\}_{\lambda\in\Sigma}$ --- the
resolution of the identity --- that satisfies
$\lim_{\lambda\to\,\inf{\Sigma}}Q(\lambda)=0$ and
$\lim_{\lambda\to\,\sup{\Sigma}}Q(\lambda)=I$~\cite{Stone:64}. Here, $\Sigma$ is
the \emph{spectrum} of the operator $M$, while $0$ and $I$ denote the null and
identity operators. Define the \emph{complex valued} function
$\mu_{jk}(\lambda)=\langle Q(\lambda)g_j,g_k\rangle_{1,2}$, $j,k=1,\ldots,d$,
where  $g_j=(-\Delta)^{-1}u_j$ is defined in~\eqref{eq:Resolvent_Rep_Scalar}.
The real, $\Real\mu_{jk}(\lambda)$, and imaginary, $\Imag\mu_{jk}(\lambda)$,
parts of the function $\mu_{jk}(\lambda)$ are of bounded variation, and
therefore have Stieltjes measures $\Real\mu_{jk}$ and $\Imag\mu_{jk}$ associated
with them~\cite{Stone:64}. The function $\mu_{kk}(\lambda)$ is positive hence
$\mu_{kk}$ is a positive measure, while $\Real\mu_{jk}$ and $\Imag\mu_{jk}$,
$j\neq k$, are signed measures. Given certain regularity
conditions~\cite{Murphy:ADSTPF-2017} on the components $u_j$ of the fluid
velocity field $\vecu$, the functional formulas for $\Sg^*_{jk}$ and
$\Ag^*_{jk}$ in~\eqref{eq:Eff_Diff_Resolvent_Sobolev} have the following
Radon--Stieltjes integral representations, for all $0<\varepsilon<\infty$
%
\begin{align}\label{eq:Integral_Rep_kappa*}
  \Sg^*_{jk}=
  \varepsilon\left(\delta_{jk}
  +\int_{-\infty}^\infty
  \frac{\d\Real\,\mu_{jk}(\lambda)}{\varepsilon^2+\lambda^2}\right),
  \qquad
  \Ag^*_{jk}=
  -\int_{-\infty}^\infty
  \frac{\lambda\,\d\Imag\,\mu_{jk}(\lambda)}{\varepsilon^2+\lambda^2}\,.         
\end{align}
The integration in~\eqref{eq:Integral_Rep_kappa*} is over the spectrum
$\Sigma\subseteq\mathbb{R}$ of the self-adjoint operator $M=-\imath
A$~\cite{Stone:64,Reed-1980}. In the setting of a time-independent flow,
$\vecu=\vecu(\vecx)$, the operator $A=(-\Delta)^{-1}[\vecu\bcdot\bnabla]$ and
the self-adjoint operator $M=-\imath A$ is
compact~\cite{Bhattacharya:AAP:1999:951}. Therefore, the spectrum  
$\Sigma$ is discrete outside a neighborhood of 
$\lambda=0$ with a limit point at
$\lambda=0$~\cite{Stakgold:BVP:2000}. In the setting of a
time-dependent flow, $\vecu=\vecu(t,\vecx)$, $M=-\imath A$ is an unbounded
operator~\cite{Murphy:ADSTPF-2017,Stone:64}. Therefore, in general, 
the spectrum $\Sigma$ can be an unbounded subset of $\mathbb{R}$, it
can have discrete and continuous components, and can
even coincide with $\mathbb{R}$ itself~\cite{Stone:64}.

\section{Iterative moment method}\label{sec:Iterative_moments}
In this section we provide an iterative method which may be used to 
calculate, in principle, an arbitrary number of moments for spectral 
measures associated with the effective diffusivity for spatially and 
space-time periodic fluid velocity fields.
The spectral theorem shows that the mass $\mu^0_{jk}$ and the moments 
$\mu^n_{jk}$, $n=1,2,3,\ldots$, of the spectral measure $\mu_{jk}$ are given by
\begin{align}
	\mu^0_{jk}=\int_{-\infty}^{\infty} \d\mu_{jk}(\lambda)
		   =\langle g_j,g_k\rangle_{1,2}\,,
	\qquad
	\mu^n_{jk}=\int_{-\infty}^{\infty} \lambda^n \;\d\mu_{jk}(\lambda)
		   =\langle M^n g_j,g_k\rangle_{1,2}\,, 
	\notag
\end{align}
where $M=-\imath A$ and the operator $M^n$ is defined through
composition. Antisymmetric properties of differentiation and symmetric 
properties of the operator $(-\Delta)^{-1}$ in the sesquilinear 
$L^2(\Tc\times\Vc)$ inner-product
$\langle\cdot,\cdot\rangle_{2}$  
show that the mass $\mu^0_{jk}$ of
the measure $\mu_{jk}$ is given by~\cite{Murphy:ADSTPF-2017}   
\begin{align}\label{eq:measure_mass}
\mu^0_{jk}=\langle g_j,g_k\rangle_{1,2}
	  = \langle\bnabla(-\Delta)^{-1}u_j\bcdot\bnabla(-\Delta)^{-1}u_k\rangle
	  = \langle(-\Delta)^{-1}u_j,u_k\rangle_2
          = \langle g_j,u_k\rangle_2\,.
\end{align}
Similarly, denoting the 
\emph{material derivative}
$D_t=\partial_t+\vecu\bcdot\bnabla$ for time-dependent $\vecu$ and
$D_t=\vecu\bcdot\bnabla$ for time-independent $\vecu$, with
$A=(-\Delta)^{-1}D_t$, the anti-symmetry of $A$ in the inner-product
$\langle\cdot,\cdot\rangle_{1,2}$ yields~\cite{Murphy:ADSTPF-2017}   
\begin{align}\label{eq:General_Moments_Functional_1_and_2}
\mu^{1}_{jk}&=\langle M g_j,g_k\rangle_{1,2}
              =-\imath\,\langle D_t g_j, g_k\rangle_{2}\,,
\\
\mu^{2}_{jk}&=\langle M^2 g_j,g_k\rangle_{1,2}
               =\langle A g_j,A g_k\rangle_{1,2}
	       =\langle D_t g_j,A g_k\rangle_{2}\,.
\notag               
\end{align}
Higher moments are found in a similar way, for $n=1,2,3,\cdots$,
\begin{align}\label{eq:Moments_Functional}  
\mu^{2n+1}_{jk}&=\langle M^{2n+1} g_j,g_k\rangle_{1,2}
               =-\imath\,\langle A^{n+1} g_j,A^n g_k\rangle_{1,2}
	       =-\imath\,\langle D_t A^{n} g_j,A^n g_k\rangle_{2}\,,
\\
\mu^{2n+2}_{jk}&=\langle M^{2n+2} g_j,g_k\rangle_{1,2}
               =\langle A^{n+1} g_j,A^{n+1} g_k\rangle_{1,2}
	       =\langle D_t A^{n} g_j,A^{n+1} g_k\rangle_{2}\,.
\notag
\end{align}
From equation~\eqref{eq:Moments_Functional} and the asymmetry of $A$, we have
$\mu^{2n+1}_{jk}=0$ for all $n=0,1,2,\ldots$~\cite{Murphy:ADSTPF-2017}.
Moreover, $\mu^{2n}_{kj}=\mu^{2n}_{jk}$ as $g_k$, $D_tg_k$, $A^n g_k$, and $D_t
A^n g_k$ are real-valued functions for all $n=1,2,\ldots$. In summary,
\begin{align}
  \mu^{2n}_{kj}=\mu^{2n}_{jk}\,,
  \qquad
  \mu^{2n+1}_{jk}=0\,.
\end{align}

Two key properties of the operators $(-\Delta)^{-1}$, $D_t$, and
$A=(-\Delta)^{-1}D_t$ are that they are \emph{linear} and complex exponentials
are eigenfunctions of these operators. We consider spatially and space-time
periodic fluid velocity fields that can be written in terms of a finite number
of Fourier modes. For space-time periodic fluid velocity fields
$\vecu=\vecu(t,\vecx)$, we utilize the complex, orthonormal Fourier
basis~\cite{Folland:99:RealAnalysis} for $L^2(\Tc\times\Vc),$
$\phi_{\ell,\veck}(t,\vecx)= \exp[\imath(\ell t + \veck\bcdot\vecx)]$ where
$\ell\in\mathbb{Z}$ and $\veck\in\mathbb{Z}^d$. In the setting of a
time-independent flow, $\vecu=\vecu(\vecx)$, we will instead utilize the basis
functions $\phi_{\veck}(\vecx)= \exp(\imath\,\veck\bcdot\vecx)$, so that
$\veca_{\ell,\veck}=\veca_{0,\veck}=\veca_{\veck}$. The basis functions
$\{\phi_{\ell,\veck}\}_{\ell,\veck}$ are \emph{eigenfunctions} of the operators
$\partial_t$, $\bnabla$, and $(-\Delta)^{-1}$, with~\cite{Murphy:ADSTPF-2017}
%
\begin{align}\label{eq:Operator_Eigenfunctions}
  \partial_t\phi_{\ell,\veck}=\imath\ell\,\phi_{\ell,\veck}\,,
  \qquad
  \bnabla\phi_{\ell,\veck}=\imath\veck\,\phi_{\ell,\veck}\,,
  \qquad
  (-\Delta)^{-1}\phi_{\ell,\veck}=|\veck|^{-2}\phi_{\ell,\veck}\,.
\end{align}
Similarly, the basis functions $\{\phi_{\veck}\}_{\veck}$ are eigenfunctions of
the operators $\bnabla$ and $(-\Delta)^{-1}$, with
$\bnabla\phi_{\veck}=\imath\veck\,\phi_{\veck}$ and
$(-\Delta)^{-1}\phi_{\veck}=|\veck|^{-2}\phi_{\veck}$.

In order to take advantage of these properties in our
calculation of the moments $\mu^n_{jk}$ of the measure $\mu_{jk}$, we
take $\Tc=[0,2\pi]$ and $\Vc=[0,2\pi]^d$, and write~\cite{Folland:99:RealAnalysis} 
\begin{align}\label{eq:Fourier_u}
\vecu=\sum_{\ell,\veck}
            \veca_{\ell,\veck}\,
            \phi_{\ell,\veck},
\qquad            
            \phi_{\ell,\veck}(t,\vecx)=
            \exp[\imath(\ell t + \veck\bcdot\vecx)]\,,
\quad
\veca_{\ell,\veck}=\langle\vecu,\phi_{\ell,\veck}\rangle_2\,,
\end{align}
where $\ell\in\mathbb{Z}$, $\veck\in\mathbb{Z}^d$, 
$\veca_{\ell,\veck}=(a^1_{\ell,\veck},\ldots,a^d_{\ell,\veck})$,
and the average
$\langle\vecu,\phi_{\ell,\veck}\rangle_2$ is understood to be 
component-wise, so 
$(\veca_{\ell,\veck})_i=\langle u_i,\phi_{\ell,\veck}\rangle_2$.
Similarly, for a time-independent flow, $\vecu=\vecu(\vecx)$, we write
$\vecu=\sum_{\veck}\veca_{\veck}\,\phi_{\veck},$
$\veca_{\veck}=(a^1_{\veck},\ldots,a^d_{\veck})$,
where $(\veca_{\veck})_i=\langle u_i,\phi_{\veck}\rangle_2\,,$.
The orthonormal basis vectors satisfy
\begin{align}\label{eq:orthonormality}
\langle\phi_{\ell,\veck},\phi_{\ell^\prime,\veck^\prime}\rangle_2
=
\langle\phi_{\ell-\ell^\prime,\veck-\veck^\prime}\rangle
=
\delta_{\ell,\ell^\prime}\,\delta_{\veck,\veck^\prime}\,,
\end{align}
where $\langle\cdot\rangle$ denotes averaging over $\Tc\times\Vc$ and
$\delta_{i,j}$ is the Kronecker delta, with
$\delta_{\veck,\veck^\prime}=\prod_i\delta_{k_i,k_i^\prime}$. Similarly, the
orthonormal basis vectors satisfy
$\langle\phi_{\veck},\phi_{\veck^\prime}\rangle_2 =
\langle\phi_{\veck-\veck^\prime}\rangle = \delta_{\veck,\veck^\prime}\,.$ For
the rest of this section, we will focus on the case of a time-dependent flow,
$\vecu=\vecu(t,\vecx)$, as the case of a time-independent flow,
$\vecu=\vecu(\vecx)$, is a special case of the time-dependent setting. We
emphasize that we assume there are a \emph{finite} number of terms in the
Fourier expansion of $\vecu$.

Denoting $g_k^{[n]}=A^n g_k$, $g_k^{[0]}=g_k$, and writing equation 
\eqref{eq:Moments_Functional} as 
\begin{align}\label{eq:Moments_Functional_itteration}  
\mu^{2n+1}_{jk}=-\imath\,\langle D_t g_j^{[n]},g^{[n]}_k\rangle_{2}\,,
\qquad
\mu^{2n+2}_{jk}=\langle D_t g_j^{[n]},g^{[n+1]}_k\rangle_{2}\,,
\end{align}
it is clear that the mass and moments of the spectral measure $\mu_{jk}$ can be
expressed \emph{itteratively} in terms of just the Fourier coefficients 
$\veca_{\ell,\veck}$ of $\vecu$, using 
equations~\eqref{eq:measure_mass}--\eqref{eq:orthonormality}. 
By linearity, it suffices to 
understand how the operators  
$(-\Delta)^{-1}$, $D_t=\vecu\bcdot\bnabla$, and $A=(-\Delta)^{-1}D_t$ map a
single Fourier mode
$p_{\ell^\prime,\veck^\prime}\phi_{\ell^\prime,\veck^\prime}$ to potentially
multiple modes $\sum_{\ell^\prime,\veck^\prime}
p^\prime_{\ell^\prime,\veck^\prime} \phi_{\ell^\prime,\veck^\prime} $. By
equations \eqref{eq:Operator_Eigenfunctions} and \eqref{eq:Fourier_u} we have 
\begin{align}\label{eq:mode_mappings}
(-\Delta)^{-1}\phi_{\ell^\prime,\veck^\prime}
&= 
|\veck^\prime|^{-2}
\phi_{\ell^\prime,\veck^\prime}\,,
\\\notag
  D_t\phi_{\ell^\prime,\veck^\prime}
     &=
     (\partial_t+\vecu\bcdot\bnabla)     
     \phi_{\ell^\prime,\veck^\prime}
     \\\notag
     &= \imath\ell^\prime\,\phi_{\ell^\prime,\veck^\prime}
     +[\vecu\,\bcdot\,\imath\veck^\prime]\,\phi_{\ell^\prime,\veck^\prime}  
     \\\notag
     &= \imath\ell^\prime\,\phi_{\ell^\prime,\veck^\prime}
     -\sum_{\ell,\veck}
     \imath[\veca_{\ell,\veck}\,\bcdot\,\veck^\prime\,]
     \,\phi_{\ell^\prime+\ell,\veck^\prime+\veck}  
     \\\notag
  A\phi_{\ell^\prime,\veck^\prime}
     &=\imath\ell^\prime|\veck^\prime|^{-2}
     \,\phi_{\ell^\prime,\veck^\prime}
     -     
     \sum_{\ell,\veck}
     \imath     
     [\veca_{\ell,\veck}\,\bcdot\,\veck^\prime\,]
     \,|\veck^\prime+\veck|^{-2}
     \,\phi_{\ell^\prime+\ell,\veck^\prime+\veck}\,. 
\end{align}
Here the $-\imath$ comes from the sesquilinearity of the dot product. 

It follows from equations~\eqref{eq:measure_mass}, \eqref{eq:Fourier_u},
and \eqref{eq:mode_mappings}, and the orthonormality of the
$\phi_{\ell,\veck}$ that
\begin{align}\label{eq:Mass_Fourier}
&\mu^0_{jk}
=\langle(-\Delta)^{-1}u_j,u_k\rangle_2
=\sum_{\ell,\veck}
    |\veck|^{-2}\,a^{\,j}_{\ell,\veck}\,
    \bar{a}^{\,k}_{\ell,\veck}\,,
\end{align}
where $\bar{a}^{\,j}_{\ell,\veck}$ is the complex conjugate of
$a^{\,j}_{\ell,\veck}$. Similarly, the first $\mu^1_{jk}$ and second
$\mu^2_{jk}$ moments of the measure $\mu_{jk}$ can be expressed explicitly in
terms of just the Fourier coefficients $\veca_{\ell,\veck}$ of $\vecu$. However,
these equations expressed in terms of just $\veca_{\ell,\veck}$ can be
complicated and those for higher moments $\mu^n_{jk}$, $n=3,4,\ldots$, expressed
in terms of just the $\veca_{\ell,\veck}$ become unmanageable, by hand. However,
many moments can be found exactly, in closed form using symbolic mathematics
software like Maple or Python-SymPy, or as a floating-point approximation using
numerical linear algebra software like MATLAB or Python-NumPy.

To make this explicit, we write
\begin{align}\label{eq:uj_gj_Dtgj_Agj_Fourier}
  &u_j=\sum_{\ell,\veck}
    a^{\,j}_{\ell,\veck}\,
    \phi_{\ell,\veck}\,,
  &g_j=\sum_{\ell,\veck}
    b^{\,j}_{\ell,\veck}\,
    \phi_{\ell,\veck}\,,
    \\\notag
  &D_tg_j=\sum_{\ell,\veck}
    c^{\,j}_{\ell,\veck}\,
    \phi_{\ell,\veck}\,,
  &Ag_j=\sum_{\ell,\veck}
    d^{\,j}_{\ell,\veck}\,
    \phi_{\ell,\veck}\,.
\end{align}
Therefore, 
equations~\eqref{eq:measure_mass},
\eqref{eq:General_Moments_Functional_1_and_2},
and~\eqref{eq:uj_gj_Dtgj_Agj_Fourier}, and the  
orthonormality of the $\phi_{\ell,\veck}$ yield
\begin{align}\label{eq:Mass_Moments_1_and_2_Fourier}
  \mu^0_{jk}
      =\sum_{\ell,\veck}
       b^{\,j}_{\ell,\veck}\,
       \bar{a}^{\,k}_{\ell,\veck}\,,
   \qquad
   \mu^1_{jk}
      =-\imath
      \sum_{\ell,\veck}
       c^{\,j}_{\ell,\veck}\,
       \bar{b}^{\,k}_{\ell,\veck}\,,
   \qquad
   \mu^2_{jk}
      =
      \sum_{\ell,\veck}
       c^{\,j}_{\ell,\veck}\,
       \bar{d}^{\,k}_{\ell,\veck}\,.
\end{align}
In a similar way, for each $n=1,2,3,\ldots\,$, denote
\begin{align}\label{eq:Dt_An_Fourier} 
  D_tg_j^{[n]}=\sum_{\ell,\veck}
    c^{\,j\,,\,n}_{\ell,\veck}\,
    \phi_{\ell,\veck}\,,
    \qquad
  g_j^{[n]}=\sum_{\ell,\veck}
    d^{\,j\,,\,n}_{\ell,\veck}\,
    \phi_{\ell,\veck}\,,
\end{align}
with $d^{\,j\,,\,1}_{\ell,\veck}=d^{\,j}_{\ell,\veck}$ and
$d^{\,j\,,\,n}_{\ell,\veck}=|\veck|^{-2}c^{\,j\,,\,n}_{\ell,\veck}$. Then, from
equations~\eqref{eq:Moments_Functional}
and~\eqref{eq:Moments_Functional_itteration}, we have  
\begin{align}\label{eq:Higher_Moments_Fourier}
  \mu^{2n+1}_{jk}
      =-\imath
      \sum_{\ell,\veck}
       c^{\,j\,,\,n}_{\ell,\veck}\,
       \bar{d}^{\,k\,,\,n}_{\ell,\veck}\,,
   \qquad
   \mu^{2n+2}_{jk}
      =\sum_{\ell,\veck}
       c^{\,j\,,\,n}_{\ell,\veck}\,
       \bar{d}^{\,k\,,\,n+1}_{\ell,\veck}\,.
\end{align}

To illustrate the method, in Appendix \ref{sec:moment_calculations_detailed} we
provide detailed calculations of all the moments for shear flow as well as the
mass and first moment for example 2D steady and dynamic flows. In Appendix
\ref{sec:moment_calculations_numerical} we use the iterative mappings in
\eqref{eq:mode_mappings} and formulas for the spectral measure mass and moments
in equations \eqref{eq:Mass_Moments_1_and_2_Fourier} and
\eqref{eq:Higher_Moments_Fourier} to calculate many moments for some steady and
dynamic, periodic fluid flows in both 2D and 3D. In particular, we consider the
fluid velocity field for the 2D steady BC-flow, given by
\begin{align}\label{eq:BC-flow}
\vecu(\vecx)=(C\cos y,B\cos x)\,,
\end{align}
where $B, C \geq 0$ control the streamline geometry of the flow, and the
corresponding stream function is $\Psi_{BC}(x,y) = B\sin x - C\sin
y$~\cite{Biferale:PF:2725}. The flow geometry transitions between two limiting
regimes as $B$ and $C$ vary. When $B = 0$ or $C = 0$, BC-flow reduces to a
unidirectional shear flow in the $x$- or $y$-direction, respectively; in this
case the spectral measure $\mu_{11}$ concentrates at the spectral origin and
$\mu_{22} \equiv 0$ (or vice versa)~\cite{Avellaneda:CMP-339}. When $B = C$, the
streamlines of $\Psi_{BC}$ form closed cellular structures, and the flow is a
cell flow for which $\Dg^*_{kk} \sim \varepsilon^{1/2}$ as $\varepsilon \to
0$~\cite{Fannjiang:1994:SIAM_JAM:333}. For $0 < B \neq C$, the flow transitions
between these two extremes, with asymmetric cellular structures; this asymmetry
follows from the stream function symmetry $\Psi_{BC}(x,y;B,C) =
-\Psi_{BC}(y,x;C,B)$, which implies that interchanging $B \longleftrightarrow C$
recovers the original flow under a $90^\circ$ rotation~\cite{Murphy:ADSRF-2019}.

The fluid velocity field for the 2D steady cat's eye flow is given by
\begin{align}\label{eq:cat_eye_flow}
\vecu(\vecx)=(-\sin x\cos y+A\cos x\sin y,\cos x\sin y-A\sin x\cos y)\,,
\end{align}
where $A \in [0,1]$ is a free parameter controlling the flow geometry,
and the corresponding stream function is $\Psi_{CE}(x,y) = \sin x\sin y
+ A\cos x\cos y$~\cite{Fannjiang:1994:SIAM_JAM:333}. The flow is
incompressible ($\bnabla\bcdot\vecu_{CE} = 0$) for all $A$, and
satisfies the steady Euler equations, since the vorticity $\omega =
-\Delta\Psi_{CE} = \Psi_{CE}$ satisfies
$\vecu_{CE}\bcdot\bnabla\omega = 0$. When $A = 0$, the streamlines of
$\Psi_{CE}$ form closed cellular structures and the flow is a cell
flow for which $\Dg^*_{kk}\sim\varepsilon^{1/2}$ as $\varepsilon\to
0$~\cite{Fannjiang:1994:SIAM_JAM:333}. When $A = 1$, the streamlines
become open channels directed along the diagonal $y = x$, and the flow
is a shear flow~\cite{Murphy:ADSRF-2019}. For $0 < A < 1$, the flow
transitions between these two regimes; the streamlines are symmetric
about the line $y = x$ for all $A$, which implies $\mu_{11} =
\mu_{22}$~\cite{Murphy:ADSRF-2019}.

The fluid velocity field for the 2D time-periodic flow is given by
\begin{align}\label{eq:space_time_periodic_u}
\vecu(t,\vecx)=(\cos y+A\cos t\sin y,\cos x+A\cos t\sin x)\,.
\end{align}
The steady part $(\cos y, \cos x)$ of this flow is BC-cell flow with $B=C=1$,
subject to a time-periodic perturbation $A\cos t\,(\sin y, \sin x)$ with
amplitude $A \geq 0$. The addition of the time-periodic term gives rise to a
transition to Lagrangian chaos for $A > 0$, in contrast to the integrable
streamline geometry of the steady BC-cell flow 
\cite{Biferale:PF:2725,ZCX_2015,Lag18}.
In the context of \emph{residual
diffusivity}~\cite{Biferale:PF:2725,ZCX_2015,Lyu:2017:NMTMA:10:351,Lyu:2018:CMS:16:2033},
the dynamic ($A > 0$) and steady ($A = 0$) flows exhibit markedly different
large-$\Pen$ behavior: the effective diffusivity for the steady cell flow
satisfies $\Dg^*_{kk}\sim\varepsilon^{1/2}$ as $\varepsilon\to
0$~\cite{Fannjiang:1994:SIAM_JAM:333}, while the time-periodic perturbation
gives rise to residual diffusivity, with $\Dg^*_{kk}$ flattening to an $O(1)$
value as $\varepsilon\to 0$~\cite{Biferale:PF:2725,Lyu:2017:NMTMA:10:351,
Murphy:ADSTPF-2017,Lag18,XinLyu:2020super}.

The fluid velocity field for the 3D steady ABC-flow is given by
\begin{align}\label{eq:ABC-flow}
\vecu(\vecx)=(A\sin z+C\cos y,B\sin x+A\cos z,C\sin y+B\cos x)\,.
\end{align} 
The Arnold--Beltrami--Childress (ABC)
flow~\cite{Arnold:1965:CRAS:261:17,Dombre:1986:JFM:167:353} belongs to the class
of \emph{Beltrami flows}, characterized by the curl identity
$\bnabla\btimes\vecu = \vecu$. This identity holds componentwise
for~\eqref{eq:ABC-flow} for all values of the parameters $A$, $B$, $C$, as may
be verified directly. The Beltrami condition together with incompressibility
($\bnabla\bcdot\vecu = 0$) imply $\Delta\vecu = -\vecu$ and
$(\vecu\bcdot\bnabla)\vecu = \bnabla(|\vecu|^2/2)$, so the steady incompressible
Euler equations $(\vecu\bcdot\bnabla)\vecu = -\bnabla p$ are satisfied with
pressure $p = -|\vecu|^2/2$; thus, the ABC-flow~\eqref{eq:ABC-flow} is an exact
steady solution to the three-dimensional inviscid Euler equations for all values
of $A$, $B$, $C$. 
For generic parameter values with $A$, $B$, $C$ all nonzero, Lagrangian particle
trajectories of the ABC-flow exhibit chaotic
behavior~\cite{Dombre:1986:JFM:167:353} with co-existence of ballistic orbits
\cite{ABC_16a,ABC_16b}. The latter coherent structures are mainly responsible
for enhancing 
diffusion relative to molecular diffusion, making the ABC-flow a canonical
three-dimensional benchmark for effective diffusivity in chaotic (but
non-ergodic) velocity fields~\cite{Kao:2022:SIAM:MMS:20:107}. Numerical
computation of $\Dg^*_{kk}$ for the steady and time-dependent ABC-flow has been
carried out using stochastic structure-preserving Lagrangian schemes
\cite{Lag18,Lag20,Lag21,Lag22}.

The fluid velocity field for the 3D steady Kolmogorov flow~\cite{Childress:Gilbert:1995:FastDynamo} is given by
\begin{align}\label{eq:Kolmogorov_flow}
\vecu(\vecx)=(\sin z,\sin x,\sin y)\,.
\end{align}
The flow~\eqref{eq:Kolmogorov_flow} is incompressible ($\bnabla\bcdot\vecu =
0$), and each component is an eigenfunction of $-\Delta$ with eigenvalue $1$, so
$\Delta\vecu = -\vecu$. 
The cyclic coordinate structure of \eqref{eq:Kolmogorov_flow} — in which each
component depends on a distinct spatial variable shifted by one index — is
preserved under the simultaneous cyclic permutation of coordinates and velocity
components, which implies $\Dg^*_{11} = \Dg^*_{22} = \Dg^*_{33}$ by symmetry.
The 3D Kolmogorov flow serves as a standard benchmark for effective diffusivity
in three-dimensional flows, and has been studied numerically using stochastic
structure-preserving
schemes \cite{Lag18,Lag21,Lag22}
and semi-Lagrangian methods~\cite{Kao:2022:SIAM:MMS:20:107}.

The fluid velocity field for the 3D time-periodic Kolmogorov flow is given by
\begin{align}\label{eq:space_time_periodic_3D_u}
\vecu(t,\vecx)=(\sin z+\theta\cos t\cos z,\sin x+\theta\cos t\cos x,\sin y+\theta\cos t\cos y)\,.
\end{align} 
The flow~\eqref{eq:space_time_periodic_3D_u} is incompressible.
At $\theta = 0$, \eqref{eq:space_time_periodic_3D_u} reduces to
the steady 3D Kolmogorov flow~\eqref{eq:Kolmogorov_flow}; for $\theta > 0$, the
time-periodic modulation $\theta\cos t\,(\cos z, \cos x, \cos y)$ perturbs the
steady flow and gives rise to Lagrangian chaos \cite{Lag18}.
The cyclic coordinate symmetry of the steady part is preserved
in~\eqref{eq:space_time_periodic_3D_u}: the flow is invariant under the
simultaneous cyclic permutation $(x,y,z) \to (y,z,x)$ of coordinates and
velocity components, which implies $\Dg^*_{11} = \Dg^*_{22} = \Dg^*_{33}$ by
symmetry. Numerical computation of $\Dg^*_{kk}$ for the time-dependent 3D
Kolmogorov flow has been carried out using stochastic structure-preserving
Lagrangian schemes \cite{Lag21,Lag22}.

\section{Bounds for effective diffusivity}
\label{sec:bounds}
In this section we discuss Pad\'{e} approximant bounds for Stieltjes functions
with a positive measure, and how the bounds are determined by the moments of the
measure. Pad{\'e} approximants $\{[N-1/N]\}_{N=1}^\infty$ and
$\{[N/N]\}_{N=1}^\infty$ form rigorous lower and upper bounds for the Stieltjes
function $f(z)$   
\begin{align}\label{eq:Pade_bounds}
[N-1/N](z)\le f(z) \le [N/N](z)\,,
\qquad
f(z)=\int_{-\infty}^\infty\frac{\d \mu(\lambda)}{1+z\lambda^2}\,,      
\end{align}
with \emph{positive} measure $\mu$, which can converge for certain values of
$z$~\cite{Baker:1996:Book:Pade}. To simplify notation, for the remainder of this
section, we will focus on a single diagonal component of the (symmetric part of
the) effective diffusivity, $\Dg^*=\Dg^*_{kk}$, in
\eqref{eq:Integral_Rep_kappa*} and the associated Stieltjes function $f(z)$,
with $z=\varepsilon^{-2}$,
$\Dg^*=\varepsilon(1+\varepsilon^{-2}f(\varepsilon^{-2}))$, and positive
spectral measure $\mu=\mu_{kk}$ with moments $\mu^n$.
Since all of the terms of the expression in
equation~\eqref{eq:Integral_Rep_kappa*} for $\Dg^*$ are positive, we have the
following rigorous lower and upper bounds for
$\Dg^*_{kk}$~\cite{Baker:1996:Book:Pade}
\begin{align}\label{eq:Pade_bounds_diagonal_component} 
  \varepsilon(1+\varepsilon^{-2}[N-1/N](\varepsilon^{-2}))
  \le \Dg^*(\varepsilon) \le
  \varepsilon(1+\varepsilon^{-2}[N/N](\varepsilon^{-2}))\,.
\end{align}

The theory of Pad{\'e} approximants for $f(z)$ follows by expanding
$1/(1+z\lambda^2)$ in a geometric series and writing $f(z)$ as a
\emph{series of Stieltjes}~\cite{Baker:1996:Book:Pade} 
\begin{align}\label{eq:Series_Stieltjes}
f(z)=\sum_{n=0}^\infty c_n\,z^n\,,
\qquad
c_n=(-1)^n\,\mu^{2n},
\qquad
\mu^{2n}=\int_{-\infty}^\infty \lambda^{2n} \,\d\mu(\lambda)\,.
\end{align}
The $L$, $N$ Pad{\'e} approximant
of $f(z)$ is given by
%
\begin{align}\label{eq:Pade_polynomials}
  [L/N](z)=\frac{P^{[L/N]}(z)}{Q^{[L/N]}(z)}\,,
\end{align}
where
$P^{[L/N]}(z)$ is a polynomial of degree of at most $L$ and
$Q^{[L/N]}(z)$ is a polynomial of degree of at most $N$. The
\emph{formal} power series for $f(z)$ in
equation~\eqref{eq:Series_Stieltjes} and the condition
$f(z)-P^{[L/N]}(z)/Q^{[L/N]}(z)=O(z^{L+N+1})$
determines the coefficients of $P^{[L/N]}(z)$ and $Q^{[L/N]}(z)$ and these 
polynomials can be written in terms of determinants involving the $c_n$ 
\cite{Baker:1996:Book:Pade}, which are given in terms 
of the measure moments $\mu^{2n}$.

The iterative moment method described in Section \ref{sec:Iterative_moments}
enables many moments, hence Pad\'e approximant bounds to be calculated for
$\Dg^*$. In Appendix \ref{sec:moment_calculations_numerical} we describe our
numerical implementation of  
the iterative moment method. Moments for the fluid velocity fields in equations
\eqref{eq:BC-flow}--\eqref{eq:space_time_periodic_3D_u} were calculated exactly
in closed form using the symbolic mathematics software Maple and Python-SymPy,
and to floating-point precision using MATLAB and Python-NumPy, with relative
differences between the two implementations less than $10^{-14}$ for all moments
computed. In Section \ref{sec:numerical_implementation_pade} we utilize the
moments calculated in Appendix \ref{sec:moment_calculations_numerical} to
calculate Pad\'{e} approximant bounds for Stieltjes functions associated with
$\Dg^*_{kk}$. The resultant software repository, Janus, is publicly available on
GitHub. We discuss these results.

\subsection{Bound accuracy for steady and dynamic flows} 
\label{sec:bound_accuracy}   

The Pad\'{e} bounds presented in Sections
\ref{sec:bounds_2D_steady_flows}–\ref{sec:bounds_space_time_flows} below are
computed from spectral measure moments $\mu^{2n}$, $2n = 0, \ldots , 60$,
obtained via the iterative algorithm of Section \ref{sec:Iterative_moments}. For
steady flows, the operator $M$ is compact and the moments decay exponentially
with order (Figure \ref{fig:moments_decimal_steady_flows}); for space-time
periodic flows, $M$ is unbounded and the moments grow exponentially (Figure
\ref{fig:moments_decimal_dynamic_flows}). This qualitative distinction has
important consequences for the accuracy of the Pad\'{e} bounds: steady flows
yield well-conditioned moment sequences amenable to high-order approximation,
while the exponential growth for dynamic flows limits the effective Pad\'{e}
order that can be robustly computed. To stabilize the numerical computation of
Pad\'{e} approximants, we employ the robust SVD-based algorithm
\texttt{padeapprox} \cite{Gonnet:2013:55:1:101:110853236} together with a moment
scaling procedure that reduces the dynamic range of the moment sequence. Full
details of the moment computation, complexity analysis, and Pad\'{e}
stabilization procedures are given in Appendix
\ref{sec:moment_calculations_numerical}.

\begin{figure}[htbp]  
  \centering
  \includegraphics[width=0.48\textwidth]{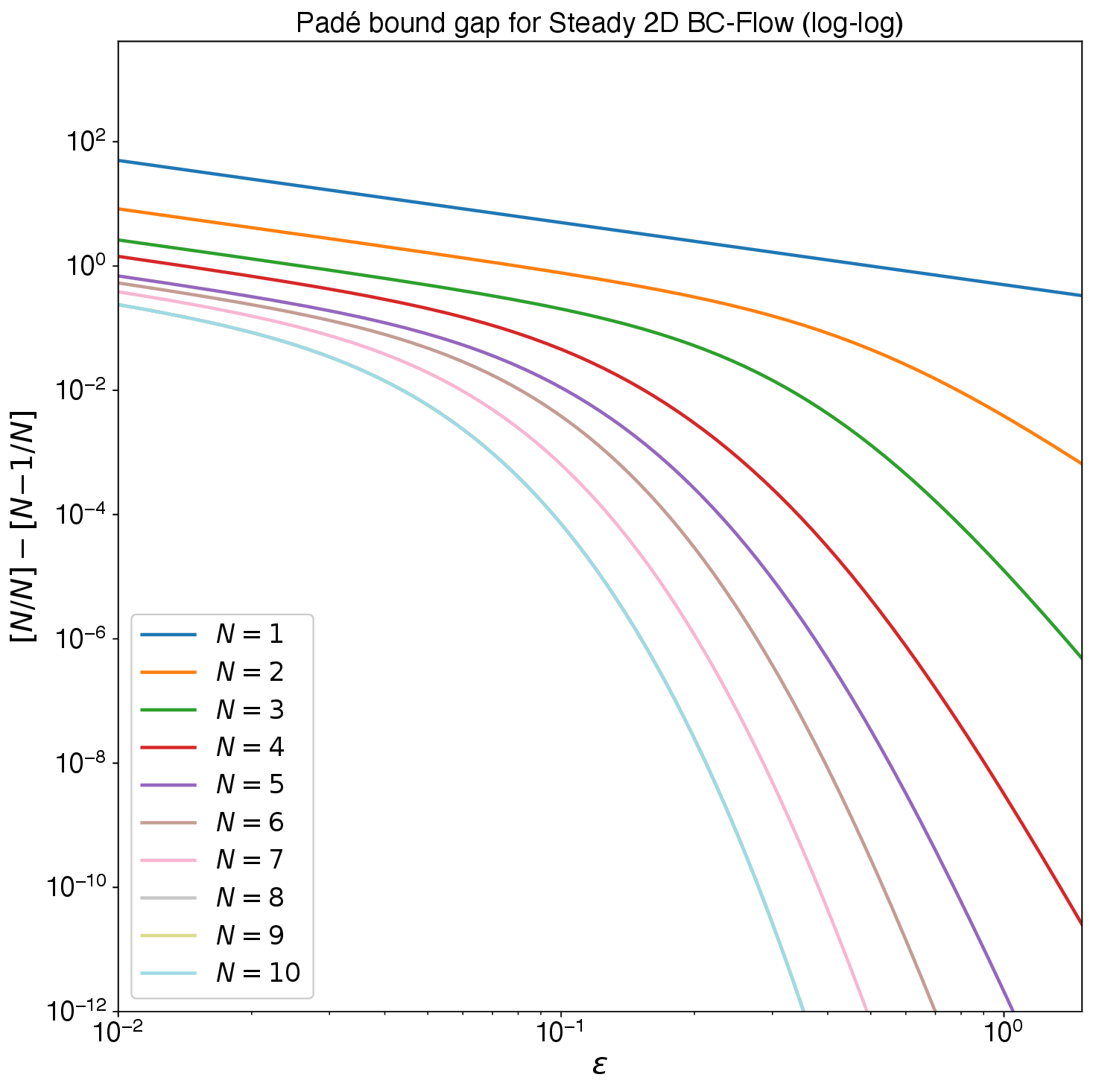}%
  \hfill
  \includegraphics[width=0.48\textwidth]{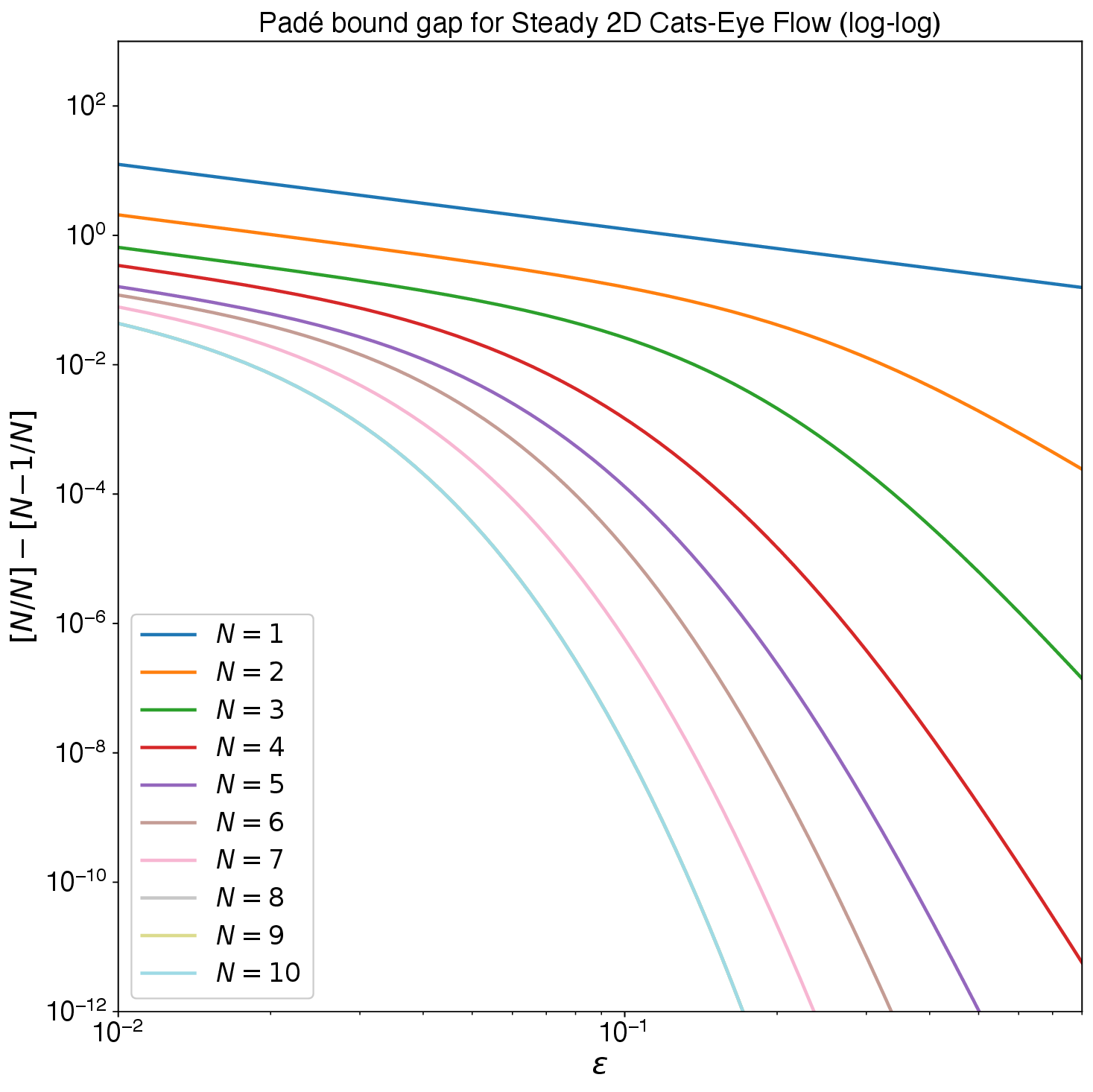}\\[6pt]
  \includegraphics[width=0.48\textwidth]{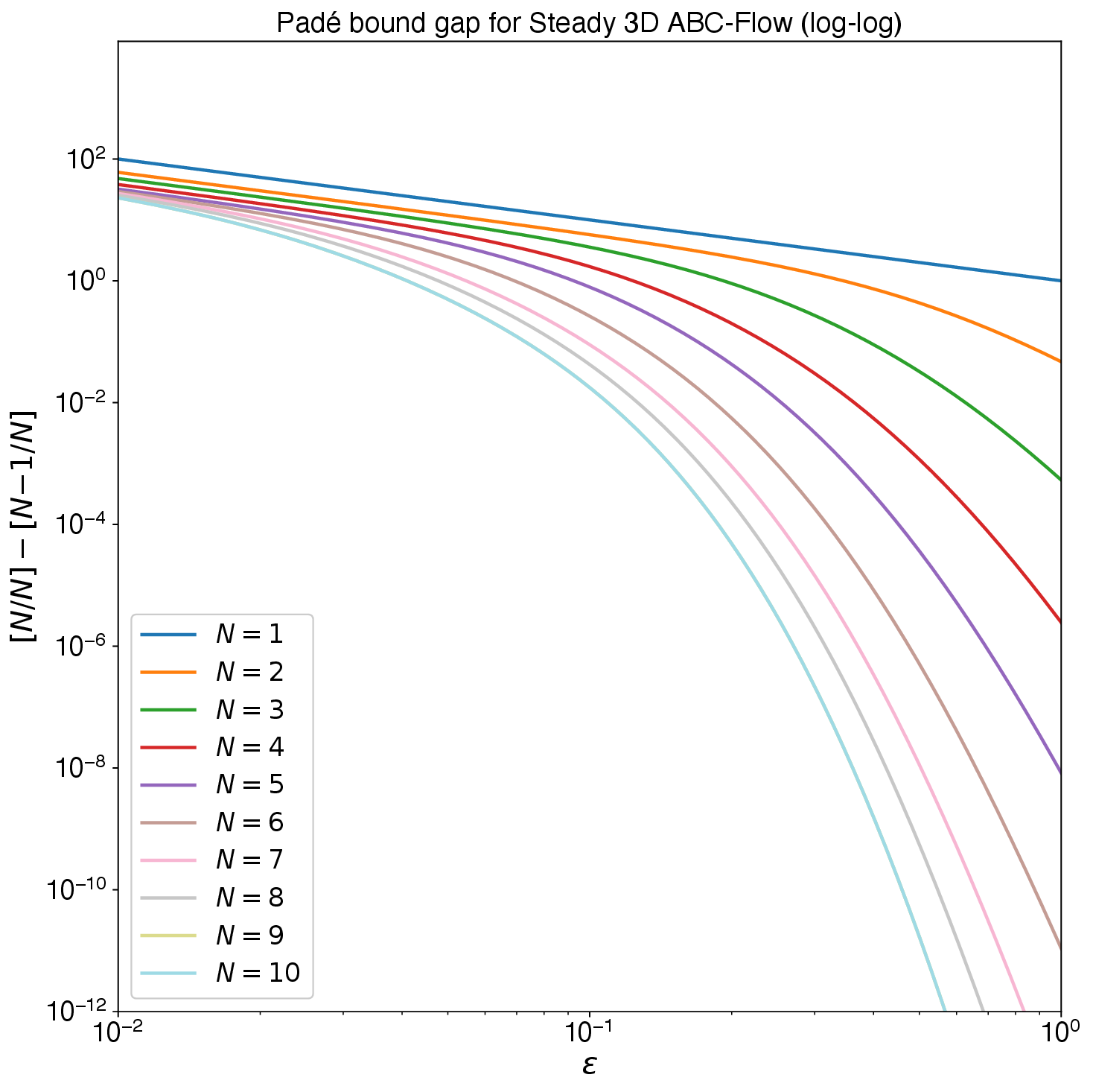}%
  \hfill
  \includegraphics[width=0.48\textwidth]{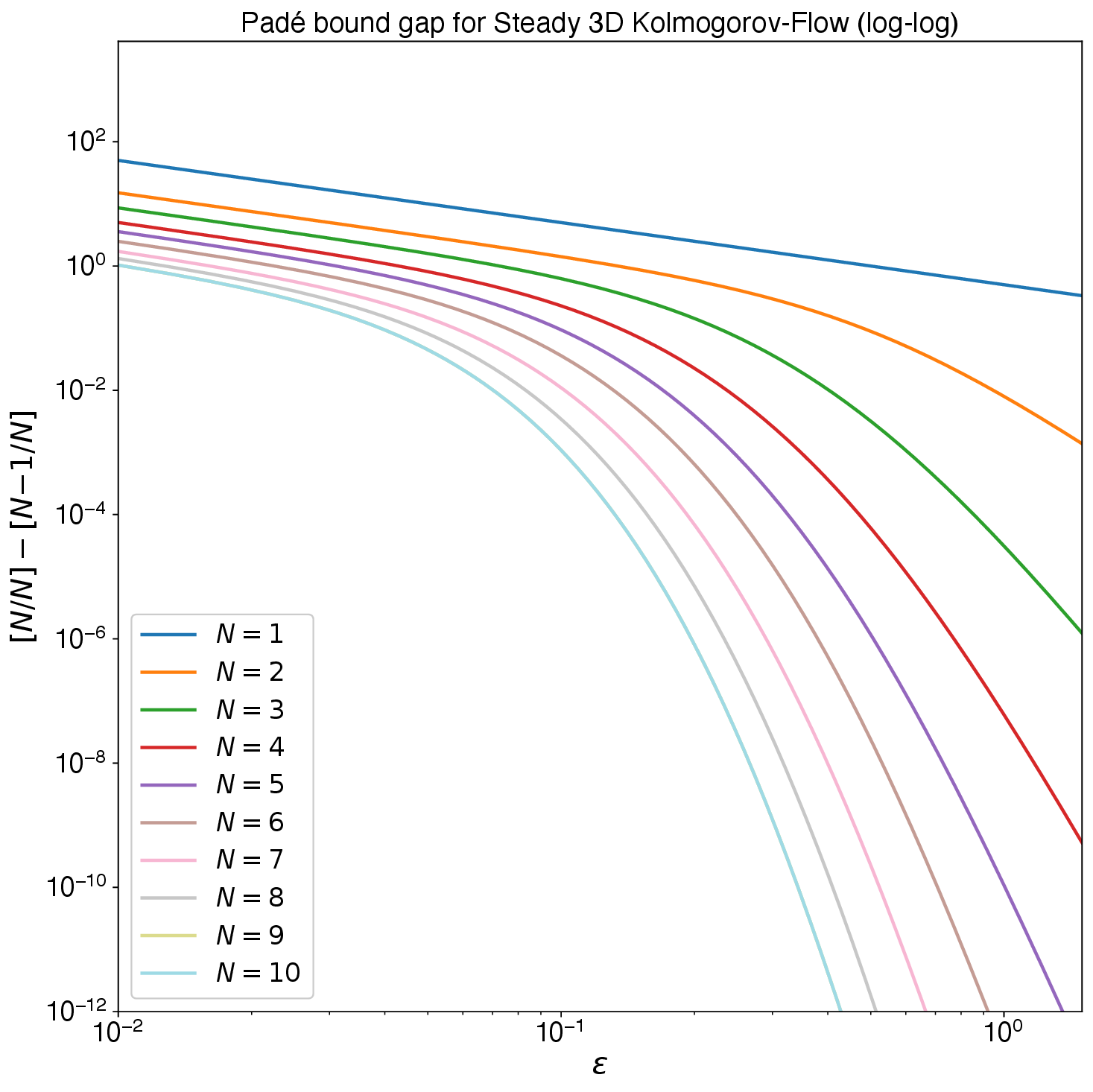}\\[6pt]
  \includegraphics[width=0.48\textwidth]{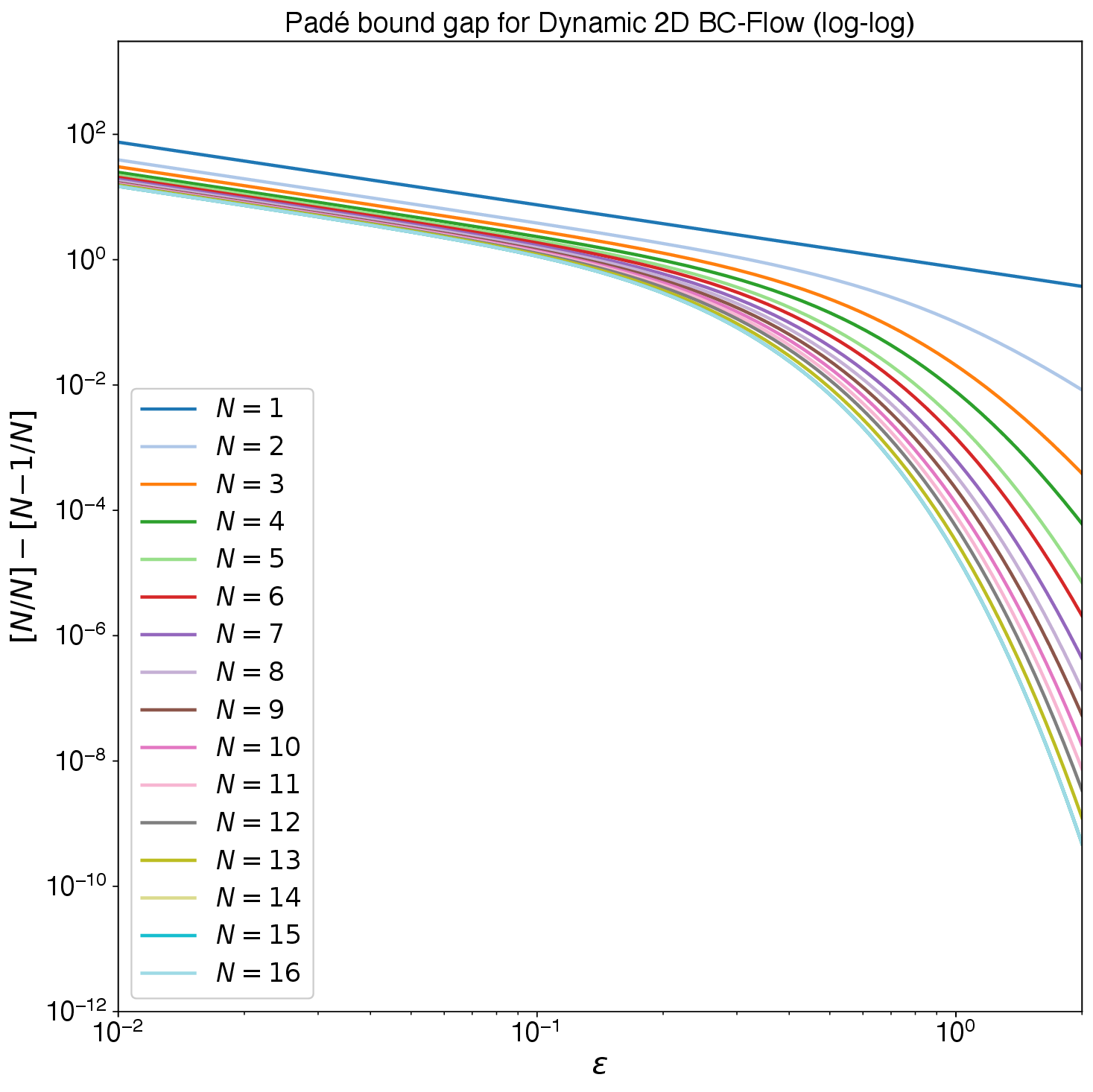}%
  \hfill
  \includegraphics[width=0.48\textwidth]{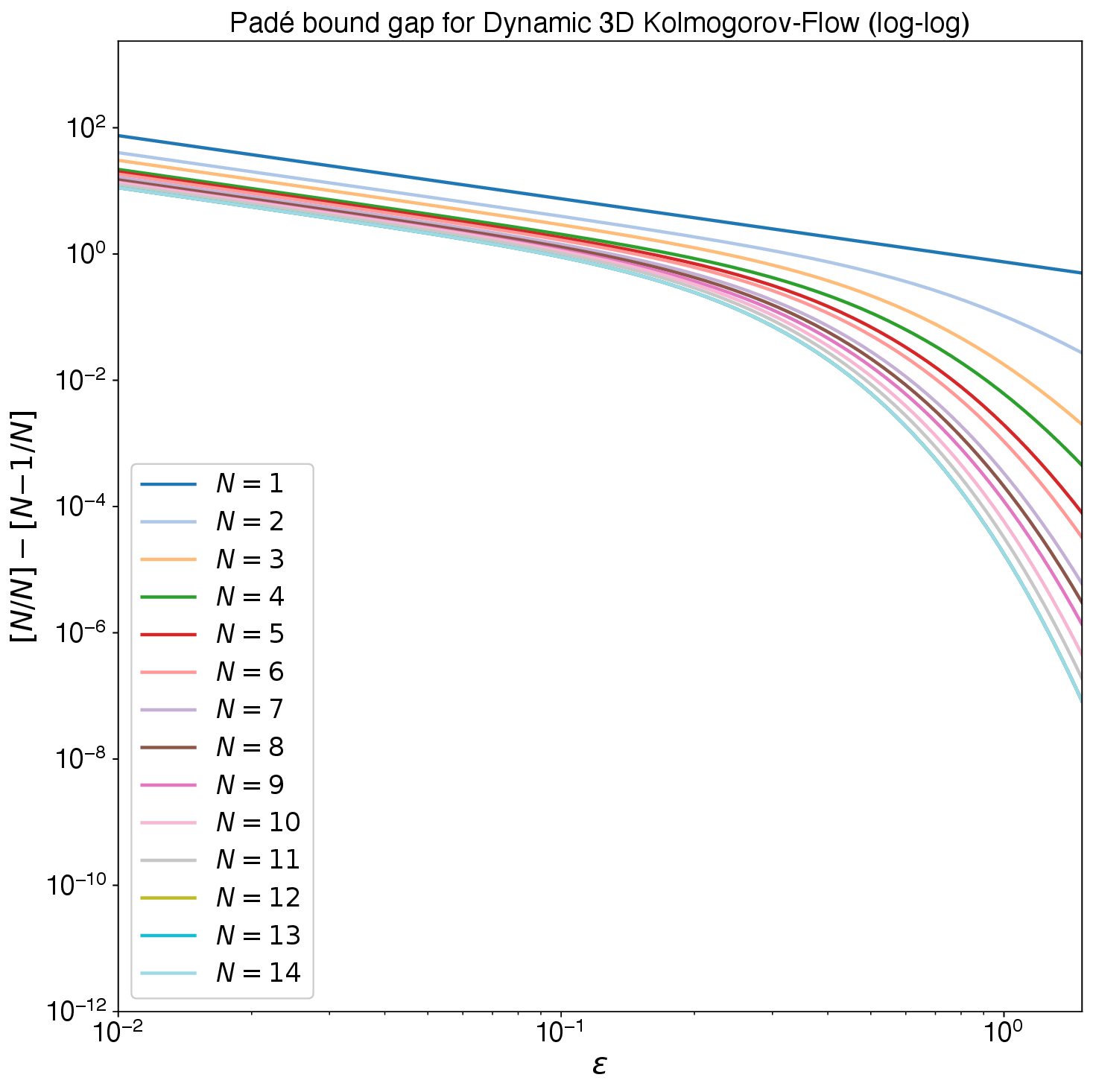}
  \caption{\textbf{Pad\'{e} bound differences for effective diffusivity.}
    Log-log plots of the differences $[N/N]-[N-1/N]$ between upper and lower
    Pad\'{e} bounds for $\Dg^*_{kk}(\varepsilon)$ as a function of
    $\varepsilon$, for various values of $N$. Results are shown for (top row)
    steady 2D BC-flow~\eqref{eq:BC-flow} and
    cats-eye flow~\eqref{eq:cat_eye_flow}, (middle row) steady 3D
    ABC-flow~\eqref{eq:ABC-flow} and 3D Kolmogorov flow, and (bottom row)
    dynamic 2D BC-flow~\eqref{eq:space_time_periodic_u} and dynamic 3D
    Kolmogorov flow~\eqref{eq:space_time_periodic_3D_u}.}
  \label{fig:Pade_gaps}
\end{figure}

Figure~\ref{fig:Pade_gaps} displays, in log-log scale, the differences
$[N/N]-[N-1/N]$ between the upper and lower Pad\'{e} bounds for
$\Dg^*_{kk}(\varepsilon)$ as a function of $\varepsilon$, for each of the six
flows considered in this paper. In every case the bound differences decrease
monotonically with increasing $N$, confirming that each additional pair of
spectral measure moments yields strictly tighter bounds for $\Dg^*_{kk}$. For
all flows the bounds are most accurate at large $\varepsilon$ (small $\Pen$),
where the differences decay steeply, and become progressively less tight as
$\varepsilon\to 0$ (large $\Pen$), as anticipated. For the two 2D steady flows
(top row), the bounds are accurate to within $10^{-2}$ for the interval
$10^{-2}\lesssim\varepsilon\lesssim10^{-1}$. We will see that this is sufficient
accuracy to capture the powerlaw behavior of these 2D cell-flows
$\Dg^*\sim\sqrt{\varepsilon}$ for $\varepsilon\ll1$
\cite{Fannjiang:1994:SIAM_JAM:333}. Successive orders are well-separated across
the full range of $\varepsilon$ shown, demonstrating each increment in $N$
produces a substantial improvement in the bounds.

The bound differences for 3D
ABC flow and 3D Kolmogorov flow (middle row) are qualitatively similar to the 2D
steady flows but only achieve an accuracy of $\sim10^{-1}$ for
$\varepsilon\sim10^{-1}$ with decreasing accuracy for smaller $\varepsilon$. The
dynamic flows (bottom row) exhibit markedly different behavior. The gap curves
for successive orders are much more tightly clustered, particularly at small
$\varepsilon$, indicating slower convergence of the approximants in the
large-$\Pen$ regime---perhaps partially due to the exponential growth of the
moments for the dynamic flows and the associated challenges in the moment
scaling shown in Figure~\ref{fig:scaling_diagnostics} and bound stabilization.
The bounds are less accurate, with an accuracy $\gtrsim10^{0}$ for all
$\varepsilon\lesssim10^{-1}$.

\subsection{Bounds for 2D steady flows} 
\label{sec:bounds_2D_steady_flows}        

\begin{figure}[htbp]
  \centering
  \includegraphics[width=0.48\textwidth]{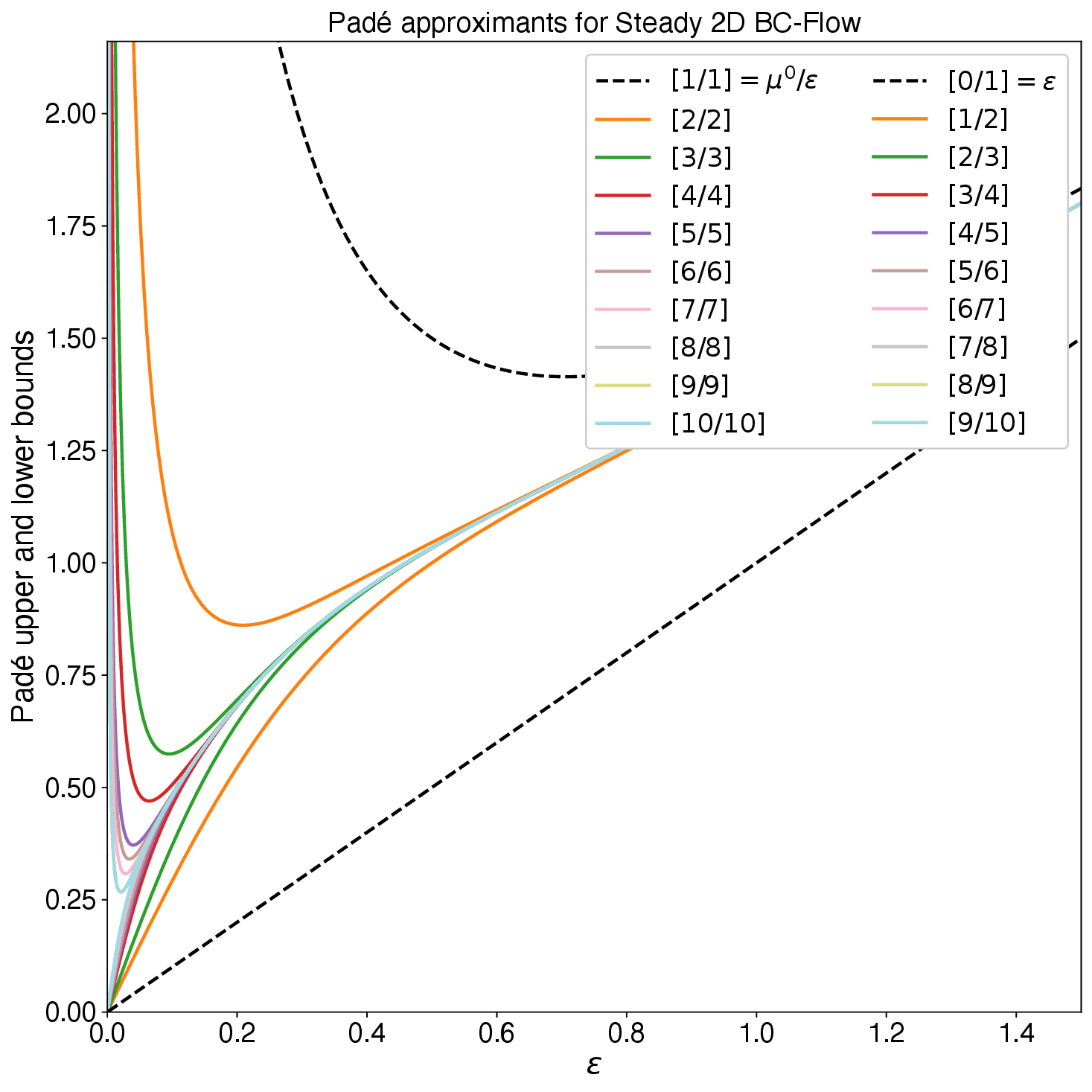}%
  \hfill
  \includegraphics[width=0.48\textwidth]{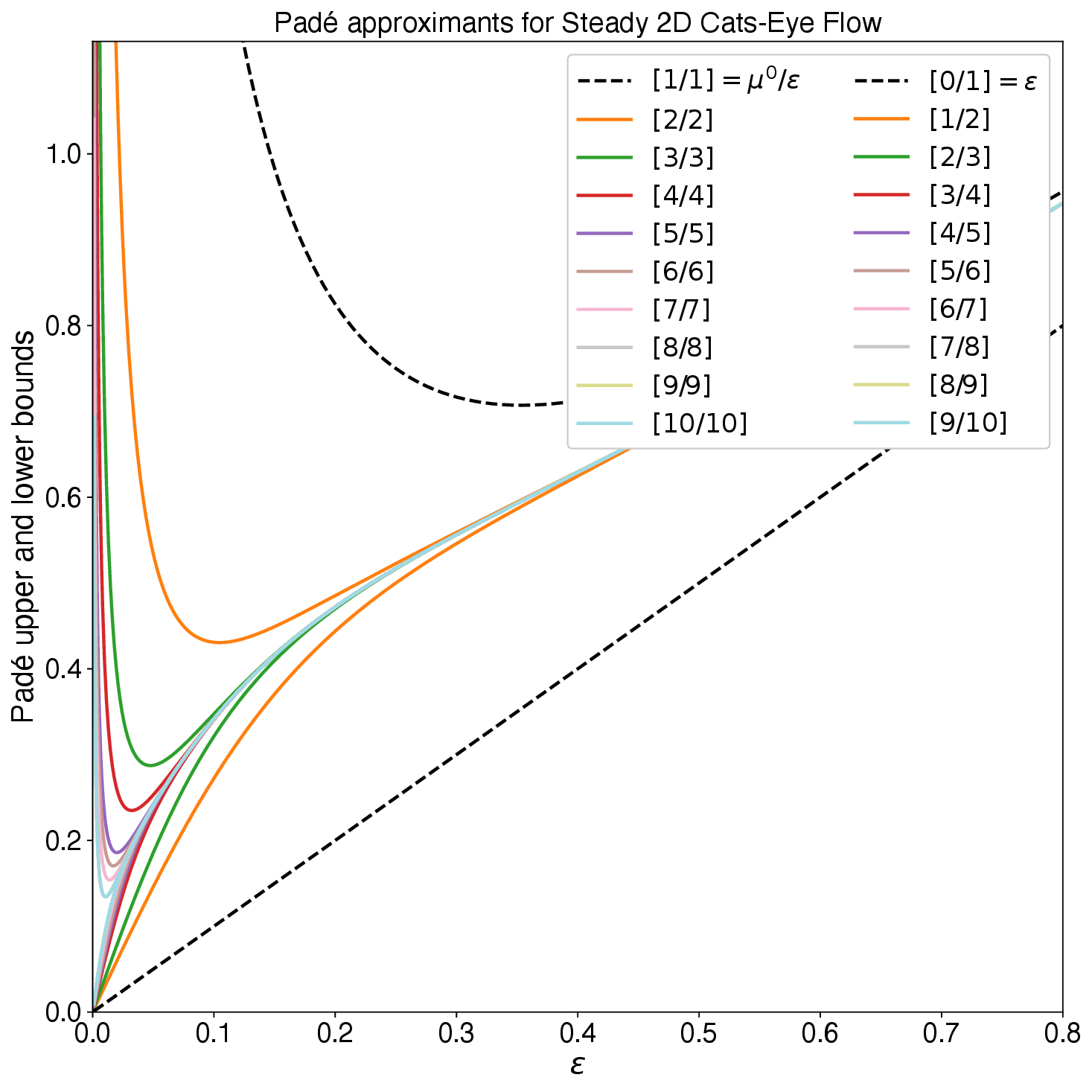}\\[6pt]
  \includegraphics[width=0.48\textwidth]{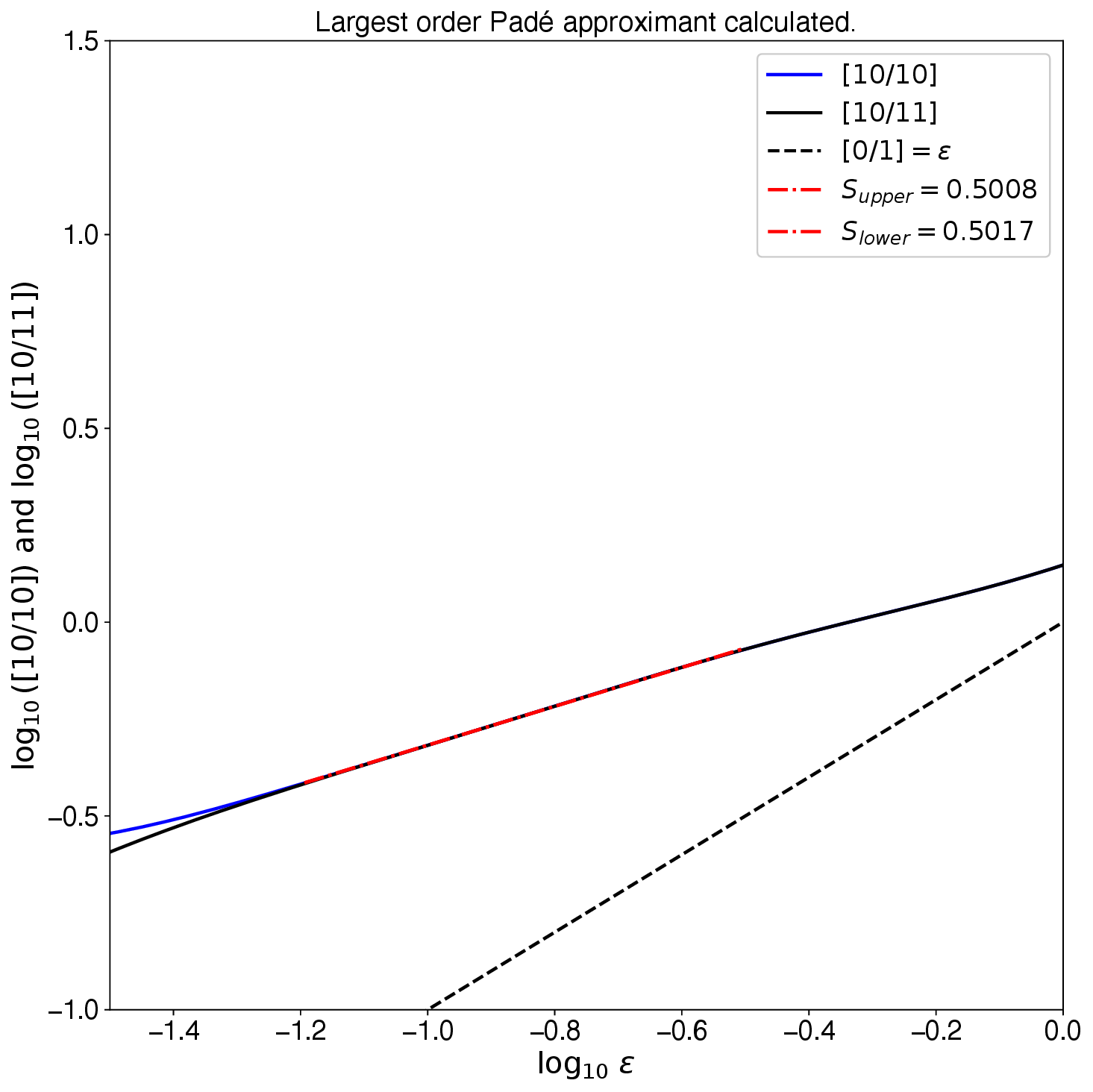}%
  \hfill
  \includegraphics[width=0.48\textwidth]{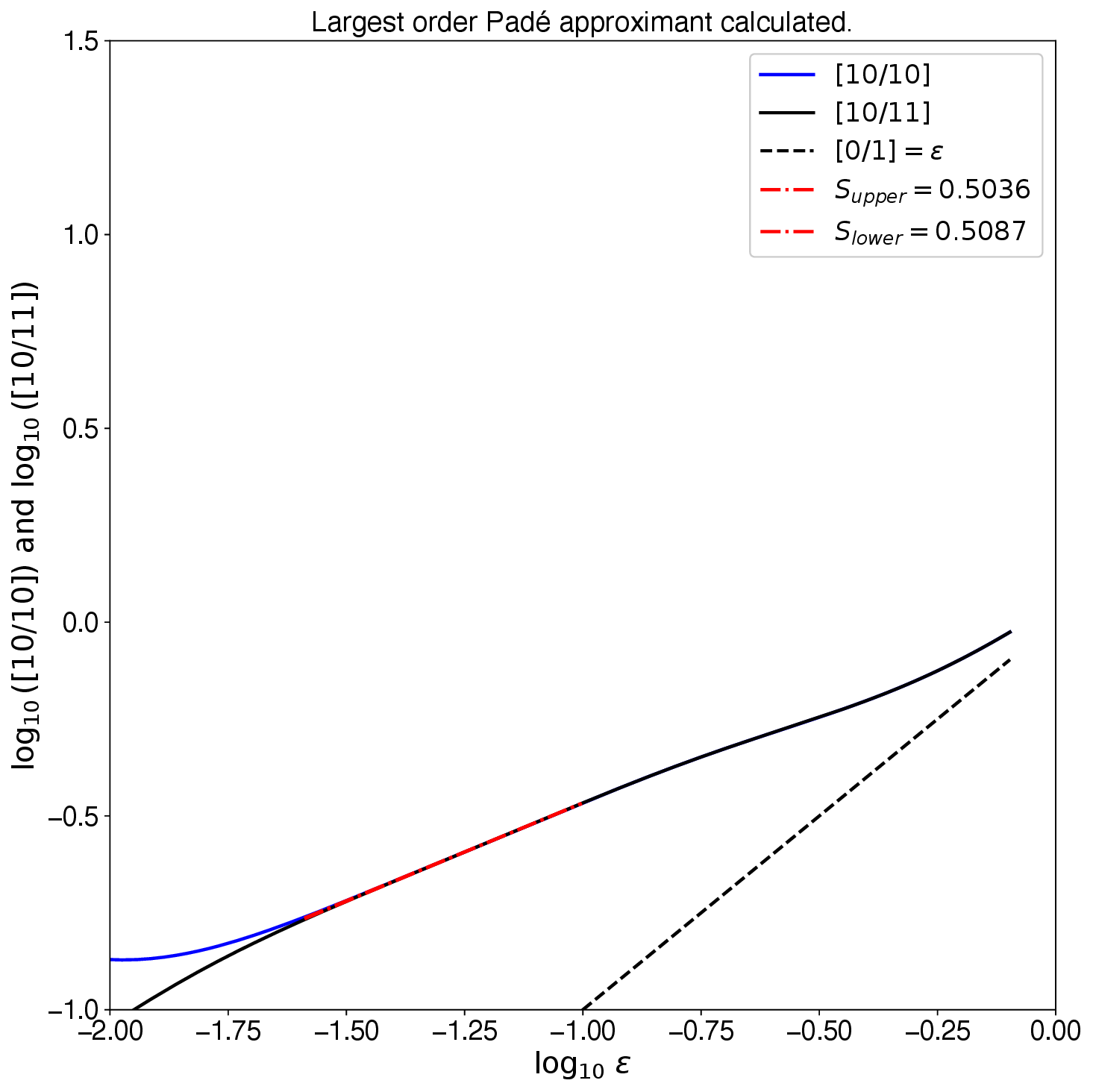}
  \caption{\textbf{Pad\'{e} bounds for 2D steady flows.} (\textit{Top row})
    Linear-scale upper $[N/N]$ and lower $[N-1/N]$ Pad\'{e} bounds for
    $\Dg^*_{kk}(\varepsilon)$ as a function of $\varepsilon$, for
    BC-flow~\eqref{eq:BC-flow} (left) and cats-eye flow~\eqref{eq:cat_eye_flow}
    (right), for successive orders $N$. (\textit{Bottom row}) Corresponding
    log-log plots with power-law fits in the large-$\Pen$ regime
    ($\varepsilon\ll 1$), capturing the known asymptotic behavior
    $\Dg^*_{kk}\sim\varepsilon^{1/2}$ for both flows
    \cite{Fannjiang:1994:SIAM_JAM:333}. Here, $S_{\text{upper}}$ and
    $S_{\text{lower}}$ denote the slopes of the linear fits in log-log scale for
    the highest-order upper and lower bounds, respectively.}
  \label{fig:steady_2D_bounds}
\end{figure}

In this section we present the Pad\'{e} approximant upper $[N/N]$ and lower
$[N-1/N]$ bounds for the effective diffusivity $\Dg^*_{kk}$ for the two steady
2D flows BC-flow~\eqref{eq:BC-flow} for $B=C=1$ and cats-eye
flow~\eqref{eq:cat_eye_flow} for $A=0$. The top row of
Figure~\ref{fig:steady_2D_bounds} shows the corresponding linear-scale upper
$[N/N]$ and lower $[N-1/N]$ bounds, with BC-flow on the left and cats-eye flow
on the right, for successive orders $N$. The bottom row shows the corresponding
log-log plots for the highest order upper and lower bounds calculated, with
linear fits in the large-$\Pen$ (small-$\varepsilon$) regime. For both flows,
the slopes $S_{upper}$ and $S_{lower}$ of the linear fits for the upper and
lower bounds, respectively, capture the known asymptotic behavior
$\Dg^*_{kk}\sim\varepsilon^{1/2}$ for $\varepsilon\ll1$ for both cell flows
\cite{Fannjiang:1994:SIAM_JAM:333}, with the estimates for the critical exponent
yielding $0.500$, rounded to in the thousanths place. 

For BC-flow and cat's eye flow, the moments $\mu^{2n}$, $2n=0,\ldots,60$,
computed via the iterative algorithm of Section~\ref{sec:Iterative_moments},
were supplied to the Pad\'{e} bound computation algorithm \texttt{padeapprox}
\cite{Gonnet:2013:55:1:101:110853236}. Due to the truncation of the algorithm,
ensuring accurate results, instead of computing bounds up to order $N=29$ bounds
only up to $N=10$ were returned, as shown in Figure~\ref{fig:Pade_orders} for BC
flow. From $\Dg^*_{kk}=\varepsilon(1+\varepsilon^{-2}f(\varepsilon^{-2}))$ with
$f\ge0$, the global lower bound is $\Dg^*_{kk}\ge\varepsilon$, shown as a dashed
black line in Figure~\ref{fig:steady_2D_bounds} (top-left panel). The global
upper bound is attained by shear flow \cite{Avellaneda:CMP-339}, discussed in
Section~\ref{sec:Moments_shear_flow}, where the spectral measure is a delta mass
at the spectral origin, $\mu=\mu^0\,\delta_0(\d\lambda)$, yielding
$\Dg^*_{kk}\le\varepsilon(1+\mu^0/\varepsilon^2)$, also shown as a dashed black
line in Figure~\ref{fig:steady_2D_bounds} (top-left panel). 

The upper and lower bounds in Figure~\ref{fig:steady_2D_bounds} (top row) get
progressively tighter with increasing $N$. For $\varepsilon\gtrsim1$ all
Pad\'{e} bounds lie virtually on top of each other. The highest-order bounds
demonstrate a significant enhancement of the effective diffusivity above
$\varepsilon$ which decreases with decreasing $\varepsilon$, with power-law
behavior $\Dg^*_{kk}\sim\varepsilon^{1/2}$ for $\varepsilon\ll1$
\cite{Fannjiang:1994:SIAM_JAM:333}. These results demonstrate that the bounds
are accurate for BC-flow up to $\log{\epsilon}~\approx-1.2$, equivalently
$\varepsilon\approx0.063$, and for cat's eye flow up to
$\log{\epsilon}~\sim-1.5$, or equivalently $\varepsilon\sim0.032$. This is made
more precisely by the bound differences $[N/N]-[N-1/N]$ in
Figure~\ref{fig:Pade_gaps} (top-left panel).

\subsection{Bounds for 3D steady flows}
\label{sec:bounds_3D_steady_flows}

\begin{figure}[htbp]
  \centering
  \includegraphics[width=0.48\textwidth]{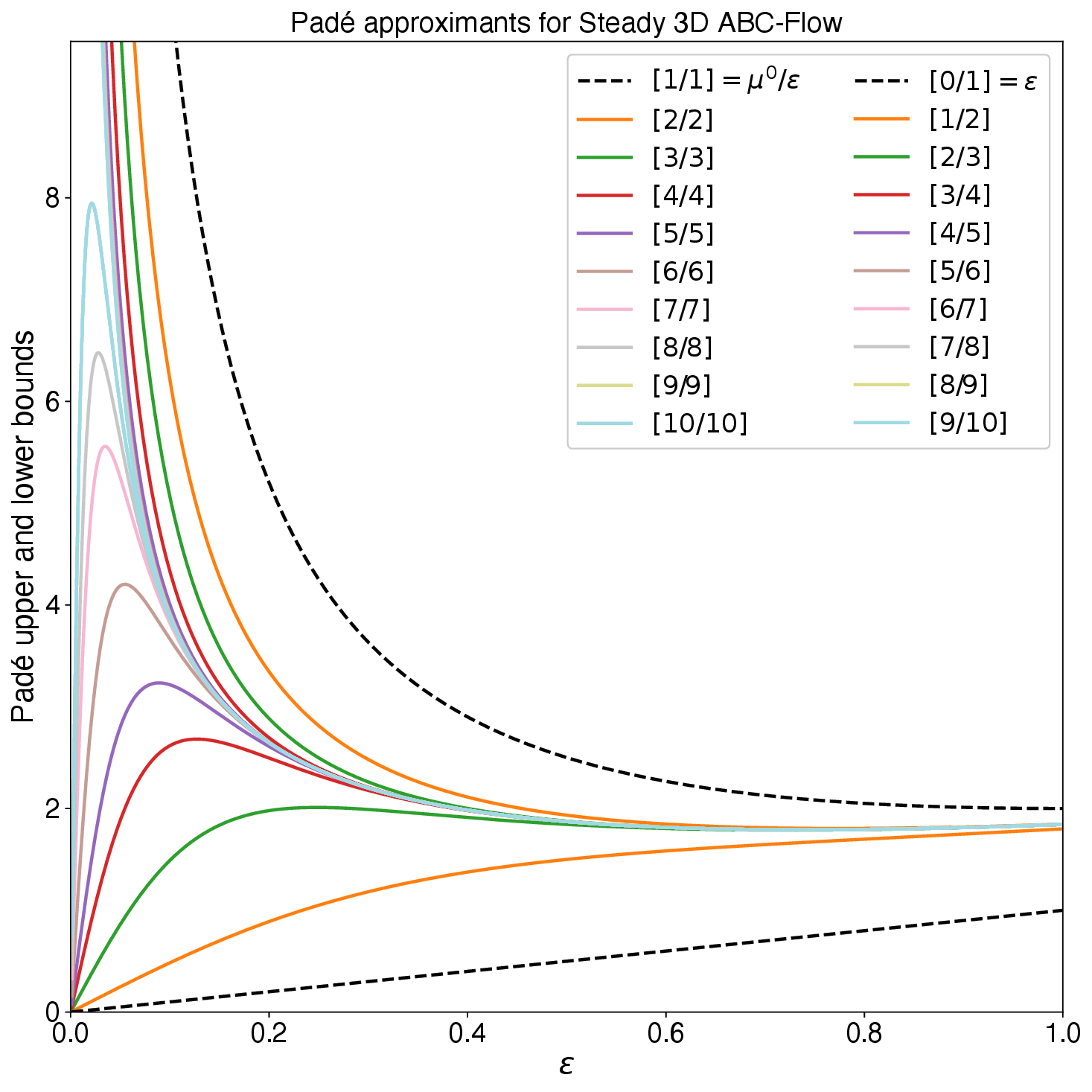}%
  \hfill
  \includegraphics[width=0.48\textwidth]{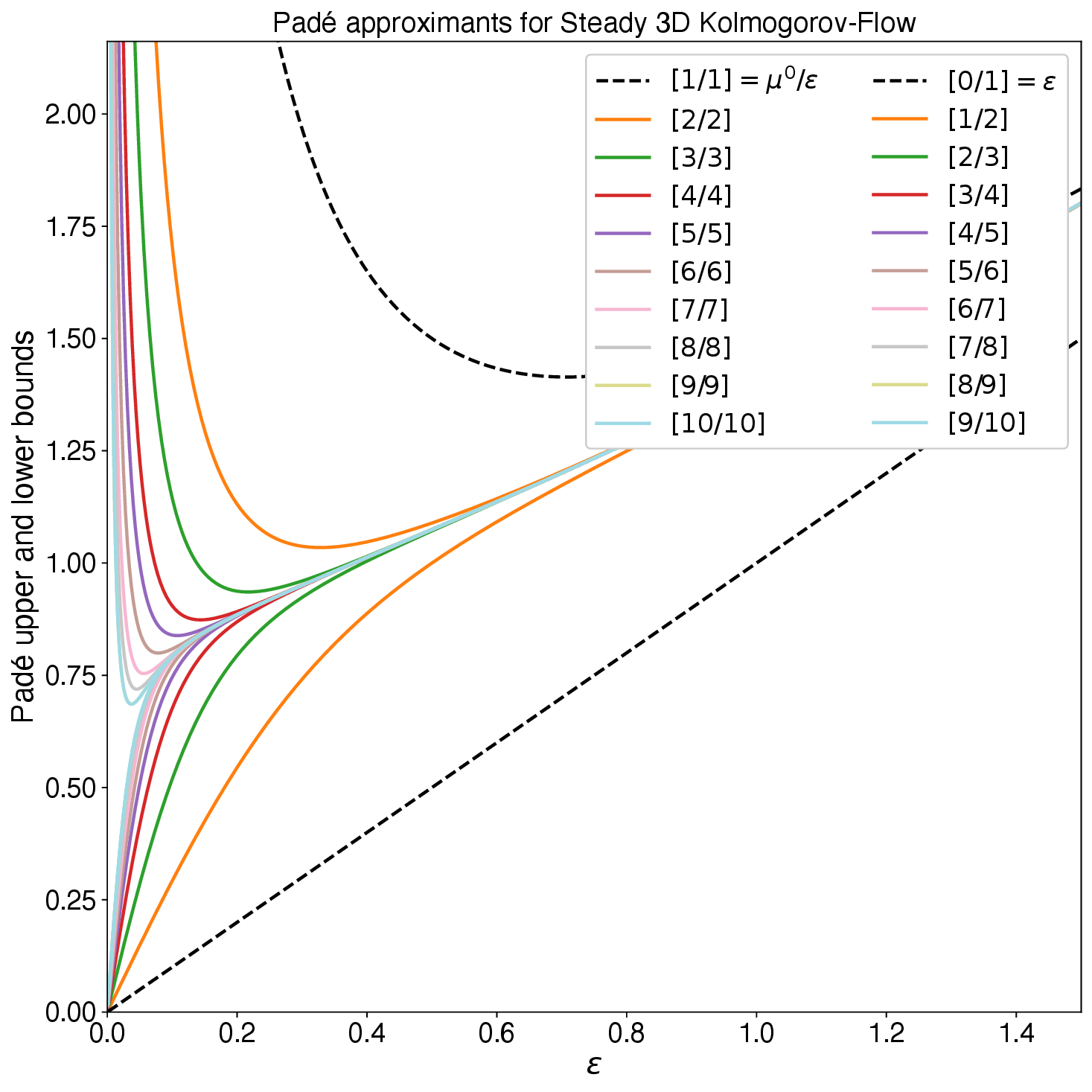}\\[6pt]
  \includegraphics[width=0.48\textwidth]{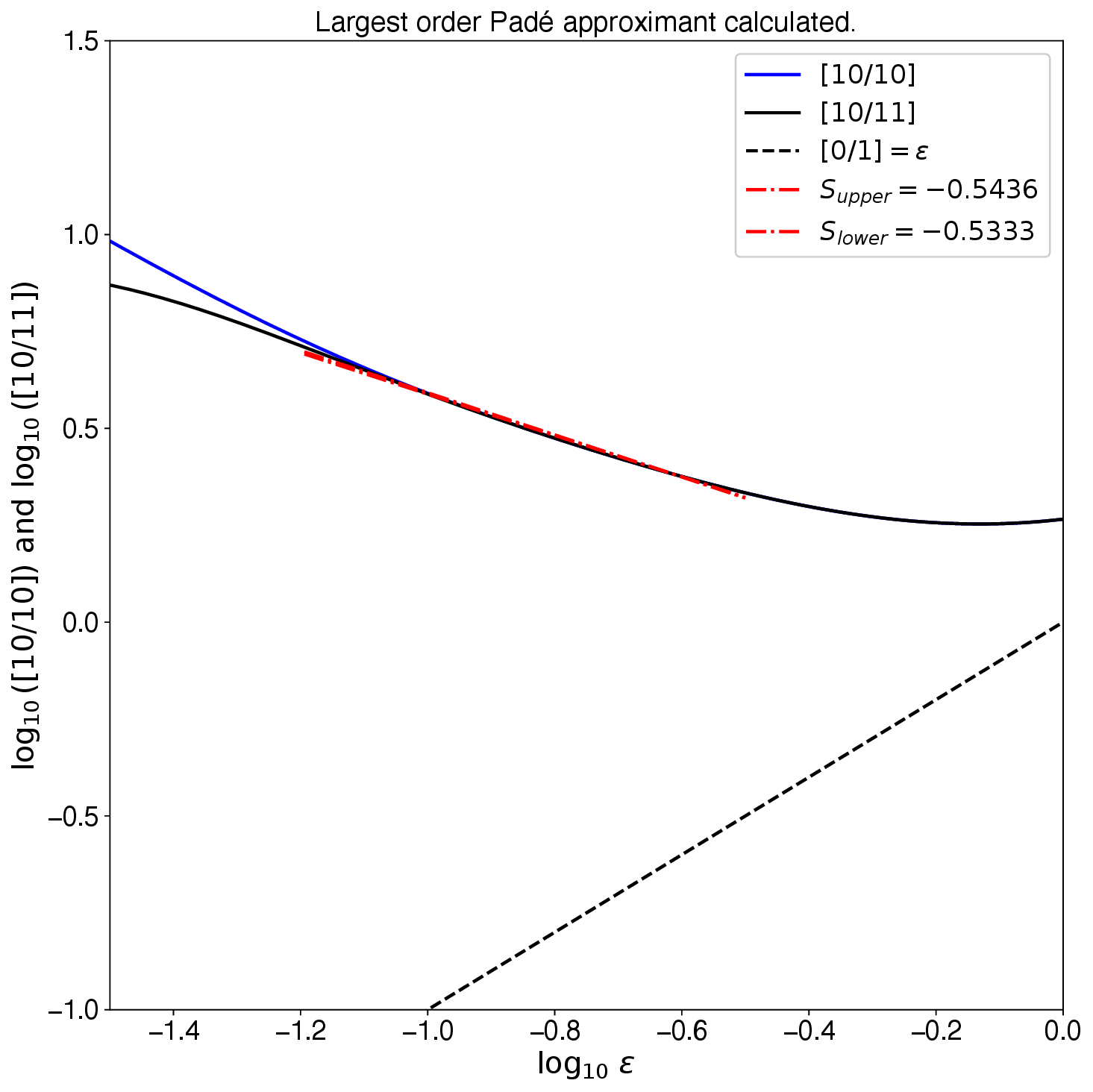}%
  \hfill
  \includegraphics[width=0.48\textwidth]{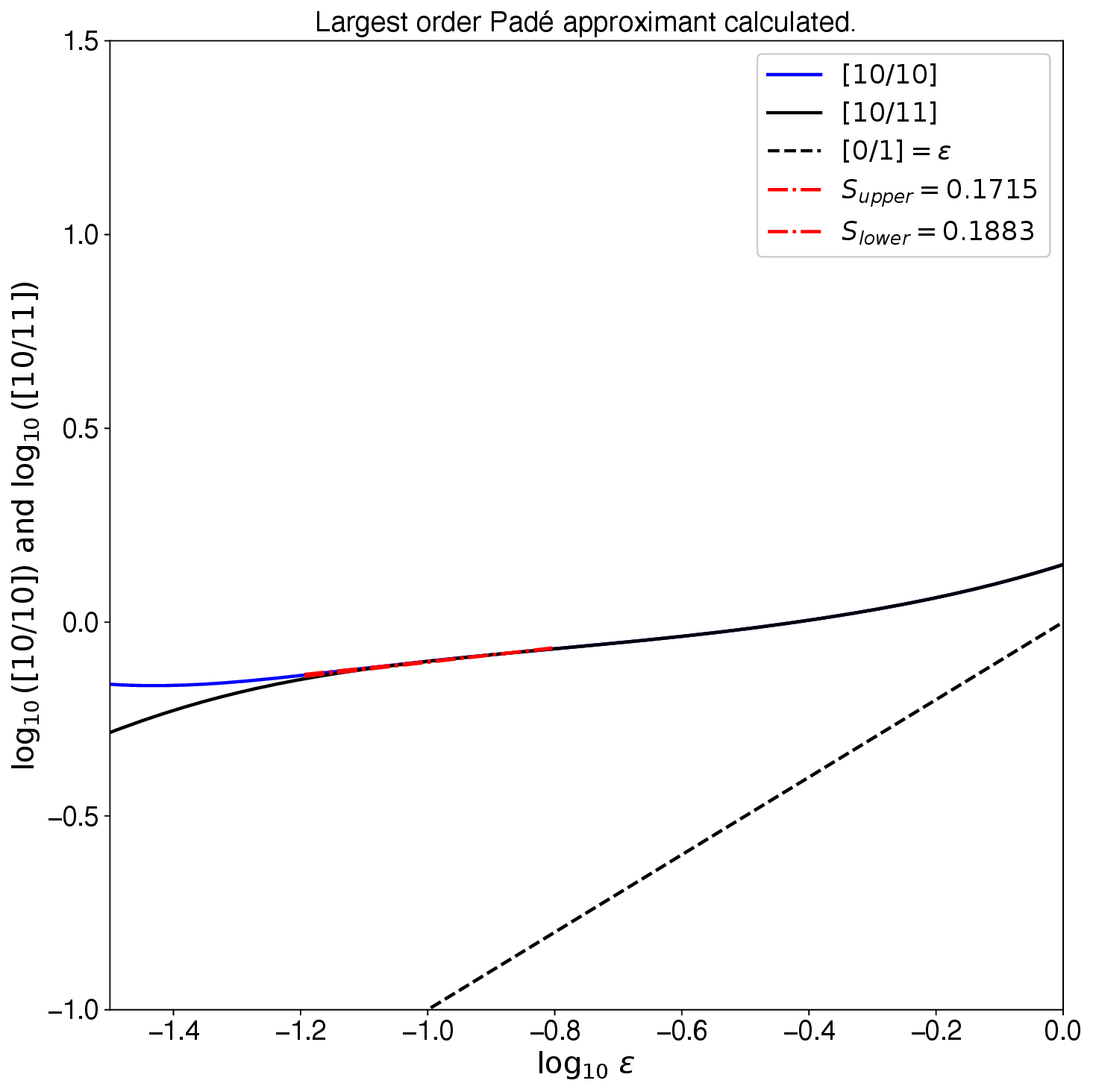}
  \caption{\textbf{Pad\'{e} bounds for 3D steady flows.} (\textit{Top row})
    Linear-scale upper $[N/N]$ and lower $[N-1/N]$ Pad\'{e} bounds for
    $\Dg^*_{kk}(\varepsilon)$ as a function of $\varepsilon$, for
    ABC-flow~\eqref{eq:ABC-flow} (left) and Kolmogorov flow (right), for
    successive orders $N$. (\textit{Bottom row}) Corresponding log-log plots
    with power-law fits in the large-$\Pen$ regime ($\varepsilon\ll 1$). Here,
    $S_{\text{upper}}$ and $S_{\text{lower}}$ denote the slopes of the linear
    fits in log-log scale for the highest-order upper and lower bounds,
    respectively.}
  \label{fig:steady_3D_bounds}
\end{figure} 

In this section we present the Pad\'{e} approximant upper $[N/N]$ and lower
$[N-1/N]$ bounds for the effective diffusivity $\Dg^*_{kk}$ for the two steady
3D flows: ABC-flow~\eqref{eq:ABC-flow} for $A=B=1$ 
and Kolmogorov flow \eqref{eq:Kolmogorov_flow}. The top row of
Figure~\ref{fig:steady_3D_bounds} shows the corresponding linear-scale upper
$[N/N]$ and lower $[N-1/N]$ bounds, with ABC-flow on the left and Kolmogorov
flow on the right, for successive orders $N$. The bottom row shows the
corresponding log-log plots for the highest order upper and lower bounds
calculated, with linear fits in the large-$\Pen$ (small-$\varepsilon$) regime.

For ABC-flow and Kolmogorov flow, the moments $\mu^{2n}$, $2n=0,\ldots,60$,
computed via the iterative algorithm of Section~\ref{sec:Iterative_moments},
were supplied to the Pad\'{e} bound computation algorithm \texttt{padeapprox}
\cite{Gonnet:2013:55:1:101:110853236}. 
Following the same procedure as for the 2D flows, bounds up to order $N=10$ 
were returned. 
The global lower bound is
$\Dg^*_{kk}\ge\varepsilon$ and the global upper bound is
$\Dg^*_{kk}\le\varepsilon(1+\mu^0/\varepsilon^2)$, both shown as dashed black
lines in Figure~\ref{fig:steady_3D_bounds} (top row). 

The upper and lower bounds in Figure~\ref{fig:steady_3D_bounds} (top row) get
progressively tighter with increasing $N$. ABC-flow shows a significantly larger
effective diffusivity enhancement than Kolmogorov flow. The asymptotic behavior
at small $\varepsilon$ differs qualitatively between the two 3D flows, as shown
in the log-log plots. For ABC-flow, the slopes $S_{\text{upper}} = -0.544$ and
$S_{\text{lower}} = -0.533$ indicate inverse-power behavior, in agreement with
\cite{Lag21}, 
while for Kolmogorov flow, the slopes $S_{\text{upper}} = 0.172$ and
$S_{\text{lower}} = 0.188$ instead indicate a power-law decay; the decay is
gentler compared to the decay of the 2D flows, with additional diffusivity
enhancement. These distinctions highlight the qualitatively different effective
diffusivity enhancement mechanisms in 3D versus 2D flows. The bound differences
$[N/N]-[N-1/N]$ are shown in Figure~\ref{fig:Pade_gaps} (middle row).

\subsection{Bounds for space-time periodic flows}
\label{sec:bounds_space_time_flows}

\begin{figure}[htbp]
  \centering
  \includegraphics[width=0.48\textwidth]{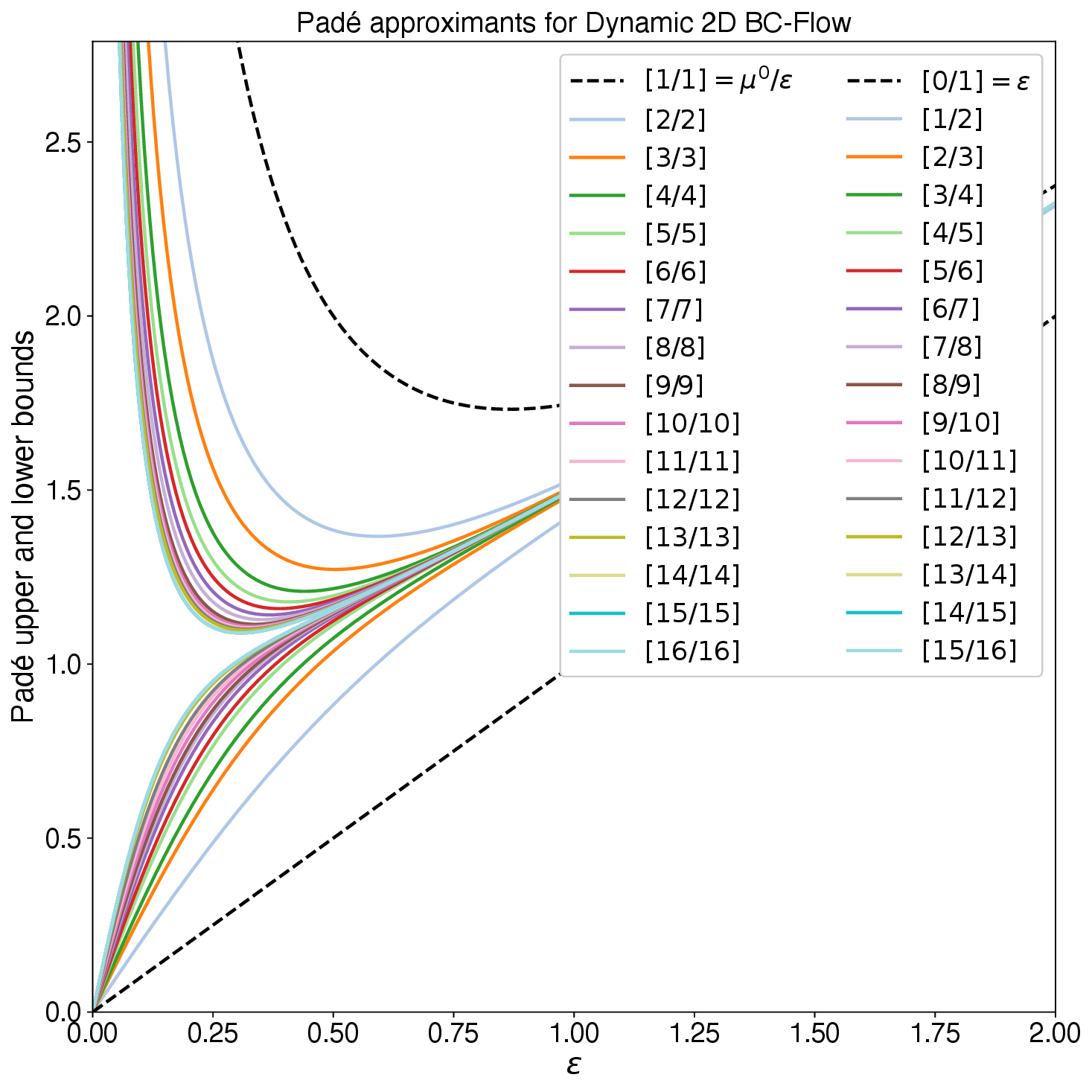}%
  \hfill
  \includegraphics[width=0.48\textwidth]{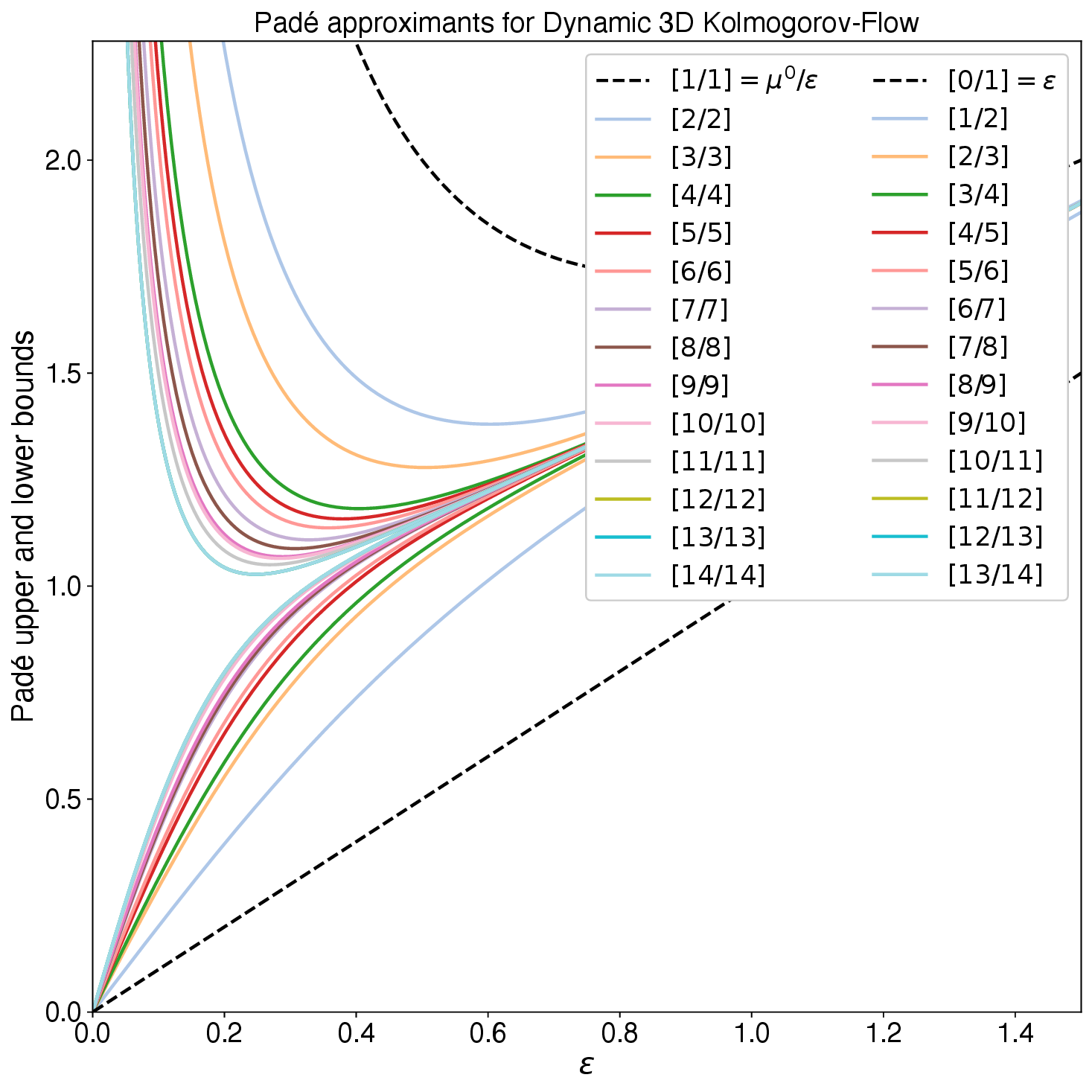}\\[6pt]
  \includegraphics[width=0.48\textwidth]{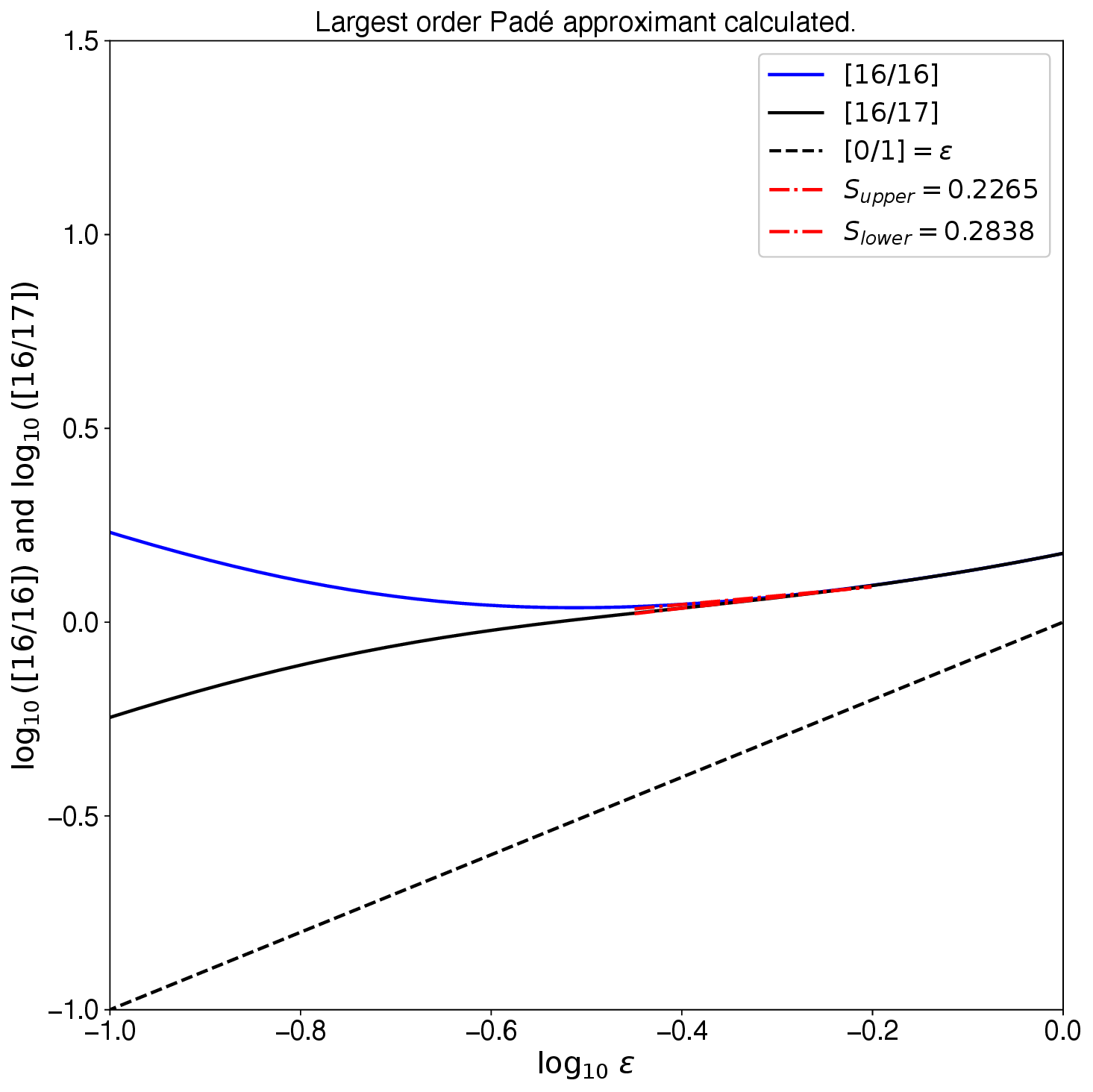}%
  \hfill
  \includegraphics[width=0.48\textwidth]{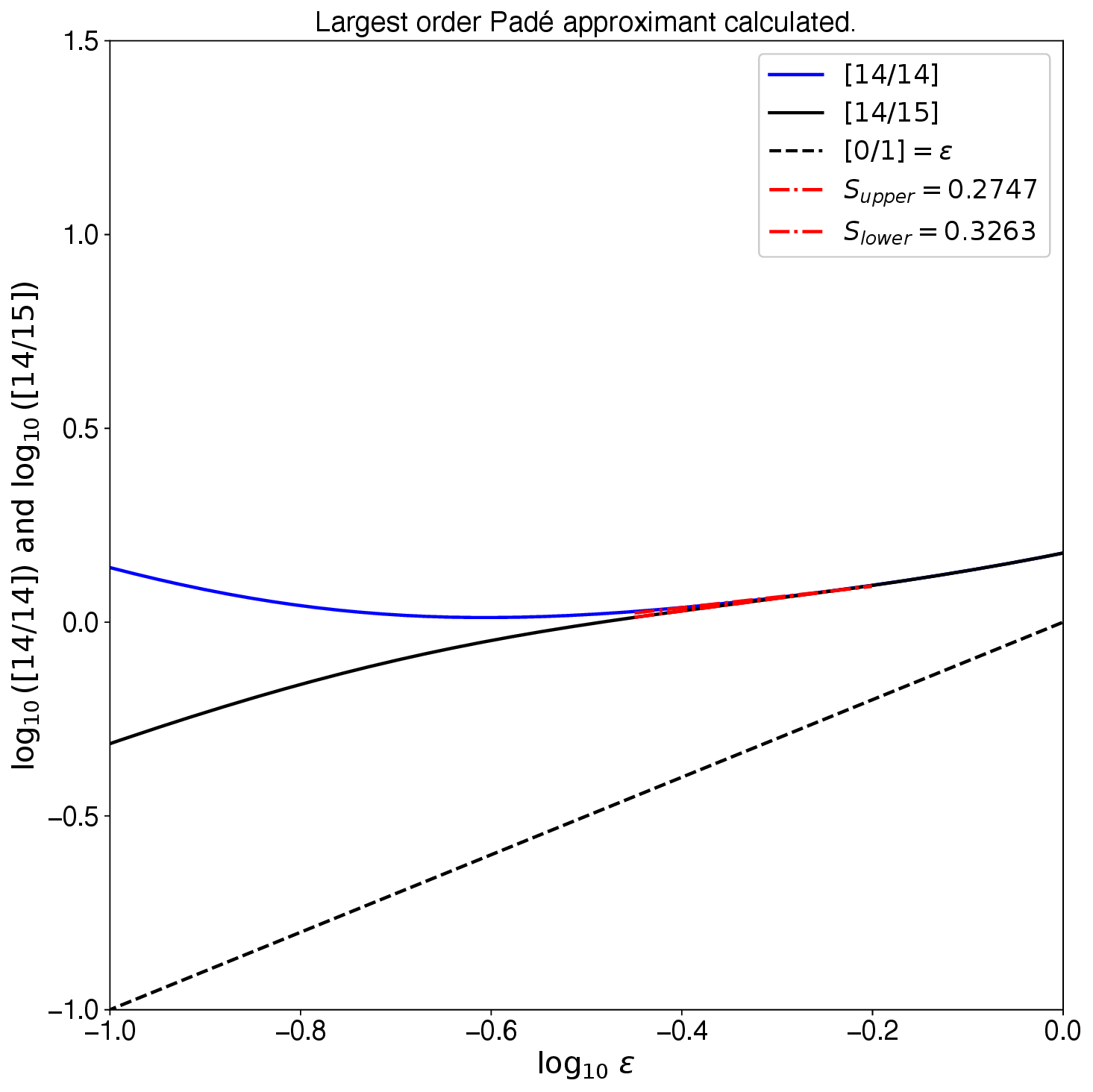}
  \caption{\textbf{Pad\'{e} bounds for space-time periodic flows.}
    (\textit{Top row}) Linear-scale upper $[N/N]$ and lower $[N-1/N]$ Pad\'{e}
    bounds for $\Dg^*_{kk}(\varepsilon)$ as a function of $\varepsilon$,
    for space-time periodic BC-flow (left) and space-time periodic Kolmogorov
    flow (right), for successive orders $N$.
    (\textit{Bottom row}) Corresponding log-log plots with power-law fits
    in the large-$\Pen$ regime ($\varepsilon\ll 1$). Here,
    $S_{\text{upper}}$ and $S_{\text{lower}}$ denote the slopes of the
    polynomial fits in log-log scale for the highest-order upper and lower
    bounds, respectively.}
  \label{fig:dynamic_bounds}
\end{figure}

In this section we present the Pad\'{e} approximant upper $[N/N]$ and lower
$[N-1/N]$ bounds for the effective diffusivity $\Dg^*_{kk}$ for the two
space-time periodic flows: the dynamic 2D
BC-flow~\eqref{eq:space_time_periodic_u} with $A=1$ and the dynamic 3D
Kolmogorov flow~\eqref{eq:space_time_periodic_3D_u} with $\theta=1$. The top row
of Figure~\ref{fig:dynamic_bounds} shows the corresponding linear-scale upper
$[N/N]$ and lower $[N-1/N]$ bounds, with the dynamic 2D BC-flow on the left and
the dynamic 3D Kolmogorov flow on the right, for successive orders $N$. The
bottom row shows the corresponding log-log plots for the highest-order upper and
lower bounds computed, with linear fits in the large-$\Pen$
(small-$\varepsilon$) regime.

For both flows, the moments $\mu^{2n}$, $2n=0,\ldots,60$, computed via the
iterative algorithm of Section~\ref{sec:Iterative_moments}, were supplied to the
Pad\'{e} bound computation algorithm \texttt{padeapprox}
\cite{Gonnet:2013:55:1:101:110853236}. For the dynamic 2D BC-flow, bounds up to
order $N_{\max}=16$ were returned; for the dynamic 3D Kolmogorov flow,
$N_{\max}=14$. The global lower bound $\Dg^*_{kk}\ge\varepsilon$ and global
upper bound $\Dg^*_{kk}\le\varepsilon(1+\mu^0/\varepsilon^2)$ are shown as
dashed black lines in Figure~\ref{fig:dynamic_bounds} (top row).

The upper and lower bounds in Figure~\ref{fig:dynamic_bounds} (top row) get
progressively tighter with increasing $N$ for $\varepsilon\gtrsim0.5$, where the
higher-order bounds cluster near a common value. However, for
$\varepsilon\lesssim0.3$ the bounds fan out dramatically with decreasing
$\varepsilon$ --- a marked contrast with the steady 2D flows in
Figure~\ref{fig:steady_2D_bounds}, where the bounds remained tight down to
$\varepsilon\sim0.03$. This divergence at moderately small $\varepsilon$ is a
fundamental challenge for space-time periodic flows. The rapid growth of the
spectral measure moments renders the associated Pad\'{e} approximants
numerically ill-conditioned at large $\Pen$, so that additional moment orders do
not yield substantially tighter bounds in the advection-dominated regime. As a
consequence, the bounds do not resolve the expected $\Dg^*_{kk}\sim1$ residual
diffusivity behavior for $\varepsilon\ll1$ established for the dynamic 2D flow
in~\cite{Biferale:PF:2725,Murphy:ADSTPF-2017}. 

The log-log fits in Figure~\ref{fig:dynamic_bounds} (bottom row) are accurate
for $\log_{10}\varepsilon\gtrsim-0.4$. For the
dynamic 2D BC-flow the fits yield slopes $S_{\text{upper}}=0.227$ and
$S_{\text{lower}}=0.284$, and for the dynamic 3D Kolmogorov flow
$S_{\text{upper}}=0.275$ and $S_{\text{lower}}=0.326$. The significant
discrepancy between upper and lower slopes in both cases --- in contrast to
the near-perfect agreement of $0.500$ for both bounds in the steady 2D flows ---
reflects the limited accuracy of the current bounds in the power-law regime for
these more complex flows. The bound differences $[N/N]-[N-1/N]$ are shown in
Figure~\ref{fig:Pade_gaps} (bottom row).

Comparing the dynamic 2D BC-flow bounds to those for the steady BC-flow in
Section~\ref{sec:bounds_2D_steady_flows} shows that the addition of the
time-dependent perturbation $\vecu_\theta=\theta\cos{t}\,(\sin{y},\sin{x})$
with $\theta=1$ yields an increase in $\Dg^*_{kk}$ at moderate $\varepsilon$.
Numerical results in~\cite{Biferale:PF:2725,Murphy:ADSTPF-2017} indicate this
enhancement grows to $\approx10^{1.5}$ times that for BC-flow when
$\varepsilon\approx10^{-4}$, with $\Dg^*_{kk}(10^{-4})\approx10^{-1/2}$
compared to $\Dg^*_{kk}(10^{-4})\approx10^{-2}$ for BC-flow. Capturing this
large-$\Pen$ enhancement rigorously with Pad\'{e} bounds remains an open
challenge requiring both higher moment orders and more stable numerical
procedures for the exponentially growing moments of dynamic flows.

\section{Conclusions}
\label{sec:conclusions}
The effective diffusivity $\Dg^*$ in spatially and space-time
periodic fluid flows has a Stieltjes function representation involving a
spectral measure $\mu$ of a self-adjoint operator and the P{\'e}clet number
$\Pen$ of the flow. 
A non-dimensionalization  \emph{separates} the geometry and dynamics of the
fluid flow (encoded in $\mu$) from the strength of the flow in $\Pen$
(reciprocal of the molecular diffusivity $\varepsilon$).
Pad{\'e} approximants for
Stieltjes functions provide rigorous nested upper and lower bounds for $\Dg^*$.
With more $\mu$-moments incorporated, the bounds get tighter and 
converge to 
$\Dg^*$ at  
$\Pen\ll 1$. 
Theoretical and numerical barriers of
Pad{\'e} approximants 
limit the accuracy of bounds for
$\Dg^*$ in the advection dominated regime ($\Pen\gg1$), an open issue since  
\cite{Avellaneda:CMP-339}.

An iterative method was introduced to compute any number of 
$\mu$-moments.
Our method was implemented in Maple and Python-SymPy to calculate  
dozens of moments of steady and space-time periodic flows in closed form.
The method was also implemented in MATLAB and Python-NumPy to compute hundreds
of moments to floating point precision, in excellent 
agreement with the exact values. 
For 2D steady (BC) flow, 
the 
$\Dg^*\sim\varepsilon^{1/2}$  ($\varepsilon\ll1$) behavior 
was 
captured by 
our bounds 
down to $\varepsilon\approx0.053$. Bound differences
indicate that $\Dg^*$ was 
computed within $0.01$ for $\varepsilon\gtrapprox0.053$, with similar type of
findings for 3D steady (ABC/Kolmogorov) flows and their space-time periodic
extensions. The iterative moment method developed here applies to spatially
periodic and space-time periodic velocity fields admitting finite Fourier series
representations; extension to random velocity fields, where statistical
symmetries may allow analogous iterative strategies for computing spectral
measure moments, is left for future research.

\section{Acknowledgements}
\label{sec:acknowledgements}
We gratefully acknowledge support from the Applied and Computational Analysis
Program 
at the U.S. Office of Naval
Research (ONR) through grants N00014-18-1-2552, N00014-21-1-2909, 
and N00014-26-1-2114.
We are also grateful for support from
the Division of Mathematical Sciences 
at the
U.S. National Science Foundation (NSF) through grants 
DMS-1715680, DMS-2111117, DMS-2136198, DMS-2206171, and DMS-2309520.


\appendix

\section{Moment calculations}
\label{sec:moment_calculations_detailed}

In this section we analytically and numerically calculate moments
$\mu^{2n}_{jk}$, for $n=0,1,2,\ldots$, for each of the spectral measures
$\mu_{jk}$, $j,k=1,\ldots,d$. For shear flow we analytically calculate all of
the moments. These results show that each of the spectral measures is a
$\delta$-function centered at the spectral origin. Detailed calculations for the
first few moments for 2D steady BC-flow and a related time-dependent 2D flow are
given in Appendix \ref{sec:moments_bc_flow} and
\ref{sec:moments_2_time_periodic_flow}, respectively. A numerical implementation
of the iterative method developed in Section \ref{sec:Iterative_moments} is
discussed in Appendix \ref{sec:moment_calculations_numerical}, where the method
is applied to various steady and dynamic flows in both 2D and 3D. We implement
the iterative method to calculate dozens of moments in closed form using the
Python-SymPy symbolic math toolbox. We also implement the iterative moment
method in Python-NumPy and compute hundreds of moments to floating point
precision. Our numerical implementation of Pad\'e approximants which incorporate
the positive moment values $\mu^{2n}_{kk}$ is described in Appendix
\ref{sec:numerical_implementation_pade}. In Section \ref{sec:bounds} the
Pad\'{e} approximant bounds for the diagonal components $\Dg^*_{kk}$ of the
effective diffusivity matrix $\Dg^*$ are discussed for each steady and dynamic
fluid velocity field.

\subsection{Moments of shear flow} \label{sec:Moments_shear_flow}
We now consider the special case of shear flow. The fluid velocity field for
shear flow is time-independent. In 2D shear flow in the $x$-direction has a
fluid velocity $\vecu=(0,\zeta(x))$ and $\vecu=(\xi(y),0)$ for flow in the
$y$-direction, where in the current context $\xi$ and $\zeta$ are arbitrary
mean-zero functions which are expressible by finite Fourier series, as shown in
equation in~\eqref{eq:Fourier_u}. Examples of 3D shear flow fluid velocity
fields are $\vecu=(\xi(y,z),0,0)$, $\vecu=(0,\zeta(x,z),0)$, and
$\vecu=(0,0,\gamma(x,y))$, where $\gamma$ is also an arbitrary function of the
same type as $\xi$ and $\zeta$.

The key property of these simple shear flows is that only one component of the
fluid velocity field is non-zero, the $i$th component say, and that component is
a function which is independent of $x_i$, the $i$th component of $\vecx$. This
property and equations~\eqref{eq:Operator_Eigenfunctions}
and~\eqref{eq:uj_gj_Dtgj_Agj_Fourier} imply $\bnabla g_j=0$ for all
$j=1,\ldots,d$, hence $D_tg_j=\vecu\bcdot\bnabla g_j=0$ for all $j=1,\ldots,d$
which, in turn, implies that $Ag_j=(-\Delta)^{-1}D_tg_j=0$ for all
$j=1,\ldots,d$ which, in turn,  
implies that $A^ng_j=0$ and $D_tA^ng_j=0$ for all $j=1,\ldots,d$ and
$n=1,2,\ldots\,$. Therefore, by
equations~\eqref{eq:measure_mass}--\eqref{eq:Moments_Functional} we have
%
\begin{align}\label{eq:moments_shear_flow}
\mu^0_{jk} = \langle g_j,u_k\rangle_2\,,
\qquad
\mu_{jk}^n=0\,,
\quad
j,k=1,\ldots,d\,,
\quad
n=1,2,3,\ldots\;.
\end{align}
%

Let's first focus on the spectral measure $\mu_{kk}$ for some $k=1,\ldots,d$,
which is a \emph{positive} measure. It's clear that the only positive Stieltjes
measure with all moments having  value zero is a $\delta$-measure concentrated
at $\lambda=0$, $\delta_0(\d\lambda)$, where $\delta_a(\d\lambda)$ is the
$\delta$-measure concentrated at $\lambda=a$. We therefore have the following
result regarding the positive measure $\mu_{kk}$ for simple shear flow, 
\begin{align}\label{eq:mukk_delta_shear}
  \mu_{kk}=\mu_{kk}^0\,\delta_0(\d\lambda)\,.
\end{align}

We now show the signed measures $\Real\mu_{jk}$ and $\Imag\mu_{jk}$, 
$j\ne k$, also satisfy~\eqref{eq:mukk_delta_shear} in a weak sense. By the 
Jordan decomposition theorem~\cite{Folland:99:RealAnalysis} there exist
unique 
positive measures $\Real\mu_{jk}^+$ and $\Real\mu_{jk}^-$ such that
$\Real\mu_{jk}=\Real\mu_{jk}^+-\Real\mu_{jk}^-$ and
$\Real\mu_{jk}^+\perp\Real\mu_{jk}^-$, and similarly for the signed
measure $\Imag\mu_{jk}$.  
This and~\eqref{eq:moments_shear_flow} imply the moments of these measures
satisfy
\begin{align}
  [\Real\mu_{jk}^+]^n=[\Real\mu_{jk}^-]^n\,,
  \quad
  [\Imag\mu_{jk}^+]^n=[\Imag\mu_{jk}^-]^n\,,
  \quad
  n=1,2,3,\ldots\,,
\end{align}
where $ [\Real\mu_{jk}^+]^n$ and $[\Real\mu_{jk}^-]^n$ are the $n$th moments
of the measures $\Real\mu_{jk}^+$ and $\Real\mu_{jk}^-$, for example.
Consequently,~\eqref{eq:moments_shear_flow} implies for all polynomial 
$P(\lambda)$ satisfying $P(0)=0$ that 
\begin{align}\label{eq:zero_int}
  \int_\Sigma P(\lambda)\,\d\Real\mu_{jk}(\lambda)=0\,,
  \qquad
  j\ne k\,.
\end{align}
Since $\Real\mu_{jk}^+\perp\Real\mu_{jk}^-$, it is clear that this can 
only be the case if $\Real\mu_{jk}=\Real\mu_{jk}^0\,\delta_0(\d\lambda)$.
A similar argument establishes that 
$\Imag\mu_{jk}=\Imag\mu_{jk}^0\,\delta_0(\d\lambda)$. Hence, we have that
\begin{align}\label{eq:mujk_delta_shear}
\mu_{jk}=\mu_{jk}^0\,\delta_0(\d\lambda)\,,
\quad
j\neq k\,.
\end{align}

This argument can be strengthened for the setting of a time-independent 
fluid velocity field, $\vecu=\vecu(\vecx)$. In this case, the self-adjoint 
operator $M$ has bounded spectrum 
$\Sigma\subseteq[-\|M\|,\|M\|]$~\cite{Reed-1980,Stone:64}. 
The Stone
Weierstrass theorem~\cite{Folland:99:RealAnalysis} then extends the 
result in~\eqref{eq:zero_int} to functions $\xi(\lambda)$
continuous on the closed interval $[-\|M\|,\|M\|]$ satisfying
$\xi(0)=0$, thus~\eqref{eq:mujk_delta_shear} holds in this strengthened 
sense.

\subsection{Moments for BC-flow}
\label{sec:moments_bc_flow}
In this section we demonstrate how the iterative method developed in
\secref{sec:Iterative_moments} can be used to analytically calculate 
the first few moments for the spectral measure $\mu_{jk}$
associated with $BC$-flow, which is given by the fluid velocity
field~\cite{Murphy:ADSTPF-2017}. 
%
\begin{align}\label{eq:BC-flow-moments}
  \vecu=(C\cos{y},B\cos{x})\,.
\end{align}
We then display higher order moments
calculated in closed form 
by the Maple and Python-SymPy symbolic mathematics toolboxes. 

In terms of the Fourier basis
$\{\phi_{m,n}\}$ for $\Hs_{\Vc}$~\cite{Folland:99:RealAnalysis}, where
$\phi_{m,n}(x,y)=\exp(\imath(mx+ny))$, the components $u_j$, $j=1,2$, of
the fluid velocity field $\vecu=(C\cos{y},B\cos{x})$ are given by 
\begin{align}\label{eq:u_BC_Flow}
  u_1&= (C/2)\,(\phi_{0,1}+\phi_{0,-1})\,,
  \\\notag
  u_2&=(B/2)\,(\phi_{1,0}+\phi_{-1,0})\,.
\end{align}
From equations~\eqref{eq:Operator_Eigenfunctions},
\eqref{eq:uj_gj_Dtgj_Agj_Fourier}, and~\eqref{eq:u_BC_Flow} we have
%
\begin{align}\label{eq:gj_uj_BC}
  g_j=u_j\,,
  \quad
  j=1,2\,.
\end{align}
%
The operator
$D_t=\vecu\bcdot\bnabla=u_1\partial_x+u_2\partial_y$ is given by  
\begin{align}\label{eq:Dt_expanded_BC_Flow}
  D_t
   =(C/2)\,[\phi_{0,1}+\phi_{0,-1}]\,\partial_x
   +(B/2)\,[\phi_{1,0}+\phi_{-1,0}]\,\partial_y\,.
\end{align}
Since $\partial_x u_1=0$ and $\partial_y u_2=0$,
equations~\eqref{eq:Operator_Eigenfunctions}
and~\eqref{eq:u_BC_Flow}--\eqref{eq:Dt_expanded_BC_Flow} yield
\begin{align}\label{eq:Dtgj_BC_Flow}
  D_t\,g_1&=(C/2)(\imath B/2)
          (\phi_{1,1}+\phi_{-1,1}-\phi_{1,-1}-\phi_{-1,-1}),
  \\\notag
  D_t\,g_2&=(B/2)(\imath C/2)
          (\phi_{1,1}+\phi_{1,-1}-\phi_{-1,1}-\phi_{-1,-1})\,.
\end{align}
Since $A=(-\Delta)^{-1}D_t$, it follows from
equations~\eqref{eq:Operator_Eigenfunctions}
and~\eqref{eq:Dtgj_BC_Flow} that
%
\begin{align}\label{eq:Agj_BC}
  Ag_j=D_t\,g_j/2\,,
  \quad
  j=1,2\,.
\end{align}

We now use equations~\eqref{eq:u_BC_Flow}--\eqref{eq:Agj_BC} to
compute the mass and the first two moments of the 
spectral measure $\mu_{jk}$. From
equations~\eqref{eq:uj_gj_Dtgj_Agj_Fourier} and~\eqref{eq:gj_uj_BC}
we have $b_{m,n}^j=a_{m,n}^j$, where $a_{0,1}^1=a_{0,-1}^1=C/2$ and
$a_{1,0}^2=a_{-1,0}^2=B/2$, and all other coefficients
$a_{m,n}^j=0$. Consequently, from equation~\eqref{eq:measure_mass},
$\mu_{jk}^0=\langle g_j,u_k\rangle_2$, or the 
first formula in~\eqref{eq:Mass_Moments_1_and_2_Fourier}, 
$\mu_{jk}^0=\sum_{m,n}b^j_{m,n}\bar{a}^k_{m,n}$, we have
\begin{align}
  \mu_{11}^0=\frac{C^2}{2}\,,
  \qquad
  \mu_{22}^0=\frac{B^2}{2}\,,
  \qquad
  \mu_{12}^0=0\,.
\end{align}
Similarly, from the formulas in 
~\eqref{eq:General_Moments_Functional_1_and_2} 
$\mu^{1}_{jk}=-\imath\,\langle D_t g_j, g_k\rangle_{2}$ 
and
$\mu^{2}_{jk}=\langle D_t g_j,A g_k\rangle_{2}$
or the last two
formulas in equation~\eqref{eq:Mass_Moments_1_and_2_Fourier}, 
$\mu^1_{jk}
      =-\imath
      \sum_{\ell,\veck}
       c^{\,j}_{\ell,\veck}\,
       \bar{b}^{\,k}_{\ell,\veck}\,,
$
and
$ 
\mu^2_{jk}
      =
      \sum_{\ell,\veck}
       c^{\,j}_{\ell,\veck}\,
       \bar{d}^{\,k}_{\ell,\veck}\,,
$
we have $\mu^{1}_{jk}=0$ and 
\begin{align}
  \mu_{11}^2 = \mu_{22}^2 = \frac{B^2C^2}{8}\,,
  \qquad
  \mu_{12}^2 = 0\,. 
\end{align}

While the calculations of the mass and the first two moments are quite
transparent, calculations of higher order moments become complicated quickly.
Analytical closed form expressions can be obtained using Maple and Python-SymPy
symbolic math toolboxes. These calculations indicate that $\mu_{12}^{2n}=0$ for
all $2n=0,2,4,6,\ldots$ and
\begin{align}
&\mu^4_{11} = 
(1/320)B^2C^4 + (11/320)B^4C^2\,,
\\\notag
&\mu^6_{11} = \frac{3B^2C^2(101B^4 + 25B^2C^2 + C^4)}{32000}\,,
\\\notag
&\mu^8_{11} = \frac{B^2C^2(567567B^6 + 233070B^4C^2 + 39610B^2C^4 + 617C^6)}{217600000}\,,
\\\notag
&\mu^{10}_{11} = \frac{C^2B^2(567567B^6 + 233070B^4C^2 + 39610B^2C^4 + 617C^6)}{217600000}\,.
\end{align}
%
The $\mu^{2n}_{22}=\mu^{2n}_{22}(B,C)$ 
satisfies $\mu^{2n}_{22}(B,C)=\mu^{2n}_{11}(C,B)$.
We were able to calculate 
$\mu^{2n}_{ij}(B,C)$ 
in closed form up to $2n=26$ for arbitrary $B,C\in(0,1]$, given the
computational resources used. Some of these are displayed in Table 
\ref{table:bc_flow_moments} for $B=C=1$.
\begin{table}[t]
\centering
\def\arraystretch{1.5}
\begin{tabular}{@{}ll@{}}
\multicolumn{1}{l|}{$2n$} & $\mu_{kk}^{2n}$                                   \\ \midrule
\multicolumn{1}{l|}{0} & $\frac{1}{2}$                                    \\
\multicolumn{1}{l|}{2} & $\frac{1}{8}$                                  \\
\multicolumn{1}{l|}{4} & $\frac{3}{80}$                                 \\
\multicolumn{1}{l|}{6} & $\frac{381}{32000}$                            \\
\multicolumn{1}{l|}{8} & $\frac{26277}{6800000}$                         \\
\multicolumn{1}{l|}{10} & $\frac{47519559}{37570000000}$                  \\
\multicolumn{1}{l|}{12} & $\frac{2960164002865793}{7127269448000000000}$ \\ 
\multicolumn{1}{l|}{14} & $\frac{56807418712571064717219}{416027270403097600000000000}$  \\ 
\multicolumn{1}{l|}{16} & $\frac{845725433928943189960402643663087}{18830209775901005048070400000000000000}$ \\ 
\multicolumn{1}{l|}{18} & $\frac{2652281628393653311493590026036436288914383079}{179505850850574462175090974199721600000000000000000}$ \\ 
\multicolumn{1}{l|}{20} & $\frac{404455666246342112121617203918794294909069461346892222329513233}{83202120549989484527334438746964410459680581766400000000000000000000}$ \\ 
\multicolumn{1}{l|}{22} & $\frac{61678397622238580001722366830219450097176873936306735282205457266492841578250541}{38564719930926020344578565530438076789148157444429069162842585600000000000000000000000}$ \\ 
\bottomrule
\end{tabular}
\caption{Exact values of the spectral measure moments $\mu_{kk}^{2n}$ for BC-
flow with fluid velocity field in equation \eqref{eq:BC-flow-moments} with
$B=C=1$, computed using the Maple and Python-SymPy symbolic math toolboxes.
Off-diagonal moments satisfy $\mu_{jk}^{2n}=0$, $j\neq k$.}
\label{table:bc_flow_moments}
\end{table}

Numerically computing higher order moments using floating point arithmetic can
be accomplished using the mapping in equation \eqref{eq:mode_mappings} as
follows. The action of the inverse Laplacian on a Fourier mode
$\phi_{m,n}(x,y)=\exp(\imath(mx+ny))$ is given by, 
\begin{align}\label{eq:inv_Laplacian_mapping_2D}
  (-\Delta)^{-1}\phi_{m,n}= \phi_{m,n}/(m^2+n^2)\,,
  \qquad
  m^2+n^2\ne0\,.
\end{align}
Also, since $\phi_{i,j}\phi_{k,l}=\phi_{i+k,j+l}$, by
~\eqref{eq:Operator_Eigenfunctions} and \eqref{eq:Dt_expanded_BC_Flow} we have
($m^2+n^2\ne0$)
\begin{align}\label{eq:Dt_mapping_BC_Flow}
D_t\phi_{m,n}
=               
(\imath mC/2)\,\phi_{m,n+1}
+
(\imath mC/2)\,\phi_{m,n-1}
+
(\imath nB/2)\,\phi_{m+1,n}
+
(\imath nB/2)\,\phi_{m-1,n}
\end{align}
with $A\phi_{m,n}=(-\Delta)^{-1}D_t\phi_{m,n}$.
Equations~\eqref{eq:Mass_Moments_1_and_2_Fourier}
and~\eqref{eq:Higher_Moments_Fourier} together
with~\eqref{eq:inv_Laplacian_mapping_2D}  
and~\eqref{eq:Dt_mapping_BC_Flow} define an iterative
method that can be used to compute Fourier coefficients 
for the functions in the iterative mapping chain:
\begin{align}\label{eq:mapping_chain}
  u_j\mapsto g_j\mapsto
  D_t\,g_j\mapsto Ag_j\mapsto
  D_tAg_j\mapsto A^2g_j\mapsto
  D_tA^2g_j\mapsto A^3g_j\mapsto\ldots\,.
\end{align}
This can be used to numerically compute an arbitrary 
number of moments $\mu_{jk}^n$ for $BC$-flow, the maximum number 
limited only by numerical accuracy and computational resources.

\subsection{Moments for a 2D time-periodic flow}
\label{sec:moments_2_time_periodic_flow}
In this section we demonstrate how the iterative method developed in
\secref{sec:Iterative_moments} can be used to analytically calculate 
the first few moments for the spectral measure $\mu_{jk}$
associated with the space-time periodic
flow~\cite{Murphy:ADSTPF-2017,Biferale:PF:2725} 
\begin{align}\label{eq:space_time_periodic_u_BCtheta}
  \vecu=(C\cos{y},B\cos{x})+\theta\cos{t}\,(\sin{y},\sin{x}).  
\end{align}
We then display higher order moments
calculated by the Maple and Python-SymPy symbolic mathematics toolboxes.

Denote $\vecu=\vecu_{BC}+\vecu_\theta$, where $\vecu_{BC}=(C\cos{y},B\cos{x})$
is the fluid velocity field in equation~\eqref{eq:u_BC_Flow} and
$\vecu_\theta=\theta\cos{t}\,( \sin{y},\sin{x})$. The components $u_j$, $j=1,2$,
of the fluid velocity field $\vecu$ in~\eqref{eq:space_time_periodic_u_BCtheta}
expressed in terms of the Fourier basis $\{\phi_{\ell,m,n}\}$ for
$\Hs_{\Tc\Vc}$~\cite{Folland:99:RealAnalysis}, where
$\phi_{\ell,m,n}(x,y,t)=\exp(\imath(\ell t+mx+ny))$, are given by  
\begin{align}\label{eq:time_periodic_velocity_fields_phimn}
  [\vecu_{BC}]_1 &= (C/2)\,(\phi_{0,0,1} + \phi_{0,0,-1})\,,
  \\\notag
  [\vecu_{BC}]_2 &= (B/2)\,(\phi_{0,1,0} + \phi_{0,-1,0})\,,             
       \\\notag
  [\vecu_\theta]_1
    &=(\theta/4\imath)
    (\phi_{1,0,1} + \phi_{-1,0,1}
   - \phi_{1,0,-1} - \phi_{-1,0,-1}
           )\,,
  \\\notag         
  [\vecu_\theta]_2
  &=(\theta/4\imath)
    (\phi_{1,1,0}  + \phi_{-1,1,0}
   - \phi_{1,-1,0} - \phi_{-1,-1,0})\,.
\end{align}
Writing $\vecg=\vecg_{BC}+\vecg_\theta$ with 
$\vecg=(-\Delta)^{-1}\vecu$, using the notation that $(-\Delta)^{-1}$
operates component-wise on $\vecu$, we have
from~\eqref{eq:time_periodic_velocity_fields_phimn}
\begin{align}
  \vecg=\vecu_{BC}+\vecu_\theta\,.
\end{align}
In this time-dependent setting, 
$D_t=\partial_t+\vecu\bcdot\bnabla$, hence by linearity 
\begin{align}
D_t=\partial_t+\vecu_{BC}\bcdot\bnabla
+\vecu_{\theta}\bcdot\bnabla\;.
\end{align}

From equation~\eqref{eq:Operator_Eigenfunctions}, we have
\begin{align}\label{eq:time_derivative_map}
  \partial_t\phi_{\ell,m,n}= \imath\ell\phi_{\ell,m,n}\,,
  \quad
  \ell\ne0.
\end{align}
We have the following analogue of
equation~\eqref{eq:Dt_mapping_BC_Flow}, for $m^2+n^2\ne0$,
%
\begin{align}\label{eq:Dt_mapping_time_periodic_BC_Flow}
&\vecu_{BC}\bcdot\bnabla\phi_{\ell,m,n}
=
\\
&(\imath mC/2)\,\phi_{\ell,m,n+1}
+(\imath mC/2)\,\phi_{\ell,m,n-1}
+(\imath nB/2)\,\phi_{\ell,m+1,n}
+(\imath nB/2)\,\phi_{\ell,m-1,n}\,.
\notag
\end{align}
From equation~\eqref{eq:time_periodic_velocity_fields_phimn} we have, for
$m^2+n^2\ne0$, 
\begin{align}\label{eq:Dt_mapping_beta_gamma_Flow}  
\vecu_\theta\bcdot\bnabla\phi_{\ell,m,n}
&=
(\imath m c/4\imath)\,\phi_{\ell+1,m,n+1}
+(\imath m c/4\imath)\,\phi_{\ell-1,m,n+1}
-(\imath m c/4\imath)\,\phi_{\ell+1,m,n-1}
\\\notag
&-(\imath m c/4\imath)\,\phi_{\ell-1,m,n-1}
+(\imath nb/4\imath)\,\phi_{\ell+1,m+1,n}
+(\imath nb/4\imath)\,\phi_{\ell-1,m+1,n}
\\\notag
&-(\imath nb/4\imath)\,\phi_{\ell+1,m-1,n}
-(\imath nb/4\imath)\,\phi_{\ell-1,m-1,n}\,.               
\end{align}
Similarly, we have the following analogue of
equation~\eqref{eq:inv_Laplacian_mapping_2D} 
%
\begin{equation}\label{eq:inv_Laplacian_mapping_3D}
(-\Delta)^{-1}\phi_{\ell,m,n}=\phi_{\ell,m,n}/(m^2+n^2)
  \qquad
  m^2+n^2\ne0.
\end{equation}
%
Equations~\eqref{eq:Mass_Moments_1_and_2_Fourier}
and~\eqref{eq:Higher_Moments_Fourier} together with equations
~\eqref{eq:Dt_mapping_time_periodic_BC_Flow}--\eqref{eq:inv_Laplacian_mapping_3D}
enable all the Fourier coefficients in the iterative sequence shown in
equation~\eqref{eq:mapping_chain} to be obtained analytically in closed form.

\begin{table}[t]
\centering
\begin{tabular}{@{}lll@{}}
\toprule
\multicolumn{3}{c}{Dynamic BC flow}                                                      \\ \midrule
\multicolumn{1}{l|}{$2n$} & $\mu_{kk}^{2n}$                                          & $\mu_{jk}^{2n}$                  \\ \midrule
\multicolumn{1}{l|}{0} & \multicolumn{1}{l|}{3/4}                    & 0                 \\
\multicolumn{1}{l|}{2} & \multicolumn{1}{l|}{35/64}                  & 0                 \\
\multicolumn{1}{l|}{4} & \multicolumn{1}{l|}{1069/1280}              & 5/32              \\
\multicolumn{1}{l|}{6} & \multicolumn{1}{l|}{282223/163840}          & 473/800           \\
\multicolumn{1}{l|}{8} & \multicolumn{1}{l|}{15778075321/3481600000} & 83952089/51200000 \\ \bottomrule
\end{tabular}
\caption{Exact values of the spectral measure moments $\mu_{kk}^{2n}$ and
$\mu_{jk}^{2n}$, $j\ne k$, for the fluid velocity field in equation
\eqref{eq:space_time_periodic_u_BCtheta} with $B=C=\theta=1$, calculated using
the Maple and Python SymPy symbolic math toolboxes.}
\label{table:dynamic_theta_flow_moments}
\end{table}

Using the Maple and Python SymPy, this 
yields the spectral measure masses
\begin{align}
    \mu_{11}^0&=\frac{C^2}{2}+\frac{\theta^2}{4},
    \qquad
    \mu_{22}^0=\frac{B^2}{2}+\frac{\theta^2}{4},
    \qquad
    \mu_{12}^0=0.
\end{align}
Recall that all odd moments are identically zero. The second, fourth, and sixth
moments are given by
\begin{align}
\mu_{11}^2
&=\frac{3 \theta^4}{64}
+\frac{(B^2 + C^2 + 4)\theta^2}{16} 
+\frac{B^2 C^2}{8}, 
\\\notag
\mu_{12}^2&=0,
\\\notag
\mu_{11}^4
&=\frac{5\theta^6}{3072}
+ \frac{(323B^2+705C^2+1090)\theta^4}{92160}
\\\notag &+ \frac{(180 + 22B^4 + (366C^2 + 245)B^2 +8C^4+80C^2)\theta^2}{11520}
+\frac{11B^4C^2}{320}
+\frac{C^4B^2}{320}, 
\\\notag
\mu_{12}^4&=-\frac{13BC\theta^4}{5760}-\Big( \frac{18(B^3C+BC^3)+35BC}{11520} \Big)\theta^2,
\\\notag
\mu_{11}^6
&=\frac{20701\theta^8}{49152000}
+ \Big( \frac{31583}{4608000} + \frac{4861B^2}{3686400} + \frac{3119C^2}{1228800}\Big)\theta^6
\\\notag & + \Big( \frac{49}{2304} + \frac{163871B^2}{9216000} + \frac{4343B^4}{3072000} 
+ \frac{48329C^2}{3072000} + \frac{108793B^2C^2}{9216000} + \frac{161C^4}{204800}\Big) \theta^4 
\\\notag &+ \Big(\frac{1}{64} + \frac{337B^2}{9216} + \frac{33191B^4}{2304000}+\frac{101B^6}{192000}+\frac{C^2}{576} 
\\\notag
&-\frac{2569B^2C^2}{2304000}+\frac{773B^4C^2}{48000} + \frac{11C^4}{16000} + \frac{137B^2C^4}{36000} + \frac{C^6}{48000}\Big)\theta^2 
\\\notag & + \frac{303B^6C^2}{32000} + \frac{3B^4C^4}{1280} + \frac{3B^2C^6}{32000},
\\\notag
\mu_{12}^6
&=\frac{437BC\theta^6}{4608000} + \Big(\frac{6509BC}{1536000}+\frac{521B^3C}{576000} +\frac{521BC^3}{576000}\Big)\theta^4 
\\ \notag 
&+ \Big( \frac{65BC}{9216}-\frac{12841B^3C}{2304000}-\frac{9B^5C}{32000} -\frac{12841BC^3}{2304000}
+\frac{91B^3C^3}{28800}-\frac{9BC^5}{32000}\Big)\theta^2\,,
\end{align}
where $\mu_{22}^{2n}$ as a function of $B$, $C$, and $\theta$, 
$\mu_{22}^{2n}(B,C,\theta)$ satisfies 
$\mu_{22}^{2n}(B,C,\theta)=\mu_{11}^{2n}(C,B,\theta)$.
Exact rational values for $2n = 0,2,\ldots, 8$ are displayed in Table 
\ref{table:dynamic_theta_flow_moments} for $B=C=\theta=1$.

\section{Numerical implementation of moment calculations}\label{sec:moment_calculations_numerical} 
In Section \ref{sec:Iterative_moments} we provided a theoretical foundation for
how the moments $\mu^n_{jk}$ of the spectral measure $\mu_{jk}$ can be
calculated iteratively in terms of the Fourier coefficients $\veca_{\ell,\veck}$
of the fluid velocity field $\vecu$. To illustrate the method, in Appendix
\ref{sec:moment_calculations_detailed} we provided detailed calculations for all
the moments of shear flow, as well as the mass and first few moments for 2D
steady BC flow and a time-modulated BC flow. In this section, we discuss our
results utilizing the iterative moment method to calculate many moments for some
2D and 3D time-independent (steady) flows and time-dependent (dynamic) flows.

The moments were calculated exactly, in closed form using the symbolic
mathematics software tools Maple and Python-SymPy, and calculated to floating-point
precision using the numerical linear algebra software tools MATLAB and NumPy,
comparing the results to ensure consistency. The resultant GitHub software
repository, Janus, is publically available. The Python library SymPy has the
symbolic capabilities of Maple within an efficient programmatic environment not
provided by Maple. The library Numpy contains the numerical capabilities of
MATLAB with an efficient programmatic environment that is more flexible than
MATLAB. For this reason, Python was chosen as the primary implementation, while
the Maple and MATLAB results are used for verification of the Python results. We
refer to the NumPy implementation as \texttt{decimal} and the SymPy
implementation as \texttt{heuristic}.

Using the Python library, the moments were calculated for all six fluid velocity
fields: 2D steady BC-flow, 2D steady cats-eye flow, 3D steady ABC-flow, 3D
steady Kolmogorov flow, 2D space-time periodic BC-flow, and 3D space-time
periodic Kolmogorov flow described above. The code was programmed to either
calculate 60 moments or stop after the previous moment took longer than 5 hours
on a Macbook laptop with an Apple M3 Pro chip and 18 GB of RAM. The relative
differences between the \texttt{decimal} and \texttt{heuristic} results were
less than $10^{-14}$ for all moments calculated by both methods, confirming
consistency between the two implementations. The moments calculated in closed
form using Maple were the same as those calculated using Python-SymPy and SymPy
was able to calculate dozens more than Maple. Those computed using MATLAB were
consistent with the Python results to within $10^{-14}$ relative difference for
all moments calculated. The computational efficiency of the \texttt{decimal}
algorithm enabled even more moments to be calculated, while monitoring roundoff
error by ensuring the odd moments, which are theoretically zero, remained
sufficiently small etc.

\subsection{Moments for steady flows}\label{sec:MomentsSteadyFlows}
\begin{figure}[t]
\centering
\includegraphics[width=0.48\textwidth]{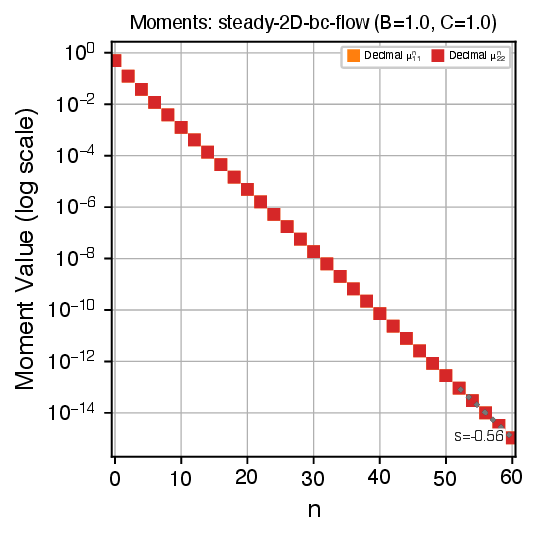}\hfill
\includegraphics[width=0.48\textwidth]{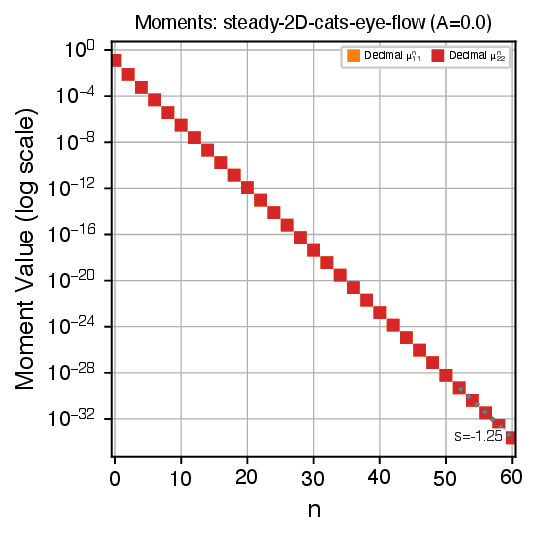}
\\[\smallskipamount]
\includegraphics[width=0.48\textwidth]{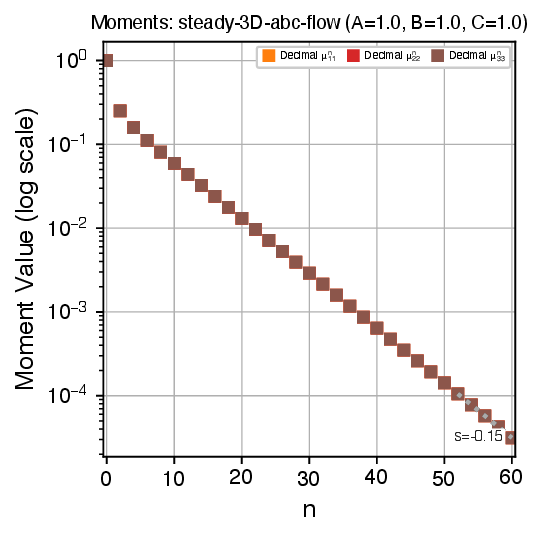}\hfill
\includegraphics[width=0.48\textwidth]{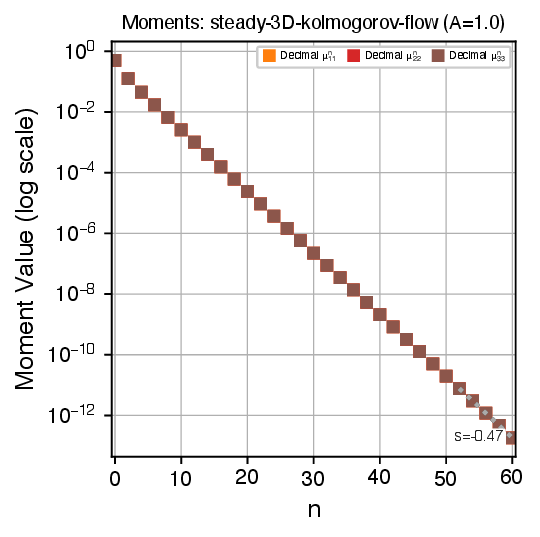}
\caption{ 
\textbf{Spectral measure moments for 2D and 3D steady fluid velocity fields.}
Decimal (floating-point) computation of spectral measure moments $\mu^{2n}$ in
semi-log scale for (top row, left to right) steady 2D BC-flow and steady 2D
cats-eye flow; (bottom row) steady 3D ABC-flow and steady 3D Kolmogorov flow.
All flows are computed with unit amplitude parameters. Consistent with the
symmetry of the flows for the unit amplitude parameters, the moments satisfy
$\mu^{2n}_{jj}= \mu^{2n}_{kk}$, $j,k=1,\ldots,d$. Consistent with the
compactness of the operator $M$ for the steady flows, the moments decrease
exponentially with order. The asymptotic behavior of the moments is indicated by
a linear fit to
the last five moments, shown as a dashed line in each panel with slope (top row,
left to right) $s=-0.56$ and $s=-1.25$ and (bottom row, left to right) $s=-0.15$
and $s=-0.47$ corresponding to the decay rate $\mu^{2n}\sim10^{2sn}$.} 
\label{fig:moments_decimal_steady_flows} 
\end{figure}

In Figure \ref{fig:moments_decimal_steady_flows} we display $2n=60$ computed
diagonal components of the spectral measure moments, $\mu^{2n}_{kk}$,
$k=1,\ldots,d$, using the \texttt{decimal} implementation for 2D steady BC-flow,
2D steady cats-eye flow, 3D steady ABC-flow, and 3D steady Kolmogorov flow.
Since the operator $M$ is compact for steady flows, the moments decrease
exponentially with order, with exponential decay rate $s$ shown. The moments are
plotted in semi-log scale to illustrate their exponential decay $\mu^{2n}_{kk}
\sim 10^{2sn}$. The exponential decay rate $s$ is shown in each panel, which is
determined by a linear fit to the logarithm of the last five moments computed.  
Consistent with the flow symmetries for the unit amplitude parameters, the
moments satisfy $\mu^{2n}_{jj}= \mu^{2n}_{kk}$, $j,k=1,\ldots,d$.

The computational cost of the iterative moment method for 2D steady BC-flow is
determined by the growth of the Fourier mode count at each step. BC-flow has
velocity components $u_1 = C\cos(y)$ and $u_2 = B\cos(x)$, each having $K_v = 2$
Fourier modes after conversion to complex exponential form. Writing $b^j_n = A^n
g_j$ for $A = (-\Delta)^{-1}D_t$ and $g_j = (-\Delta)^{-1}u_j$, the iterates are
initialized with $b^j_0 = g_j$, which has $N_0 = 2$ nonzero Fourier coefficients
(both with eigenvalue $|\veck|^{-2}=1$). Let $N_n$ denote the number of nonzero
Fourier coefficients of the function $b^j_n = A b^j_{n-1}$ at step $n$; since
$(-\Delta)^{-1}$ acts diagonally on Fourier modes and does not change the mode
count, $N_n$ is also the number of nonzero coefficients of $c_n^j = D_t
b^j_{n-1}$. The Janus \texttt{heuristic} procedure, which uses exact symbolic
arithmetic, confirms the closed-form growth law
\begin{align}\label{eq:BCflow_mode_count}
  N_n = n(n+3), \qquad n \ge 1.
\end{align}
%
This quadratic growth law, $N_n = O(n^2)$, arises from the separability of
BC-flow: $u_1$ depends only on $y$ and $u_2$ only on $x$, so each $D_t$
application shifts modes $(i,j)$ by $(\pm1, 0)$ or $(0, \pm1)$, populating an
expanding diamond-shaped region in Fourier space after cancellations.

At each step $n$, three operations each cost $O(N_n) = O(n^2)$: (i)
the material-derivative $c_n^j = D_t b^j_{n-1}$, a Fourier-space
convolution of $N_{n-1}$ modes with $K_v = 2$ velocity modes; (ii) the
pointwise inverse Laplacian $d_n^j = (-\Delta)^{-1}c_n^j$, which divides each
mode coefficient by $|\veck|^2$; and (iii) the inner products computing
$\mu^{2n}_{jk}$ and verifying $\mu^{2n-1}_{jk} = 0$.
The total cost of step $n$ is therefore $O(n^2)$.

The \texttt{decimal} and \texttt{heuristic} procedures share this loop structure
but differ in arithmetic. To mitigate roundoff error, in the \texttt{decimal}
procedure all operations use multi-precision floating-point (\texttt{evalf} with
50 significant digits by default), and Fourier modes with coefficients below
$10^{-30}$ are pruned after each step; because each coefficient is a
fixed-width float whose precision does not grow with $n$, every arithmetic
operation costs $O(1)$, and the $O(n^2)$ such operations at step $n$ therefore
contribute $O(n^2)$ total work. In the \texttt{heuristic}
procedure all arithmetic is exact rational (SymPy expressions); no mode pruning
occurs, but rational coefficient denominators grow rapidly with $n$ (the
denominators of $\mu^{26}$, for example, exceed $10^{120}$), so each rational
multiply costs $O(L)$ or $O(L\log L)$ where $L = O(n^2)$ is the bit-length,
making the heuristic cost per step $O(n^4 \log n)$---paying a significant
computational cost for exact symbolic arithmetic. 

Both procedures execute $M$ iteration steps to compute the $M+1$ even moments
$\mu^0, \mu^2, \ldots, \mu^{2M}$ (the mass $\mu^0$ is computed before the
main loop; each loop step $k=1,\ldots,M$ computes $\mu^{2k}$ and verifies
that $\mu^{2k-1}=0$). The total floating-point work for the \texttt{decimal}
procedure is therefore $\sum_{k=1}^{M} O(k^2) = O(M^3)$,
the same asymptotic scaling as eigendecomposition of an $M\times M$ real
symmetric matrix. The from-scratch cost to compute a single even moment
$\mu^{2n}$ is $O(n^3)$; the odd moment $\mu^{2n-1}$ is a byproduct of the
same step at no additional asymptotic cost.

The cases $B=0$ or $C=0$ are degenerate, reducing BC-flow to a unidirectional
shear flow in the $x$- or $y$-direction, respectively. When $C=0$, the velocity
simplifies to $u_1=0$, $u_2=B\cos(x)$, and the material derivative satisfies
$D_t g_j = 0$ identically (since $\partial_y\cos(x)=0$), so all moments
$\mu^{2n}_{jk}=0$ for $n\geq 1$, consistent with the analysis of shear flow in
Section~\ref{sec:Moments_shear_flow}. The case $B=0$ is analogous.

The complexity analysis for 2D steady cat's eye flow with amplitude parameter
$A=0$ is structurally identical, using the same \texttt{steady\_2D} module and
the same three-operation loop. The key difference is that each velocity
component has $K_v = 4$ Fourier modes (versus $K_v = 2$ for BC-flow), since $u_1
= -\sin(x)\cos(y)$ and $u_2 = \cos(x)\sin(y)$ each expand into four complex
exponential terms. The initial iterates $b^j_0 = (-\Delta)^{-1}u_j$ therefore
also have $N_0 = 4$ modes. Direct computation confirms that the mode count $N_n$
of $b^j_n = A b^j_{n-1}$ satisfies
\begin{align}\label{eq:catseye_mode_count}
  N_n = 
  \begin{cases} 
    n(n+1) & n \text{ odd}, \\ n(n+2) & n \text{ even}, 
  \end{cases}
  \qquad n \ge 1.
\end{align} 
%
These values are strictly smaller than the corresponding BC-flow counts $N_n =
n(n+3)$ at every $n \ge 1$, despite the larger $K_v$, because the non-separable
structure of cats-eye flow produces more algebraic cancellations after
zero-stripping.  Both odd- and even-$n$ mode counts are $O(n^2)$, so the
per-step cost is $O(n^2)$, the from-scratch cost of $\mu^{2n}$ is $O(n^3)$, and
the total cost of all $M+1$ even moments is $O(M^3)$ --- identical asymptotic
complexity to BC-flow.

The amplitude $A=1$ cats-eye flow is a degenerate case leading to a shear flow
with streamlines directed along the diagonal $y=x$. The velocity simplifies to
$u_1 = u_2 = \sin(y-x)$, which has only two Fourier modes $(\pm1,\mp1)$, and the
material derivative satisfies $D_t g_j = 0$ identically, so all moments
$\mu^{2n} = 0$ for $n \ge 1$, consistent with the analysis of shear flow in
\secref{sec:Moments_shear_flow}.

For $A\in(0,1)$, the velocity components~\eqref{eq:cat_eye_flow} retain $K_v=4$
Fourier modes and the same three-operation loop applies. Direct computation
confirms that $N_n$ grows as $O(n^2)$ for all $A\in(0,1)$, with mode counts
independent of the specific value of $A$ within this interval. The per-step cost
is therefore $O(n^2)$, the from-scratch cost of $\mu^{2n}$ is $O(n^3)$, and the
total cost of all $M+1$ even moments is $O(M^3)$, the same asymptotic scaling as
the $A=0$ case.

For 3D steady ABC-flow the same three-operation loop is executed by the Python
module \texttt{steady\_3D}. The velocity components $u_1 = A\sin(z)+C\cos(y)$,
$u_2 = B\sin(x)+A\cos(z)$, $u_3 = C\sin(y)+B\cos(x)$ each have $K_v = 4$ Fourier
modes, giving $N_0 = 4$ initial modes per iterate $b^j_0 = (-\Delta)^{-1}u_j$.
Compared to the module \texttt{steady\_2D}, $D_t$ now includes the third
velocity component $u_3\partial_z$, and the inverse Laplacian eigenvalue is
$|\veck|^2 = k_1^2+k_2^2+k_3^2$ for mode $\veck = (k_1,k_2,k_3)$. Direct
computation using the Janus software confirms that mode counts grow faster than
$O(n^2)$; a log-log fit to the first twenty computed values
yields $N_n \approx 5.3\,n^{2.34}$, reflecting the non-separable
three-dimensional mode coupling. Since each step of the \texttt{decimal}
procedure costs $O(N_n)$, the total cost to compute $M+1$ even moments is
$\sum_{k=1}^{M}O(k^{2.34}) = O(M^{3.34})$, modestly above the $O(M^3)$ scaling
of the 2D flows.

For 3D steady Kolmogorov flow the same \texttt{steady\_3D} module and
three-operation loop apply, but each velocity component $u_1 = \sin(z)$, $u_2 =
\sin(x)$, $u_3 = \sin(y)$ has $K_v = 2$ Fourier modes after conversion to
exponential form, giving $N_0 = 2$ initial modes per iterate $b^j_0 =
(-\Delta)^{-1}u_j$ --- half the initial count of ABC-flow. A log-log fit to the
first twenty computed values
yields $N_n \approx 2.5\,n^{2.58}$, giving a total cost of
$\sum_{k=1}^{M}O(k^{2.58}) = O(M^{3.58})$. Although the growth exponent (2.58)
exceeds that of ABC-flow (2.34), the smaller prefactor (2.5 versus 5.3) means
that mode counts for Kolmogorov flow are strictly smaller than their ABC-flow
counterparts for all computed $n \le 20$.

\begin{figure}[t]
\centering
\includegraphics[width=0.48\textwidth]{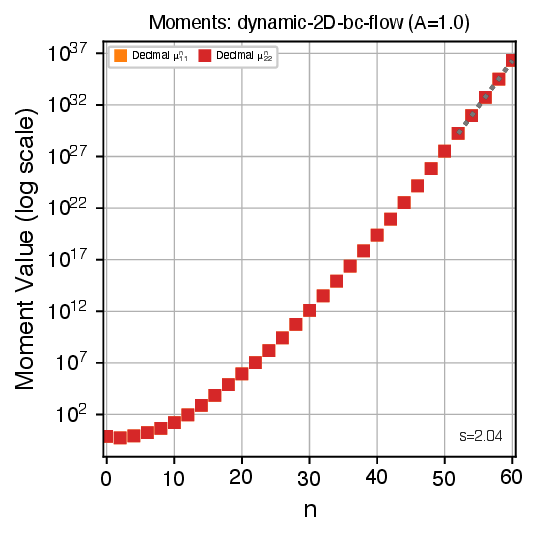}\hfill
\includegraphics[width=0.48\textwidth]{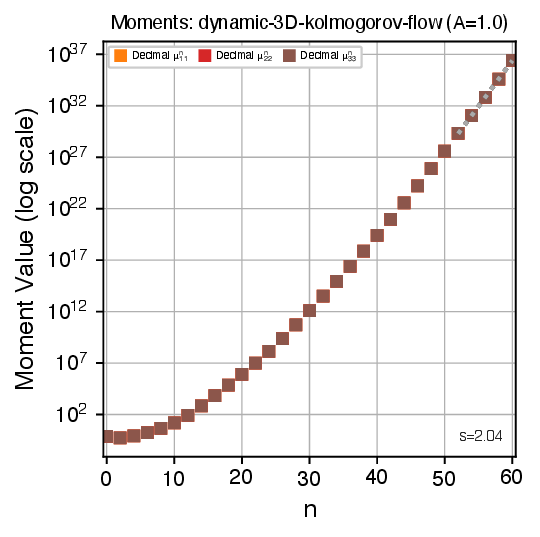}
\caption{  
\textbf{Spectral measure moments for 2D and 3D space-time periodic fluid velocity fields.}
Decimal (floating-point) computation of spectral measure moments $\mu^{2n}$ in
semi-log scale for (left) 2D space-time periodic BC-flow and (right) 3D
space-time periodic Kolmogorov flow. All flows are computed with unit amplitude
parameters. Consistent with the symmetry of the flows for the unit amplitude
parameters, the moments satisfy $\mu^{2n}_{jj}= \mu^{2n}_{kk}$,
$j,k=1,\ldots,d$. Consistent with the unboundedness of the operator $M$ for the
space-time periodic flows, the moments increase exponentially with order. The
asymptotic behavior of the moments is indicated by a linear fit to the last five
moments, shown as a dashed line in each panel with slope $s=2.04$ for both flows
corresponding to the growth rate $\mu^{2n}\sim10^{2sn}$.}
\label{fig:moments_decimal_dynamic_flows}
\end{figure}  

\subsection{Moments for dynamic flows}\label{sec:MomentsDynamicFlows}
Figure~\ref{fig:moments_decimal_dynamic_flows} displays $2n=60$ \texttt{decimal}
moments for 2D space-time periodic BC-flow and 3D space-time periodic Kolmogorov
flow. In contrast to the steady flows, the moments increase exponentially with
order, since the operator $M$ is unbounded for space-time periodic flows; the
growth rate $s$ (semi-log scale, $\mu^{2n}_{kk}\sim 10^{2sn}$, determined by a
linear fit to the last five moments) is shown in each panel. The exponential
growth of the moments for the space-time periodic flows is associated with the
presence of the time derivative $\partial_t$ in the material derivative $D_t$.
As the moment number $n$ increases, the time derivative Fourier coefficient
$\ell$ increases. The Fourier coefficients $\veck$ associated with the gradient
operator $\bnabla$ also increase but are offset by the greater decrease in
Fourier coefficients $|\veck|^{-2}$ associated with the inverse Laplacian
$(-\Delta)^{-1}$. While all terms are balanced for the steady flows in this way,
the unchecked increase in $\ell$ for the dynamic flows leads to the exponential
growth of the moments. 

The computational complexity of the \texttt{dynamic\_2D} module for the 2D
space-time periodic BC-flow differs from the steady case in two respects. First,
the Fourier mode index is now a triple $(i,j,\ell)\in\mathbb{Z}^3$, where $\ell$
is the temporal frequency, so each iterate $b^j_n$ is a function of $(x,y,t)$.
Second, because the velocity contains both $\cos(y)$ and $\cos(t)\sin(y)$ in
$u_1$ (and analogously in $u_2$), each component has $K_v = 6$ Fourier modes
after conversion to exponential form, giving $N_0 = 6$ initial modes. A log-log
fit to the first twenty computed values
yields $N_n \approx 14.5\,n^{2.59}$, giving a total cost of
$\sum_{k=1}^{M}O(k^{2.59}) = O(M^{3.59})$ for the \texttt{decimal} procedure.
Mode counts are substantially larger than for steady 2D BC-flow at every $n$
(by a factor ranging from 5 at $n=1$ to over 80 at $n=20$), reflecting the
additional temporal frequency dimension.

The computational complexity of the \texttt{dynamic\_3D} module for the 3D
space-time periodic Kolmogorov flow combines the mode-count effects of both the
additional spatial dimension and the temporal dimension. The Fourier mode index
is now a 4-tuple $(i,j,k,\ell)\in\mathbb{Z}^4$, where $\ell$ denotes the
temporal frequency, so each iterate $b^j_n$ is a function of $(x,y,z,t)$.
Because each velocity component contains both $\sin(z)$ and $\cos(t)\cos(z)$ in
$u_1$ (and cyclically in $u_2$ and $u_3$), each component has $K_v = 6$ Fourier
modes after conversion to exponential form, giving $N_0 = 6$ initial modes.
Unlike the 2D dynamic case, the $D_t$ convolution now involves three velocity
components (rather than two), which drives faster mode-count growth. A log-log
fit to the first twelve computed values
yields $N_n \approx 12.1\,n^{3.43}$, giving a total cost of
$\sum_{k=1}^{M}O(k^{3.43}) = O(M^{4.43})$ for the \texttt{decimal} procedure.
Mode counts are substantially larger than for steady 3D Kolmogorov flow at every
$n$ (by a factor ranging from 5 at $n=1$ to over 60 at $n=12$), reflecting the
additional temporal frequency dimension.

\section{Numerical implementation of Pad\'{e} approximants}
\label{sec:numerical_implementation_pade}

Given the large number moments $\mu^{2n}$ obtainable, exactly in closed form,
from the iterative moment method discussed in Appendix
\ref{sec:moment_calculations_numerical}, it might seem that Pad\'{e}
approximants for $\Dg^*_{kk}$ of any order could be attainable. However, even in
the absence of rounding errors on a computer, the theoretical treatment of
Pad\'{e} approximants is subject to the appearance of seemingly spurious
pole-zero pairs or “Froissart doublets” in arbitrary locations that prevent
pointwise convergence \cite{Gonnet:2013:55:1:101:110853236}. Such anomalies
become common in the presence of rounding errors or other forms of noise
\cite{Gonnet:2013:55:1:101:110853236}. Hence, computing the Pad\'{e}
approximants directly in terms of the determinants \cite{Baker:1996:Book:Pade}
is essentially ill-posed and becomes unreliable for large polynomial degree.

Pad\'{e} approximation uses information about the series of Stieltjes $f(z)$ in
\eqref{eq:Series_Stieltjes} at a single point $z=0$ in the complex plane to gain
information about regions of the complex plane away from $z=0$ --- similar to
analytic continuation --- and becomes more ill-posed further away from $z=0$.
This is a fundamental limitation of using Pad\'{e} approximant bounds to study
the effective diffusivity in the advection dominated regime, where
$\varepsilon\ll1$, i.e., $z\gg1$. Despite this, we will show upper and lower
Pad\'{e} bounds for $\Dg^*_{kk}$ approach the same value for fairly small values
of $\varepsilon$. Moreover, for the 2D steady BC-cell-flow and cat's eye
cell-flow, Pad\'{e} approximants capture the known asymptotic behavior
$\Dg^*_{kk}\sim\varepsilon^{1/2}$ for
$\varepsilon\ll1$~\cite{Fannjiang:1994:SIAM_JAM:333}.

\begin{figure}[t]
  \centering
  \includegraphics[width=0.24\textwidth]{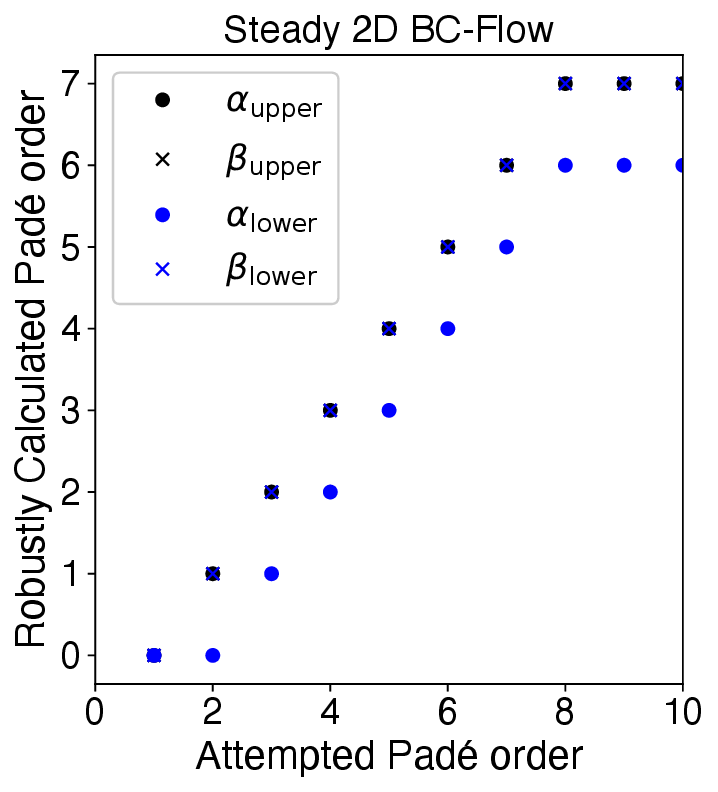}%
  \hfill
  \includegraphics[width=0.24\textwidth]{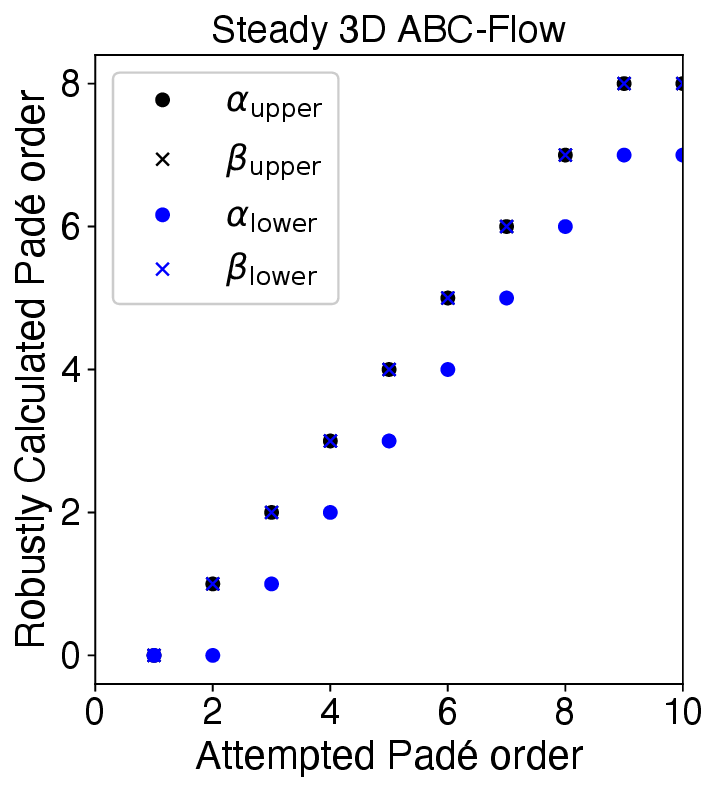}%
  \hfill
  \includegraphics[width=0.24\textwidth]{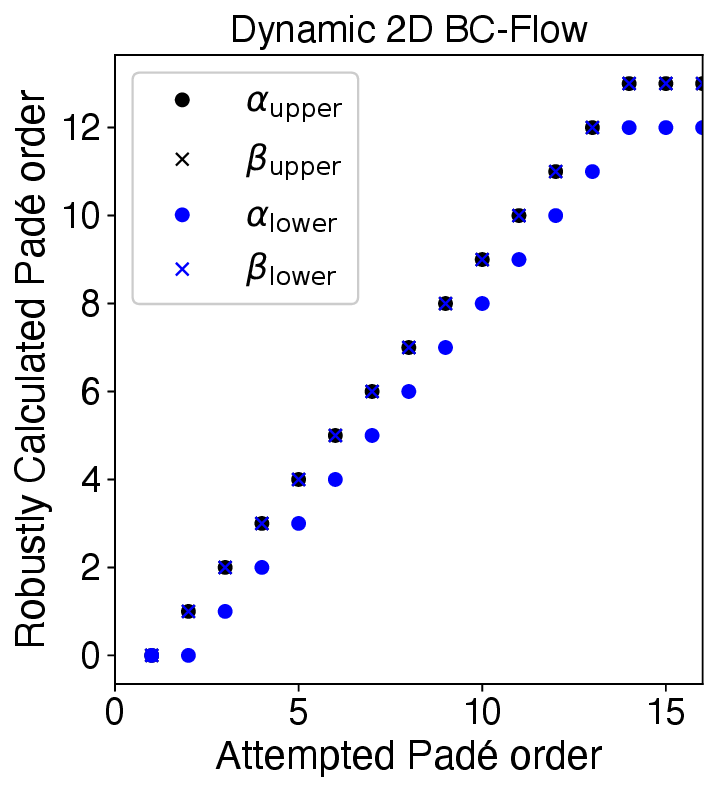}%
  \hfill
  \includegraphics[width=0.24\textwidth]{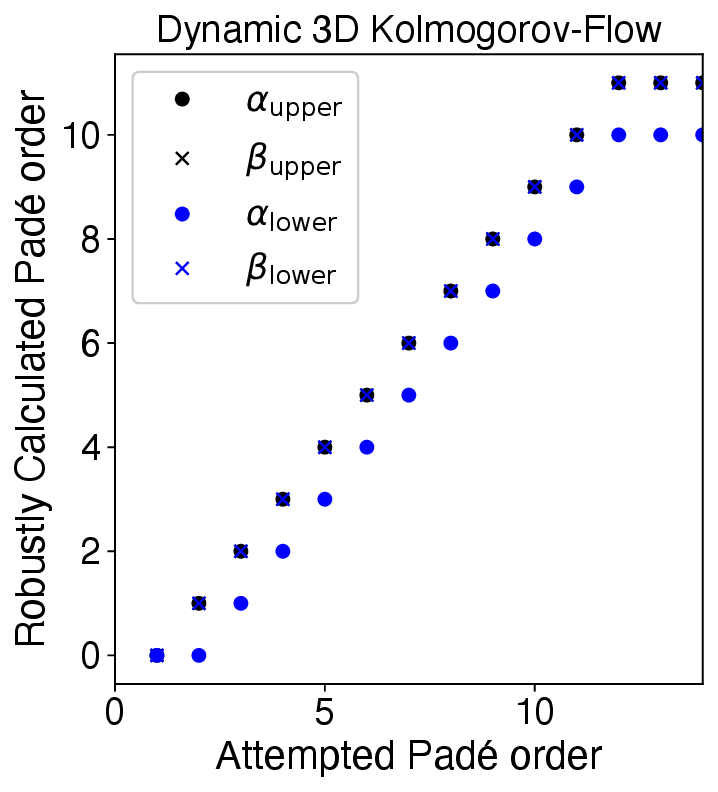}
  \caption{\textbf{Pad\'{e} bound orders for effective diffusivity.} Attempted
    and actual numerator and denominator polynomial orders for $[N/N]$ upper and
    $[N-1/N]$ lower bounds for various values of $N$. These bounds for Stieltjes
    functions associated with the effective diffusivity $\Dg^*_{kk}$ are for
    (from left to right) steady 2D BC-flow, steady 3D ABC-flow, dynamic 2D
    BC-flow, and dynamic 3D Kolmogorov flow, with fluid velocity fields given in
    equations~\eqref{eq:BC-flow}, \eqref{eq:ABC-flow},
    \eqref{eq:space_time_periodic_u}, and~\eqref{eq:space_time_periodic_3D_u},
    respectively. The actual numerator and denominator polynomial orders
    returned for the attempted $[N/N]$ upper and $[N-1/N]$ lower bounds are
    $[\alpha_{upper}/\beta_{upper}]$ and $[\alpha_{lower}/\beta_{lower}]$, with
    the leveling off of polynomial orders showing the effect of the automatic
    truncation.}
  \label{fig:Pade_orders}
\end{figure}

Such issues associated with numerical computation of Pad\'{e} approximants are
addressed in \cite{Gonnet:2013:55:1:101:110853236}, where a MATLAB function
\texttt{padeapprox} was released, which is freely available as part of
\texttt{Chebfun} \cite{ChebfunTeam:padeapprox}, which provides robust Pad\'{e}
approximants via singular value decomposition (SVD)
\cite{Gonnet:2013:55:1:101:110853236}. This numerical method does not enable
Pad\'{e} approximants of an arbitrary order to be robustly computed, but instead
truncates the Pad\'{e} approximants to an order that ensures numerical
stability, hence numerically dependable results. Incorporating additional
moments to achieve higher order Pad\'{e} approximants simply results in the same
truncated Pad\'{e} order, as shown in Figure \ref{fig:Pade_orders}. This
automatic truncation is accomplished by introducing a tolerance \texttt{tol}
that is used to zero out singular values less than \texttt{tol}. For most
purposes involving problems perturbed just by rounding errors, \texttt{tol}
$=10^{-14}$  is a reasonable value \cite{Gonnet:2013:55:1:101:110853236}. This
regularization can be circumvented by setting \texttt{tol}=0, bypassing the
numerical stability enabled by a nonzero value of \texttt{tol}
\cite{Gonnet:2013:55:1:101:110853236}. We reimplemented the \texttt{padeapprox}
function in Python, which is available in the Janus GitHub repository, and used
it to compute Pad\'{e} approximants for the effective diffusivity $\Dg^*_{kk}$
for the various flows. Extensive unit testing ensures that the MATLAB and Python
implementations produce consistent results with very low relative differences.
Inputs to the \texttt{padeapprox} function include the desired degrees of the
numerator/denominator polynomials, in our case $N/N$ and $N-1/N$, a vector of
the moments $\mu^{2n}$, and \texttt{tol}. Numerator and denominator polynomials
of actual degree $\alpha$ and $\beta$ computed are returned, as shown in Figure
\ref{fig:Pade_orders}.    

\begin{figure}[htbp]
  \centering
  \includegraphics[width=0.24\textwidth]{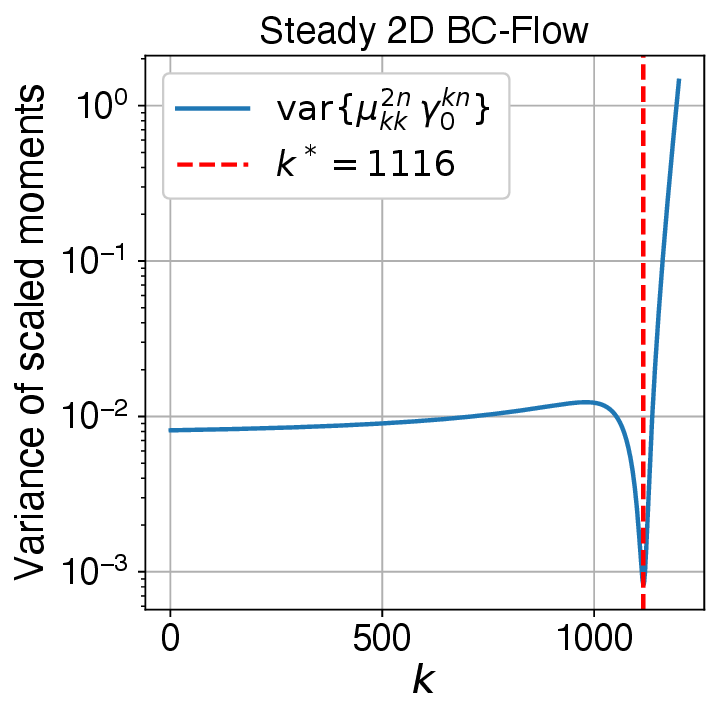}%
  \hfill
  \includegraphics[width=0.24\textwidth]{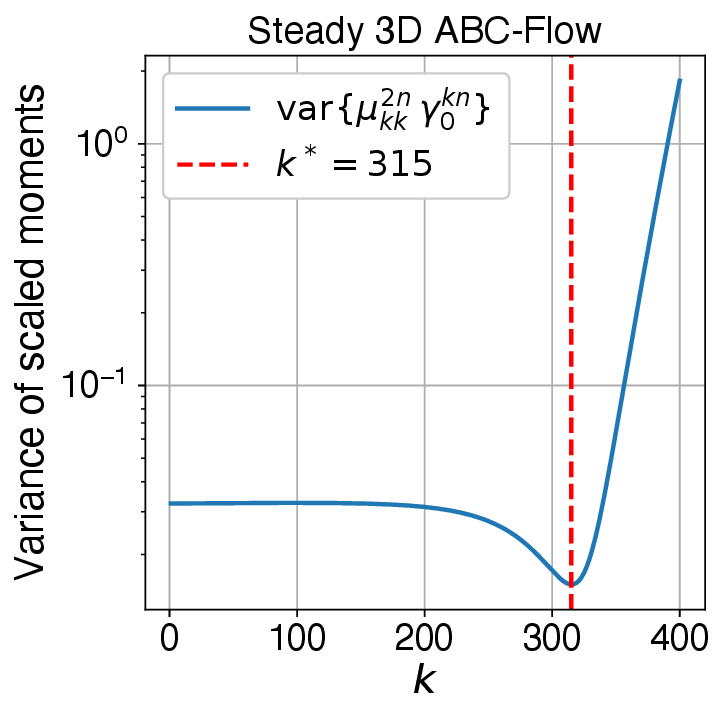}%
  \hfill
  \includegraphics[width=0.24\textwidth]{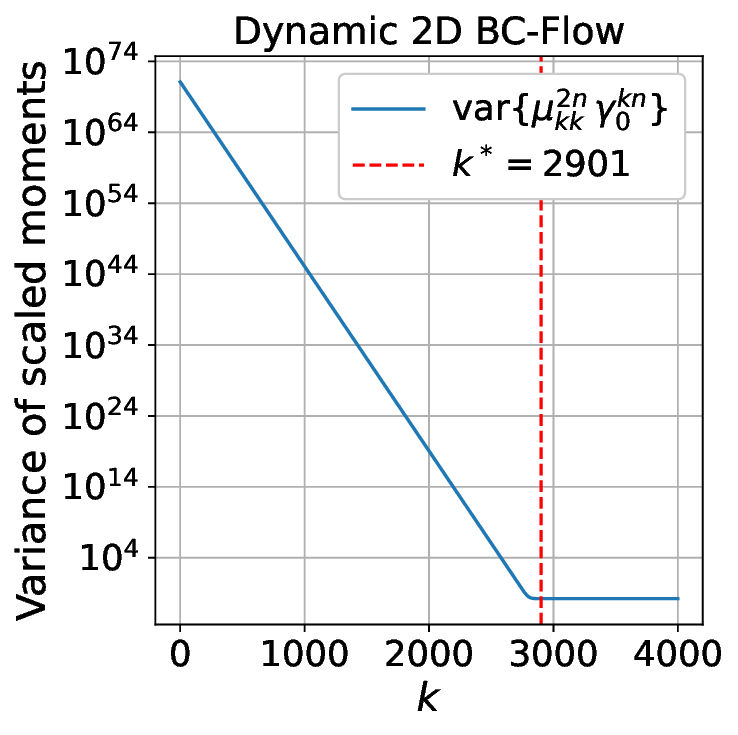}%
  \hfill
  \includegraphics[width=0.24\textwidth]{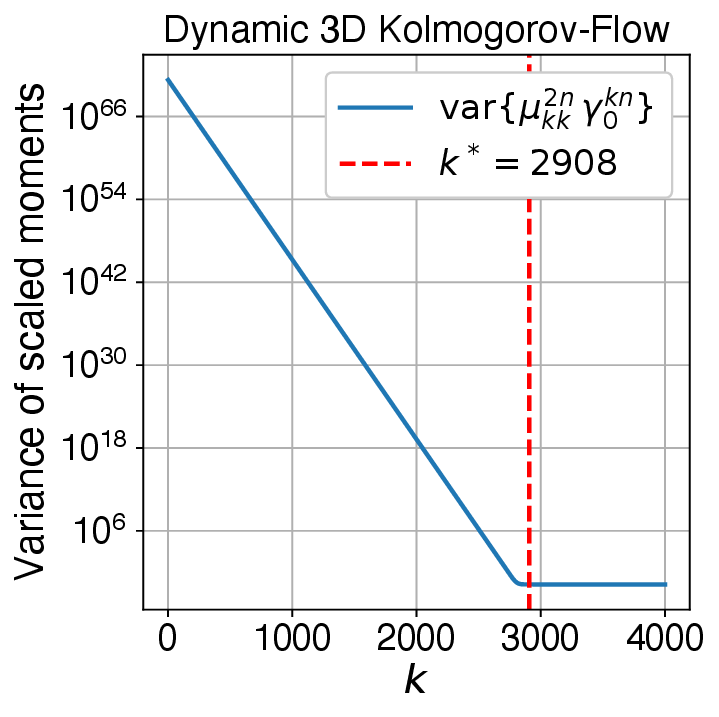}\\[4pt]
  \includegraphics[width=0.24\textwidth]{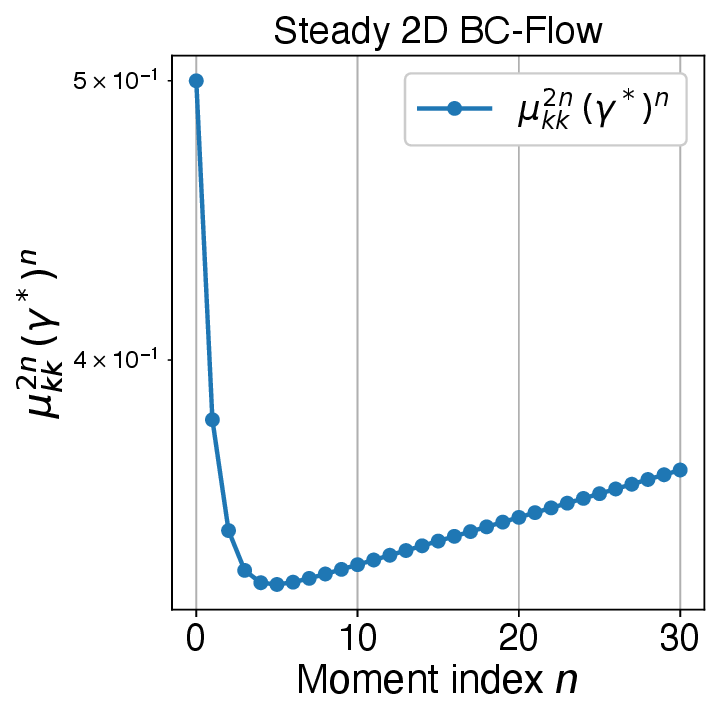}%
  \hfill
  \includegraphics[width=0.24\textwidth]{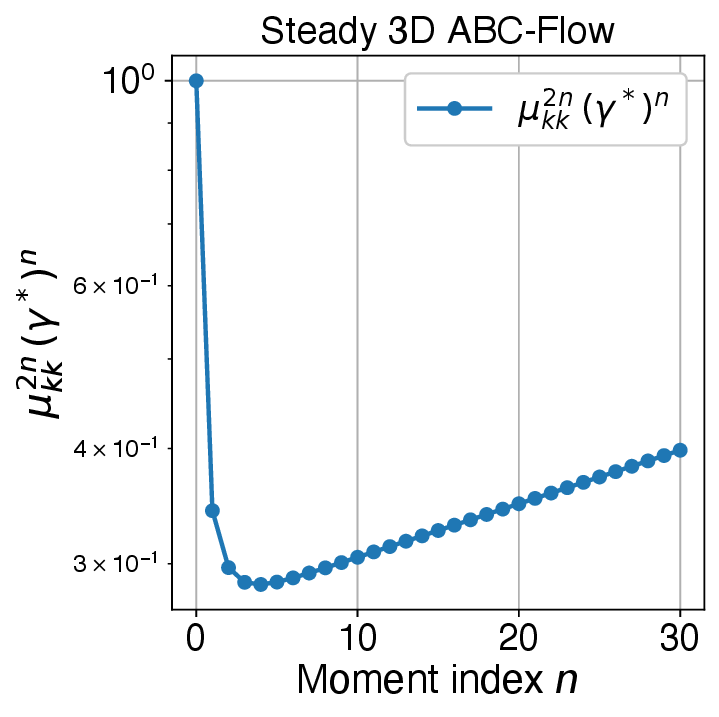}%
  \hfill
  \includegraphics[width=0.24\textwidth]{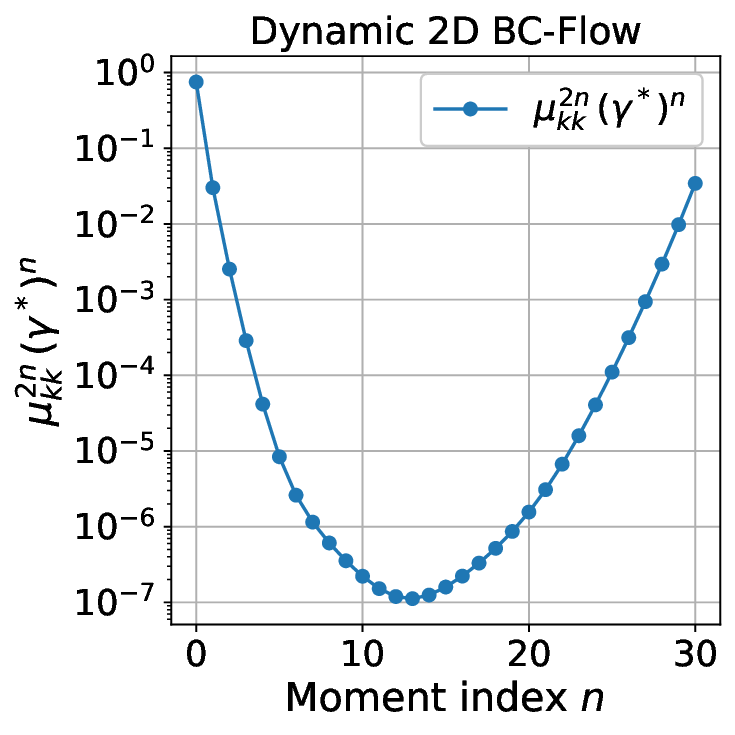}%
  \hfill
  \includegraphics[width=0.24\textwidth]{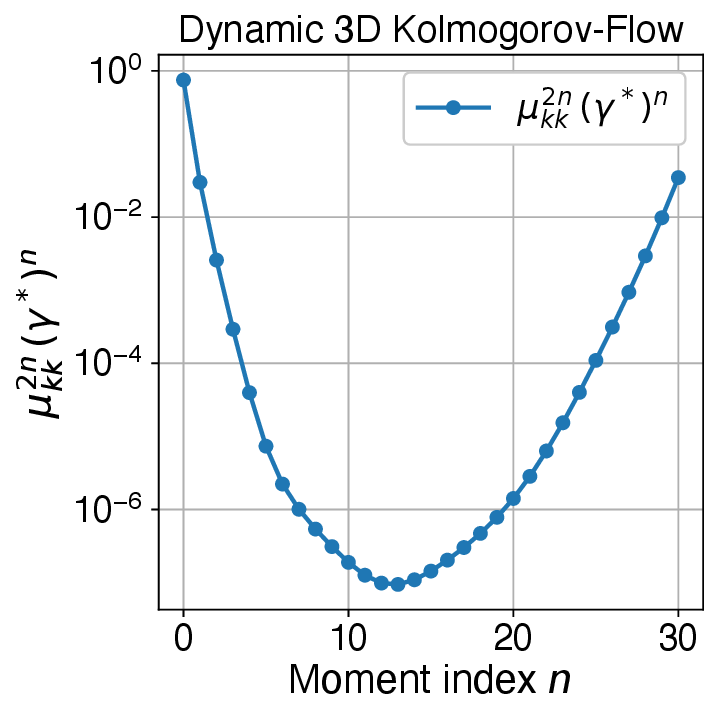}
  \caption{\textbf{Moment scaling diagnostics.}
    Scaling diagnostics for the spectral measure moments $\mu^{2n}_{kk}$,
    $n=0,1,\ldots,60$, computed via the decimal procedure for steady 2D BC-flow,
    steady 3D ABC-flow, dynamic 2D BC-flow, and dynamic 3D Kolmogorov flow, with
    fluid velocity fields given in equations~\eqref{eq:BC-flow},
    \eqref{eq:ABC-flow}, \eqref{eq:space_time_periodic_u},
    and~\eqref{eq:space_time_periodic_3D_u}, respectively. Top row: variance
    $\operatorname{var}\{\mu^{2n} \gamma_0^{kn}\}$ of the scaled moment sequence
    as a function of scaling iteration $k$ with a vertical dashed line marking
    the optimal iteration $k^*$ that minimizes the variance. Bottom row: the
    optimally scaled moments $\mu^{2n}(\gamma^*)^n$ as a function of moment
    index $n$, where $\gamma^* = \gamma_0^k$ for $k=k^*$.}
  \label{fig:scaling_diagnostics}
\end{figure}

The numerical computation of Pad\'{e} approximants can be further stabilized by
scaling the moments $\mu^{2n}$ in equation \eqref{eq:Series_Stieltjes}, choosing
$\gamma>0$ so that the scaled moments $(\mu^{2n} \gamma^n)$ in the series
$f(z)=\sum_n (-1)^n(\mu^{2n} \gamma^n)(z/\gamma)^n$ have roughly comparable
orders of magnitude --- neither decaying nor growing at a rapid rate --- for
which the algorithm \texttt{padeapprox} is most effective
\cite{Gonnet:2013:55:1:101:110853236}. We determine $\gamma$ by searching over a
discrete sequence $\gamma = \gamma_0^k$, $k = 1, 2, \ldots, k_{\max}$, and
selecting the value $k^*$ that minimizes the variance
$\operatorname{var}\{\mu^{2n}\gamma_0^{kn}\}$ of the scaled moment sequence,
$n=0, 1, \ldots$, as shown in Figure~\ref{fig:scaling_diagnostics}. The search
for $k^*$ is quite efficient, utilizing NumPy broadcasting to compute the
variance for all values of $k$ simultaneously. The behavior of this variance as
a function of $k$ differs qualitatively between steady and dynamic flows, as
shown in Figure~\ref{fig:scaling_diagnostics}.

For steady flows the unscaled moments $\mu^{2n}$ decay exponentially with order
$n$, as shown in Figure~\ref{fig:moments_decimal_steady_flows}, so we use
$\gamma_0 > 1$ to increase the magnitude of higher-order terms. At small $k$ the
high-order scaled moments remain small relative to the low-order ones,
yielding high variance. At large $k$ the high-order moments are amplified
past the low-order ones, again yielding high variance, and there is a clear
interior minimum at a point $k^*$, resembling the minimum of a quadratic well,
as shown in Figure~\ref{fig:scaling_diagnostics} for steady 2D BC-flow and 3D
ABC-flow. This phenomenon is also present for the other steady flows we studied,
including 2D cat's eye flow and 3D Kolmogorov flow, but is not shown here for
brevity. The rescaling is quite effective for steady flows, reducing the
variation in the moments from the many orders of magnitude shown in
Figure~\ref{fig:moments_decimal_steady_flows} to around one order of magnitude,
as shown in Figure~\ref{fig:scaling_diagnostics}.

For dynamic flows the unscaled moments $\mu^{2n}$ grow exponentially with order
$n$, as shown in Figure~\ref{fig:moments_decimal_dynamic_flows}, so we instead
use $\gamma_0 < 1$. Multiplying $\mu^{2n}$ by $\gamma^n=\gamma_0^{kn}$ with
$\gamma_0 < 1$ monotonically suppresses all the moments toward zero as $k$
increases. The variance decreases monotonically and asymptotes to zero rather
than forming a minimum in a pronounced well, as shown in
Figure~\ref{fig:scaling_diagnostics}. The optimal $k^*$ is nonetheless
well-defined as the minimizer of the variance curve, occurring before
$k_{\max}$. The rescaling for dynamic flows was effective in the sense that it
reduced the variation in the moments from around 36 orders of magnitude to
around 6 orders of magnitude, as shown in
Figures~\ref{fig:moments_decimal_dynamic_flows}
and~\ref{fig:scaling_diagnostics}. However, this rescaling is less effective for
dynamic flows than for steady flows, due to the exponential growth of the
moments, with the scaled moments still spanning more than six orders of
magnitude, as shown in Figure~\ref{fig:scaling_diagnostics}. This rescaling
doesn't have as dramatic an effect on the stability of the Pad\'{e} approximants
as it does for steady flows, but still enables more stable approximants to be
computed than without scaling \cite{Gonnet:2013:55:1:101:110853236}.


\section{Bounds for offdiagonal components of the effective diffusivity}
\label{sec:offdiagonal_bounds}

In this section we provide bounds for the offdiagonal components $\Dg^*_{jk}$,
$j\neq k$, of the effective diffusivity tensor. Denote the Stieltjes function
$f(z)$ in \eqref{eq:Pade_bounds} explicitly in terms of the measure by
$f(z;\mu)$. By the linearity of Stieltjes integrals
\cite{Schmudgen:2012:2012942602,Stone:64}, we have $f(z;
\mu+\nu)=f(z;\mu)+f(z;\nu)$ and the moments of the measure $\mu+\nu$ are the
sums of the moments of $\mu$ and $\nu$,
$(\mu+\nu)^n=\int\lambda^n\d(\mu+\nu)(\lambda)=\mu^n+\nu^n$. Recall from Section
\ref{app:Scalar_Fields} that the spectral measure $\mu_{jk}=\mu_{g_j,g_k}$ for
the self-adjoint operator $M$ is defined in terms of the function of bounded
variation $\mu_{jk}(\lambda)=\langle Q(\lambda) g_j,g_k\rangle_{1,2}$, where
$Q(\lambda)$ is the resolution of the identity operator,
$\langle\cdot,\cdot\rangle_{1,2}$ is the $\Hs$-inner-product, and $g_j$ and
$g_k$ are members of the Hilbert space $\Hs$. The polarization identity for
spectral measures follows from the linearity of this functional in each
argument, giving 
\begin{align}
\mu_{g_j,g_k}=\frac{1}{4}\left(
\mu_{g_j+g_k} - \mu_{g_j-g_k} + 
i(\mu_{g_j+ig_k} - \mu_{g_j-ig_k})\right),
\end{align}
where we have denoted $\mu_{\phi}=\mu_{\phi,\phi}$ for brevity.
It follows that 
\begin{align}
f(z;\mu_{g_j,g_k})&= f_r^+(z)-f_r^-(z) + i(f_i^+(z)-f_i^-(z)),
\\\notag
f_r^+(z)&=\frac{1}{4}f(z;\mu_{g_j+g_k}),
\qquad
f_r^-(z)=\frac{1}{4}f(z;\mu_{g_j-g_k}),
\\\notag
f_i^+(z)&=\frac{1}{4}f(z;\mu_{g_j+ig_k}),
\qquad
f_i^-(z)=\frac{1}{4}f(z;\mu_{g_j-ig_k}).
\end{align}

The functions $f_r^+(z)$, $f_r^-(z)$, $f_i^+(z)$, and $f_i^-(z)$ are all
Stieltjes functions associated with \emph{positive} spectral measures for the
self-adjoint operator $M$, so the Pad\'{e} bounds in \eqref{eq:Pade_bounds} hold
for each of these functions. For example 
\begin{align}
[N-1/N]^+_r\le f_r^+(z)\le[N/N]^+_r,
\qquad
[N-1/N]^-_r\le f_r^-(z)\le[N/N]^-_r.
\end{align}
It follows that
\begin{align}\label{eq:Pade_bounds_offdiagonal}
[N-1/N]^+_r-[N/N]^-_r\le f(z;\text{Re}\mu_{jk})\le[N/N]^+_r-[N-1/N]^-_r,
\\\notag
[N-1/N]^+_i-[N/N]^-_i\le f(z;\text{Im}\mu_{jk})\le[N/N]^+_i-[N-1/N]^-_i.    
\end{align}

From 
$\langle g_j+g_k,g_j+g_k\rangle=
\langle g_j,g_j\rangle+
2\text{Re}\langle g_j,g_k\rangle+
\langle g_k,g_k\rangle$ 
and 
$\langle g_j-g_k,g_j-g_k\rangle=
\langle g_j,g_j\rangle-
2\text{Re}\langle g_j,g_k\rangle+
\langle g_k,g_k\rangle$ we have
\begin{align}
\mu_{g_j+g_k}^n=\mu_{g_j}^n + 2\text{Re}\mu_{g_j,g_k}^n + \mu_{g_k}^n, 
\qquad
\mu_{g_j-g_k}^n=\mu_{g_j}^n - 2\text{Re}\mu_{g_j,g_k}^n + \mu_{g_k}^n.
\end{align}
Similarly, from 
$\langle g_j+ig_k,g_j+ig_k\rangle=
\langle g_j,g_j\rangle-
2\text{Im}\langle g_j,g_k\rangle+
\langle g_k,g_k\rangle$
and
$\langle g_j-ig_k,g_j-ig_k\rangle=
\langle g_j,g_j\rangle+
2\text{Im}\langle g_j,g_k\rangle+
\langle g_k,g_k\rangle$ we have
\begin{align}
\mu_{g_j+ig_k}^n=\mu_{g_j}^n - 2\text{Im}\mu_{g_j,g_k}^n + \mu_{g_k}^n, 
\qquad
\mu_{g_j-ig_k}^n=\mu_{g_j}^n + 2\text{Im}\mu_{g_j,g_k}^n + \mu_{g_k}^n.
\end{align} 

Consequently, the moments of the measures $\mu_{g_j+g_k}$, $\mu_{g_j-g_k}$,
$\mu_{g_j+ig_k}$, and $\mu_{g_j-ig_k}$ are given in terms of the moments
$\mu_{jj}^n$, $\mu_{kk}^n$, and $\mu_{jk}^n$. These moments were already
computed via the iterative moment method described in Appendix
\ref{sec:moment_calculations_numerical} along the way to computing the Pad\'{e}
approximant bounds for the diagonal components $\Dg^*_{jj}$, $j=1,\ldots,d$, of
the effective diffusivity tensor via equation
\eqref{eq:Pade_bounds_diagonal_component}. It follows that Pad\'{e} bounds for
$f(z;\text{Re}\mu_{jk})$ and $f(z;\text{Im}\mu_{jk})$, hence the offdiagonal
components of the symmetric and antisymmetric parts of the effective diffusivity
tensor, $\Sg^*_{jk}$ and $\Ag^*_{jk}$, $j\ne k$, in
\eqref{eq:Integral_Rep_kappa*} can be computed utilizing equations
\eqref{eq:Pade_bounds_offdiagonal} and
\eqref{eq:Pade_bounds_diagonal_component}.

\clearpage
\bibliographystyle{siam}

\bibliography{murphy.bib}

\end{document}